\documentclass[conference]{IEEEtran}
	\usepackage{graphicx}
	\usepackage{balance}
	\usepackage{amssymb, amsmath}
	\usepackage{multirow, rotating, wasysym, url}
	\usepackage[tight]{subfigure}
	\usepackage[ruled,linesnumbered,vlined]{algorithm2e}
	\usepackage{hyperref} 
	\usepackage{color}
    \usepackage{multicol}
    \usepackage{textcomp}
    \usepackage{epstopdf}
    \usepackage{graphicx}
     \usepackage{subfigure}

\graphicspath{{./DrawingFigures/},{./EvaluateFigures/}}

    \newcommand{\nop}[1]{}
	\newcommand{\presec}{\vspace{-0.1in}}
	\newcommand{\postsec}{\vspace{-0.05in}}
	\newcommand{\presub}{\vspace{-0.05in}}
	\newcommand{\postsub}{\vspace{-0.05in}}

	\newcommand{\prefig}{\vspace{-0in}}
	\newcommand{\prefigcaption}{\vspace{-0in}}
	\newcommand{\postfig}{\vspace{-0in}}

		\mathchardef\Gamma="0100 \mathchardef\Delta="0101
\mathchardef\Theta="0102 \mathchardef\Lambda="0103
\mathchardef\Xi="0104 \mathchardef\Pi="0105
\mathchardef\Sigma="0106 \mathchardef\Upsilon="0107
\mathchardef\Phi="0108 \mathchardef\Psi="0109
\mathchardef\Omega="010A

\newcommand{\outline}[1]{}

\usepackage{cite}
\usepackage{xspace}
\usepackage{url}
\usepackage{graphicx}
\usepackage{latexsym}
\usepackage{amssymb}
\usepackage{amsfonts}
\usepackage{psfrag}
\usepackage{wrapfig}
\usepackage{comment}
\usepackage{alltt}
\usepackage{color}

\newtheorem{thm}{Theorem}

\newcommand{\etc}{\emph{etc.}\xspace}

\newcommand{\Comment}[1]{}

        \newcommand{\bbb}{\noindent\textbf}
        \newcommand{\hashname}{square hashing}
	
    \newcommand{\fname}{GSS}
    \newcommand{\Paragraph}[1]{~\vspace*{-0.9\baselineskip}\\{\bf #1}}
	
	\definecolor{greener}{RGB}{0,166,0}
	\definecolor{reder}{RGB}{255,0,0}
	\definecolor{bluer}{RGB}{0,0,255}
	
\newtheorem{Def}{Definition}
\newtheorem{Exp}{Example}

\begin{document}

\title{Fast and Accurate Graph Stream Summarization}

\renewcommand{\textrightarrow}{$\rightarrow$}

\author{
\IEEEauthorblockN{Xiangyang Gou, Lei Zou, Chenxingyu Zhao, Tong Yang}

\IEEEauthorblockA{Peking University, China}
}

\maketitle

\begin{abstract}

A graph stream is a continuous sequence of data items, in which each item indicates an edge, including its two endpoints and edge weight. It forms a dynamic graph that changes with every item in the stream. Graph streams play important roles in cyber security, social networks, cloud troubleshooting systems and other fields. Due to the vast volume and high update speed of graph streams, traditional data structures for graph storage such as the adjacency matrix and the adjacency list are no longer sufficient. However, prior art of graph stream summarization, like CM sketches, gSketches, TCM and gMatrix, either supports limited kinds of queries or suffers from poor accuracy of query results. In this paper, we propose a novel \emph{G}raph \emph{S}tream \emph{S}ketch (GSS for short) to summarize the graph streams, which has the linear space cost ($O(|E|)$, E is the edge set of the graph) and the constant update time complexity ($O(1)$) and supports \emph{all} kinds of queries over graph streams with the controllable errors. Both theoretical analysis and experiment results confirm the superiority of our solution with regard to the time/space complexity and query results' precision compared with the state-of-the-art. 

\end{abstract}


\presec
\section{Introduction}\label{sec:introduction}
\postsec
\presub
\subsection{Background and Motivations}
\postsub
In the era of big data, data streams propose some technique challenges for existing systems. Furthermore, the traditional data stream is modeled as a sequence of \emph{isolated} items, and the connections between the items are rarely considered. However, in many data stream applications, the connections often play important roles in data analysis, such as finding malicious attacks in the network traffic data, mining news spreading paths among the social network. In these cases the data is organized as \emph{graph streams}. A graph stream is an unbounded sequence of items, in which each item is a vector with at least three fields (denoted by $(\langle s, d\rangle, w)$), where $\langle s, d\rangle$ represents an edge between nodes $s$ and $d$, and $w$ is the edge weight. These data items together form a dynamic graph that changes continuously and we call it \emph{streaming graph} for convenience. Below we discuss three examples to demonstrate the usefulness of streaming graph problems.

\nop{
Therefore, processing algorithms of graph streams have been a new and promising trend in many fields \cite{tcm,....}, especially in data bases and networks.
Below we show three use cases.
}

\nop{

For example, the phone calling network and network traffic data, 

Nowadays, big data is often organized as data streams, such as on-line videos, requests and responds, phone calls, messages in social networks \etc
}

\nop{
A graph stream is a common case of data streams.
It is a sequence of data items. Each item can be seen as a vector with at least three fields, and denoted by $(s, d, w)$, where $(s, d)$ represents an edge between node $s$ and node $d$, and $w$ is the weight of this edge.
}

\bbb{Use case 1:} \emph{Network traffic.} 
The network traffic can be seen as a large dynamic graph, where each edge indicates the communication between two IP addresses. With the arrival of packets in the network, the network traffic graph changes rapidly and constantly. In the network traffic graph, various kinds of queries are needed, like performing node queries to find malicious attackers, or subgraph queries to locate certain topology structures in the dynamic networks.

\bbb{Use case 2:} \emph{Social networks.}
In a social network, the interactions among the users can form a graph. The edges between different nodes 
may be weighted by the frequencies of interactions. In such a graph, queries like finding the potential friends of a user and tracking the spreading path of a piece of news are often needed.

\bbb{Use case 3:} \emph{Troubleshooting in data centers.}
Cloud systems may need to analyze communication log stream to perform real time troubleshooting. In this situation the graph stream is the sequence of communication log entries where each entry is a description of a communication from a source machine to a destination machine. In such a graph, we may perform traversal query to find out if massages created by a certain application on a source machine can reach a destination machine, or perform edge query to find the detailed information of a communication log.

These streaming graphs are very large and change fast. For example, in Twitter, there are about $100$ million user login data, with $500$ million tweets posted per day. For another example, in large ISP or data centers\cite{guha2012graph}, there could be millions of packets every second in each link. The large volume and high dynamicity make it hard to store the graph streams efficiently with traditional data structures like adjacency lists or adjacency matrices. In the context of graph streams, there are two requirements for designing a new data structure : (1) the linear space cost (2) the constant update time. To meet these two requirements, we can either apply approximated query data structures for data streams, like the CM sketch\cite{Cormode2004An}, the CU sketch \cite{estan2003new} and other sketches \cite{roy2016augmented,thomas2009efficient}, or use specialized graph summarization techniques such as gSketches \cite{Zhao2011gSketch}, TCM \cite{Tang2016Graph} and gMatrix \cite{Khan2016Query}. However, existing solutions either support limited query types or have poor query accuracy. For example, CM sketches and gSketches fail to answer queries involved with topology like reachability queries, successor queries and so on. Though TCM and gMatrix can support these queries, they have poor accuracy. More details about the related work are given in Section \ref{sec:relatedwork}. In this paper, we design a novel data structure--\emph{G}raph \emph{S}tream \emph{S}ketch (GSS for short), which can support all kinds of queries over streaming graphs with controllable errors in query results. Both theoretical analysis and experiment results show that the 
accuracy of our method outperforms state-of-the-art by orders of magnitudes.    

\nop{
Fortunately, it is not always necessary to give an exact answer in the graph stream based queries. Therefore approximate query data structures with small and controllable errors are good choices for graph stream storage. However, exiting approximate query data structures either support limited query types or have poor accuracy. In this paper we propose \fname, which has small memory usage, fast and constant update speed. It supports many kinds of queries, like edge queries, node queries and reachability queries. Moreover, in all these queries it has much higher accuracy than prior art.
}

\nop{
\subsection{Prior art and Challenges}

To design an approximate query data structure suitable for graph streams, we should overcome the following challenges:
\begin{enumerate}

\item The data structure should use less than $O(|E|)$ memory, where $|E|$ is the number of edges. Obviously, this requirment rules out the ``adjacency matrix'' due to its $O(|V|^2)$ space cost, where $|V|$ is the number of vertices.

\item  The data structure should have $O(1)$ time for updates. The tradtional adjanecy list cannot meet this requirment. 

\item The data structure should have high accuracy. 
\end{enumerate}
\nop{Usually the edges are the basic units of storage and query in the graph stream. The error rate of the edge storage in a graph stream summarization should be much lower than approximate query data structures for common data streams. Because in some topology queries, the accuracy of the result relies on a set of edge queries, error in the query of any edge in the set may lead to a wrong answer. For example, the reachability query relies on paths from the source vertex to the destination vertex, and if the query result of any edge in a path is wrong, the answer of the reachability query may be wrong. Therefore in order to guarantee a proper accuracy for such topology queries, the basic edge query must have a rather high accuracy.

4) The data structure must have resistance to the uneven distribution of node degrees. In the real world graphs, like the social network, the degrees of nodes are usually distributed very unevenly. Most edges are emitted by a few nodes with large degrees. This skewness will lead to disasters in the summarization of graph streams. For example, in adjacency lists this skewness will lead to low query speed, and in TCM it will cause a decrement in accuracy. It is a challenge to make our data structure adapt to the skewness in the distribution of the node degrees.}

\nop{In this part we will introduce the limitations of prior art and the challenges in graph stream summarization.
The traditional data structures to store graphs are adjacency lists and adjacency matrices. However, in the case of graph streams, these data structures are not sufficient. The stream graphs are usually large and sparse, with millions of nodes involved. Building an adjacency matrix may consume several TBs of memory, as the space consumption of the adjacency matrix is $O({n}^{2})$, where $n$ is the number of nodes in the graph. Such a large matrix can only be stored in disk, and access to the disk is time-consuming. It can hardly catch up with the update speed of graph streams. Moreover, as the graph is sparse, most part of the matrix will be 0s. Therefore adjacency matrices are inefficient in terms of both time and space. The space consumption of the adjacency list is linear with the number of edges in the graph. It is much more memory-efficient than adjacency matrices. However, the time consumption of adjacency lists to process an item in the graph stream is $O(d)$, where $d$ is the degree of the source vertex, in other words, the number of edges it emits. In the real world stream graphs, the node degrees are highly skewed, which means most edges are emitted by a few nodes with large degrees. The largest node degrees in a stream graph can be over thousands. Therefore, in most cases processing an item in the graph stream with adjacency lists will be time consuming.
Approximate query data structures are good choices to store graph streams, as they are fast and memory efficient, and in most applications small and controllable errors are acceptable. However, all the prior art has limitations. Classic frequency based approximate query data structures, such as the CM sketch \cite{}, and its variant, which is specialized for graphs, the gSketch\cite{}, can generate synopses for the edges in graph streams. However, they only support limited kinds of queries. TCM \cite{} is a recently proposed data structure to store graph streams. It can summarize both the attributes of edges and the topology of the graph and support most kinds of queries. However, the accuracy of TCM in reachability queries is poor. It has no false negatives, but high false positives. Moreover, when the node degrees in the graph is highly skewed, the accuracy of other queries is also not good.}
There has been several prior works on graph stream summarization. CM sketches and gSketches store each edge in the graph stream as an independent data item. They have $O(|E|)$ memory usage and $O(1)$ update time, but only support queries for the weights of edges. Queries involved with topology of the graph like reachability and successor query are all beyond their capacity. TCM and the gMatrix use hash functions to compress the streaming graph into a smaller graph which is called a \emph{graph sketch}. Then they use adjacency matrix to represent this graph sketch. They store both the edge weights and the topology of the graph. However, as the side length of the matrix has to be controlled to meet the memory usage limitation. The graph sketch it can represent is much more smaller than the streaming graph. The high compression rate makes the accuracy poor. Details about these data structures will be shown in Section \ref{sec:relatedwork}
}
\presub
\subsection{Our Solution}
\postsub
In this paper we propose \fname, which is an approximate query data structure for graph streams with linear memory usage, high update speed, high accuracy and supports all kinds of graph queries and algorithms like \cite{elkin2007streaming,braverman2013hard,feigenbaum2005graph}.
GSS can also be used in exiting distributed graph systems \cite{gonzalez2014graphx,gonzalez2012powergraph,malewicz2010pregel,low2012distributed}
\nop{Like TCM, \fname{} stores both the attributes of the edges and the topology of the graph. Therefore, it supports many kinds of queries like edge queries, sub-graph queries and traversal queries.
The basic data structure of \fname{} is composed of a matrix and several hash functions. When inserting an edge $(s, d, v)$, it computes a fingerprint pair $(f(s), f(d))$ and a hash address pair $(h(s), h(d))$, and finds $r$ buckets in the matrix, which is located with a novel technique called \hashname. Then it checks these buckets. If pair $(f(s), f(d))$ is already stored in one of the buckets, it updates the values in this bucket with $v$. Otherwise, it finds an empty one to store the fingerprint pair and the attributes. If there are no empty location for it, this edge is stored in a buffer which is a small adjacency list.
The adjacency list in the buffer area stores very few edges, usually less than $0.01\%$ of the data set, and when inserting edges we seldom need to access the buffer area. Therefore, \fname{} is much faster than a simple linked lists. The number of memory accesses are nearly constant. On the other hand, the matrix we build is much smaller than the adjacency matrix, usually no greater than $1500 \times 1500$.

By introducing fingerprints, \fname{} achieves much higher accuracy than TCM, especially in topology queries like reachability query.
In TCM, we build a matrix with size $w \times w$, and map each node $v$ in the graph into a hash address $h(v)$ in range $[0, w)$. The attributes of edge $(s, d)$ is stored in the bucket in row $h(s)$, column $h(d)$. If different edges are mapped into the same bucket, their attributes will be accumulated. In such a matrix, the nodes with the same hash values are mixed up, which will lead to low accuracy in complicated topology queries. For example, in reachability query from $s$ to $d$, we use $V(h(s))$ to represent the set of nodes mapped to $h(s)$, and use $V(h(d))$ to represent the set of nodes mapped to $h(d)$. If any node in $V(h(s))$ can reach one of the nodes in $V(h(d))$ in $G$, the report of TCM for this reachability query will be yes. There will be a lot of nodes $V(h(s))$ and $V(h(d))$, because in the graph there may be millions of nodes, while in TCM the value range of hash addresses is much smaller, usually less than $10000$, as we have to control the size of the matrix. Therefore false positive rate of TCM in reachability query will be high.
While in \fname, using fingerprints will help to further distinguish the nodes. Only the nodes with the same fingerprint and the same hash address will be mixed up. As the value range of the fingerprints is large, the number of mixed up nodes will be much smaller than TCM. Therefore the accuracy will be much higher.}

Like TCM, \fname\ uses a hash function $H(\cdot)$ to compress the streaming graph $G$ into a smaller graph $G_h$ which is named a \emph{graph sketch}. Each node $v$ in $G$ is mapped into a hash value $H(v)$. Nodes with the same hash value are combined into one node in $G_h$, and the edges connected to them are also aggregated. An example of the graph stream and the graph sketch can be referred in Fig.\ref{sample} and Fig.\ref{compress}. The compression rate can be controlled by the size of the value range of $H(\cdot)$, which we represent with $M$. The higher the compression rate is, the lower the accuracy is, as more nodes and edges will be combined. 

Different from TCM which uses an adjacency matrix to store the graph sketch $G_h$, \fname\ uses a novel data structure to store it. This data structure is specially designed for sparse graphs and can store a much bigger graph sketch with the same space. 
\nop{In TCM, an adjacency matrix is built with side length $M$. Each row/column in the adjacency matrix is corresponding to a hash value $H(v)$, and stores edges emitted/received by the node $H(v)$ in $G(h)$. In order to achieve low compression rate and high accuracy, the value range $M$ has to be large, leading to a matrix too large to store. On the other hand, as $G$ is usually very sparse,most nodes  have much fewer edges than $m$. Therefore when storing the graph sketch in \fname, }
As the graph is sparse, the number of nodes is large, but each node is connected to few edges. Therefore, different from adjacency matrix which stores edges with the same source node / destination node in one row / column, we store edges with different source nodes / destination nodes in the one row / column, and distinguish them with fingerprints. Each edge in the graph sketch is mapped to a bucket in the matrix depending on its endpoints, and marked with a fingerprint pair. If the bucket it is mapped is already occupied by other edges, we store this edge in a buffer $B$, which is composed of adjacency lists. With a $m\times m$ matrix we can represent a graph sketch with at most $m\times F$ nodes in \fname, where $F$ is the size of the value range of the fingerprint (for example, a 16-bit fingerprint has $F=65536$). On the other hand, the adjacency matrix can only store a a graph sketch with at most $m$ nodes. With a much larger graph sketch, the accuracy is also much higher compared to TCM. 

In \fname, the memory cost and update speed are greatly influenced by the size of the buffer $B$. As the buffer takes additional memory, and update speed in an adjacency list is linear with its size. In order to restrict its size, we propose a technique called \hashname. In this technique each edge is mapped to multiple buckets, and stored in the first empty one among them. This enlarges the chance that an edge finds an empty bucket. Besides, a few nodes in a sparse graph may still have very high degrees. If one node emits a lot of edges, these edges have high probability to evict each other when stored in one row. To solve this problem, In \hashname\ edges with source node $v$ are no longer mapped to one row, but $r$ rows, sharing memory with other source nodes. The higher degree a node has, the more buckets it may take. It is similar in the view of columns and destination nodes. This helps to ease the congestion brought by the skewness in node degrees. Experiments show that after this modification the buffer only stores less than $0.01\%$ of the edges in the graph stream.

The key contributions of this paper are as follows:
\begin{enumerate}

\item We propose \fname, a novel data structure for graph stream summarization. It has small memory usage, high update speed, and supports almost all kinds of queries for graphs. Most important of all, it uses a combination of fingerprints and hash addresses to achieve very high accuracy.

\item We propose a technique called \hashname{} in the implementation of \fname. It helps to decrease the buffer size, improve update speed and reduce memory cost. It also eases the influence brought by the skewness in node degrees. 

\item We define $3$ graph query primitives and give details about how \fname\ supports them. Almost all algorithms for graphs can be implemented with these primitives.

\item We carry out theoretical analysis and extensive experiments to evaluate the performance of \fname, which show that when using 1/256 memory size of the state-of-the-art graph summarization algorithm, our algorithm still significantly outperforms it for most queries.

\end{enumerate}

\presec
\section{Related Work}
\postsec
\label{sec:relatedwork}
In this part we will give a brief introduction about the related works. The prior arts of graph stream summarization can be divided into two kinds. The first kind is composed of counter arrays and stores each data item in these arrays independently, ignoring the connections between them. They only support queries for edge weights, but do not support any queries involved with topology of the graph. This kind includes CM sketches\cite{Cormode2004An}, CU sketches\cite{estan2003new}, gSketches\cite{Zhao2011gSketch} and so on. The second kind supports all queries in the streaming graph, but suffers from poor accuracy. This kind includes TCM\cite{Tang2016Graph} and gMatrix\cite{Khan2016Query}. 
Because of space limitation, in this section we only introduce the second kind which is more relevant to our work.
\nop{
\subsection{CM sketch}
The CM sketch \cite{Cormode2004An} is a classic data structure for data stream summarization. It can store the frequencies of different items in a data stream. It has $d$ arrays, which are all set to 0s initially. Each array is relative with an independent hash function ${h}_{j}(x) 1\leqslant j \leqslant d$. When inserting an item $x$, let $\forall j, 1\leqslant j \leqslant d\ count[j,{h}_{j}(x)]=count[j,{h}_{j}(x)]+1$. Similarly, when deleting the item, let $\forall 1\leqslant j \leqslant d\ count[j,{h}_{j}(x)]=count[j,{h}_{j}(x)]-1$. $count[j,{h}_{j}(x)]$ means the counter mapped by ${h}_{j}(x)$ in the ${j}_{th}$ array. When querying an item $x$, the CM sketch carries out $d$ hash functions and gets all the corresponding counters. Then it selects the minimum one as the frequency. The CM sketch only has overestimations, which means the query result of an item will be equal to or greater than the correct value.
We can use the CM sketch to store the weights of edges in a graph stream. The memory usage is $O(|E|)$ where $E$ is the edge set and the update needs only $d$ memory accesses. However, it does not store the relations between edges, in other words, the topology of the streaming graph. Therefore we can not perform any topology involved queries with it.
There are also some other data structures with similar functions, like CU sketches and Count sketches. They can also be applied to graph stream summarization, but also supports limited kinds of queries.
\nop{CM sketches can be used to summarize the attributes of edges in the data stream. When an edge $(s, d)$ appears with attribute $w$, we just join $s$ and $d$ together to get a key $x$. then we modify the value of the counter mapped by ${h}_{j}(x)$ in the ${j}_{th}$ array with $w$ using a pre-defined function $f(.)$ for all j $1\leqslant j \leqslant d$. For example if in a social network $w$ is the times user $s$ and $d$ interact, we initialize the counters with $0$ and set $f(.)$ to be addition. However, in the CM sketch, we can only query for the attributes of edges, and queries involved with topology, like reachability queries, are not supported. It is not enough for graph streams.}

 Compared to the CM sketch, gSketch \cite{Zhao2011gSketch} is specially proposed for graph streams. It aims to improve the accuracy of the CM sketch when applies to graph streams. As in CM sketches when items with small weights collide with those with large weights, in other words, mapped to the same counter, the relative error of the smaller ones will be high.
  The gSketch reduces their relative errors by classifying the edges with their weights and storing those with similar values together. The gSketch
  assumes that edges emitted by the same nodes has similar weights. It uses a small sampling of the data stream to predict of weights of edges emitted from different nodes, then divides a CM sketch into several small local sketches with different sizes, and maintains a hash table that maps the vertexes in the graph into different local sketches according to their predicted edge weights. Edges in the graph is maintained and queried in the local sketch their source vertexes are mapped to. As Only edges with similar weights will have hash collisions, the relative error of the gSketch is much smaller than the CM sketch. But like the CM sketch, it only stores the weights of edges but not any information about the topology of the graph. Therefore the gSketch also supports limited kinds of queries}


TCM \cite{Tang2016Graph} is the state-of -the-art of data structures for graph stream summarization. It is composed of an adjacency matrix that stores the compression of the streaming graph. It uses a hash function $H(\cdot)$ to compress the streaming graph $G=(V,E)$ into a smaller graph sketch $G_h$. For each node $v$ in $G$, TCM maps it to node $H(v)$ in $G_h$. For each edge $e=\overrightarrow{s,d}$ in $G$, TCM maps it to edge $\overrightarrow{H(s),H(d)}$ in $G_h$. The weight of an edge in $G_h$ is an aggregation of the weights of all edges mapped to it.
A hash table that stores the hash value and the original ID pairs can be built in this map procedure to retrieve the original node IDs for some queries. Then TCM uses an adjacency matrix to represent the graph sketch. If we represent the size of the value range of $H(\cdot)$ with $M$, we need to build an $M\times M$ adjacency matrix. Each bucket in the matrix contains a counter. The weight of edge $\overrightarrow{H(s),H(d)}$ in the graph sketch is added to the counter in the bucket in row $H(s)$, column $H(d)$.

When the memory is sufficient, we can also build multiple sketches with different hash functions, and report the most accurate value in queries. 
\nop{In order to store the relationships between different edges, it uses one hash function to map the two vertexes of each edge separately. Then it uses the two hash addresses as two dimensions to locate a mapped bucket in the matrix, and accumulates the attribute of the edge in the bucket. Usually the accumulation function is addition. Multiple hash functions can be applied in TCM. In this circumstance a TCM sketch is made up of multiple matrices, and each matrix is relevant with a hash function. These matrices work separately when updating, and report the most accurate value in them when querying.  TCM can be seen as a compressed adjacency matrix and supports all the queries an adjacency matrix can support, like queries for attributes of an edge or a subgraph, or reachability queries. The query process is also the same as adjacency matrix. }


In order to satisfy the demand on memory usage, the size of the adjacency matrix, $M\times M$ has to be within $O(|E|)$, which means $M \ll |V|$ for a sparse streaming graph where $\frac{|E|}{|V|}$ is usually within $10$. This means the graph sketch $G_h$ is usually much smaller than $G$, a lot of nodes and edges will be aggregated. As a result, the accuracy of TCM is poor. 
 \nop{However, TCM is not accurate enough for topology queries, like reachability query. The main reason for the inaccuracy of TCM is that there are usually millions of nodes in the graph stream, but in order to maintain memory efficiency, the side length $m$ of the matrix we build is much smaller than the number of nodes, usually no more than $10000$. when we map the nodes into addresses in range $[0, m)$, many nodes will get the same address. While in the matrix of TCM, we can not tell the nodes mapped to the same address apart when querying. Therefore, the result we get in a node query may be the summary of many nodes. Similarly in reachability query, the result we get is actually the reachability from a set of nodes with the same hash address, $V(h(s))$,  to another set $V(h(d))$. If node in $V(h(s))$ can reach a node in $V(h(d))$, the answer to all the node pairs from $V(h(s))$ to $V(h(d))$ will be positive. Even if there are no paths between these two sets in the graph, there is still high possibility for a positive report because of hash collisions. This leads to a very low true negative rate.

 Furthermore, like the adjacency matrix, when the distribution of the node degrees in a graph is skewed, the data will distribute very uneven in TCM. Some of the rows or columns in the matrix will be so crowed that the edges stored there will collide with many other edges, and their query result will be inaccurate. On the other hand, some areas of the matrix will be sparse and waste a lot of memory. In the original paper the author proposed to adjust the matrix from square to rectangular to adapt to this skewness, but this method is not practical. It is hard to predict the distribution of the node degrees. Therefore it is hard to decide the shape of the rectangular. If we build several matrices with different length-width ratio as suggested by the author, there are still problems, as even if we happen to build one with length-width ratio suitable for the distribution, those with wrong shape are wasted. It is a huge loss in memory.}

The gMatrix \cite{Khan2016Query} is a variant of TCM. Its structure is similar to TCM. But it uses reversible hash functions to generate graph sketches. It also extends TCM to more queries like edge heavy hitters and so on. However, different from the accurate hash tables, the reversible hash function introduces additional errors in the reverse procedure. Therefore the accuracy of gMatrix is no better than TCM, sometimes even worse. 

There are some graph algorithms for statistic graph compression \cite{raghavan2003representing,fan2012query,spielman2011graph} or specific queries in graph stream processing \cite{gao2014continuous,wang2009continuous,song2014event}. However, they are either not suitable for high dynamic graph streams or too limited in functions. We do not introduce them in detail due to space limit.
\presec
\section{Problem Definition}\label{sec:problemdef}
\postsec

\begin{Def}
\textbf{Graph Stream}: A graph stream is an unbounded timing evolving sequence of items $S=\{e_1, e_2, e_3......e_n\}$, where each item $e_i=(
\overrightarrow {s,d}; t; w)$ indicates a directed edge\footnote{The approach in this paper can be easily extended to handle undirected graphs.} from node $s$ to node $d$, with wight $w$. The timepoint $t_i$ is also referred as the timestamp of $e_i$. \nop{Two edges are connected if they share one common node.} Thus, the edge streaming sequence $S$ forms a dynamic directed graph $G=(V, E)$ that changes with the arrival of every item $e_i$, where $V$ and $E$ denote the set of nodes and the set of edges in the graph, respectively. We call $G$ a \emph{streaming graph} for convenience. 
\end{Def}

In a graph stream $S$, an edge $\overrightarrow{s,d}$ may appear multiple times with different timestamps. The weight of such edge in the streaming graph $G$ is SUM of all edge weights sharing the same endpoints. The weight $w$ can be either positive or negative. An item with $w<0$ means deleting a former data item.

\begin{figure}[htbp]
\small
\centering
\includegraphics[width=0.35\textwidth]{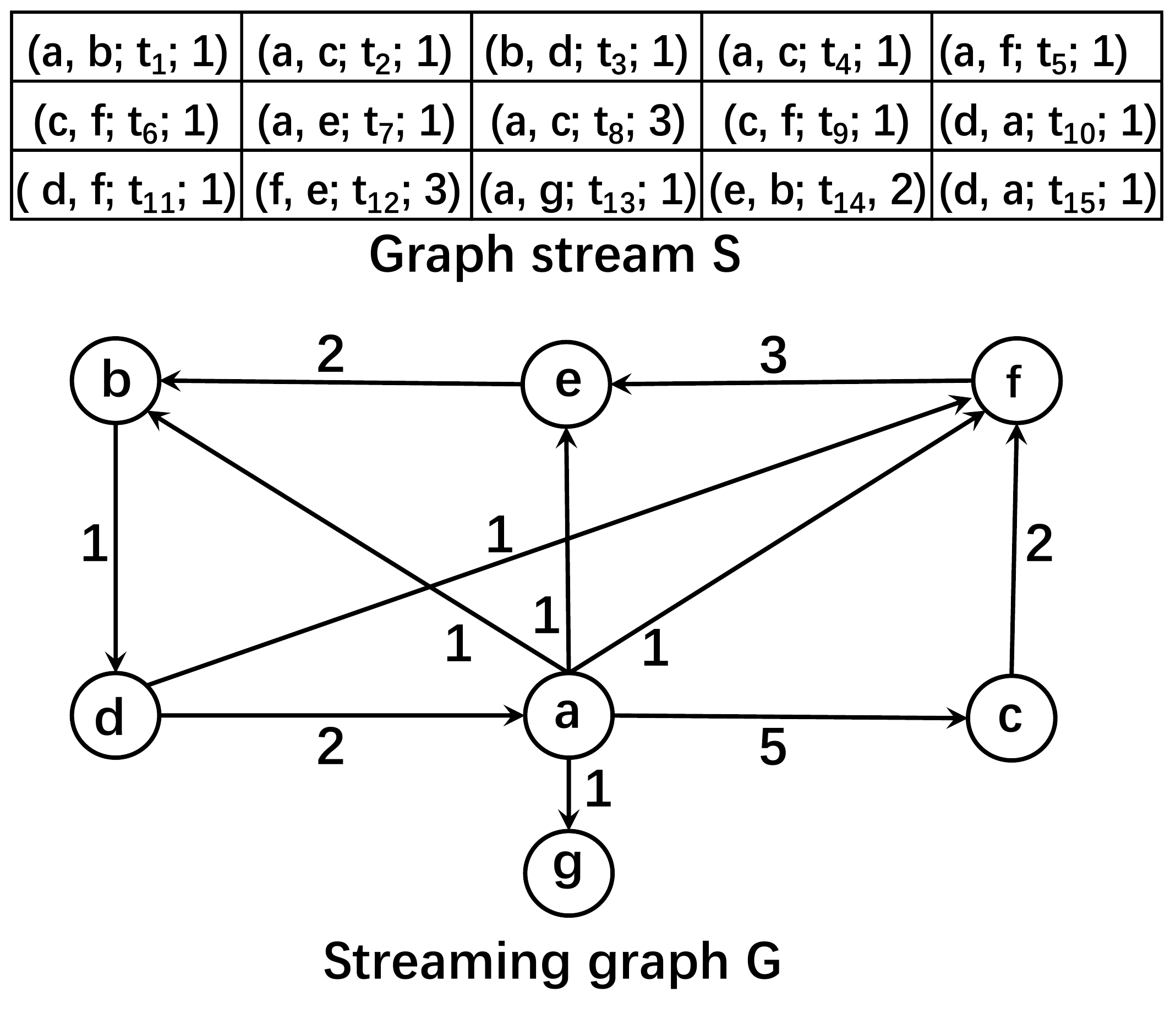}
\caption{A sample graph stream}
\label{sample}
\end{figure}

\begin{Exp}
A sample graph stream $S$ and the corresponding streaming graph $G$ are both shown in Fig. \ref{sample}. Each node has an ID that uniquely identifies itself. If an edge appears multiple times, its weights are added up as stated above.  
\end{Exp}

In practice, $G$ is usually a large, sparse and high speed dynamic graph. 
The large volume and high dynamicity make it hard to store graph streams using traditional data structures such as adjacency lists and adjacency matrices. The large space cost of $O({|V|}^2)$ rules out the possibility of using the adjacency matrix to represent a large sparse graph. On the other hand, the adjacency list has $O(|E|)$ memory cost, which is acceptable, but the time cost of inserting an edge is $O(|V|)$, which is unacceptable due to the high speed of the graph stream.

The goal of our study is to design a \emph{linear} space cost data structure with efficient query and update algorithms over high speed graph streams. To meet that goal, we allow some approximate query results but with small and controllable errors. However, traditional graph stream summarization approaches either cannot answer graph topology queries such as reachability queries (such as CM sketches \cite{Cormode2004An} and gSketches \cite{Zhao2011gSketch}) or fail to provide accurate query results (such as TCM \cite{Tang2016Graph} and gMatrix \cite{Khan2016Query}). 
\nop{More details about related work have been discussed in Section \ref{}.} Therefore, in this paper, we design a novel graph stream summarization strategy.

In order to give a definition of the graph stream summarization problem, First we define the \emph{graph sketch} as follows:
\begin{Def}
\textbf{Graph Sketch}: a graph sketch of $G=(V, E)$ is a samller graph $G_h=(V_h, E_h)$ where $|V_h|\leqslant |V|$ and $|E_h|\leqslant |E|$. A map function $H(\cdot)$ is used to map each node in $V$ to a node in $V_h$, and edge $e=\overrightarrow{s,d}$ in $E$ is mapped to edge $\overrightarrow{H(s),H(d)}$ in $E_h$. The weight of an edge in $E_h$ is the SUM of the weights of all edges mapped to it. 
\end{Def}
Formally, we define our \emph{graph stream summarization} problem as follows.

\prefig
\postfig

\begin{Def}\label{def:problem}
\textbf{Graph Stream Summarization}: Given a streaming graph $G=(V,E)$, the \emph{graph stream summarization} problem is to design a graph sketch $G_h=(V_h, E_h)$, and the corresponding data structure $DS$ to represent $G_h$, where the following conditions hold:
\begin{enumerate}
\item There is a function $H(\cdot)$ that map nodes in $V$ to nodes in $V_h$;
\nop{\item $|V_h| \leq |V|$ and $|E_h| \leq |E|$, which means that the compressed graph $G_h$ should be smaller than $G$;}
\item The space cost of $DS$ is $O(|E|)$;
\item $DS$ changes with each new arriving data item in the streaming graph and the time complexity of updating $DS$ should be $O(1)$;
\item $DS$ supports answering any query over the original streaming graph $G$ with small and controllable errors. 
\end{enumerate}

\nop{
a graph stream summarization algorithm is to store the stream graph $G$ approximately. It is composed of two parts. The first part is a function $H(.)$ that maps the stream graph $G=(V, E)$ into a compressed graph $G_h=(V_h, E_h)$. $V$ and $V_h$ are the sets of nodes in graph $G$ and $G_h$, respectively, and $E$ and $E_h$ are the sets of edges in graph $G$ and $G_h$, respectively. The second part is to design a data structure $DS$ to store $G_h$. This algorithm should satisfy the following requirements:
}
\nop{
\begin{itemize}
\item $|V_h| \leqslant |V|$, and $|E_h| \leqslant |E|$. After the compression the graph should become smaller.
\item The storage of $G_h$ should have no errors.
\item $DS$ should have memory cost less than $O(E)$, and when $G$ changes with a new arriving data item in $S$, the time cost of updating in $DS$ should be $O(1)$.
\item With $DS$ we should be able to answer queries about $G$ with small and controllable errors
\end{itemize}
}

\end{Def}

\nop{\begin{figure*}[htbp]
\subfigure[Edge Query]
{
\begin{minipage}[b]{0.3\textwidth}
\centering
\includegraphics[width=1\textwidth]{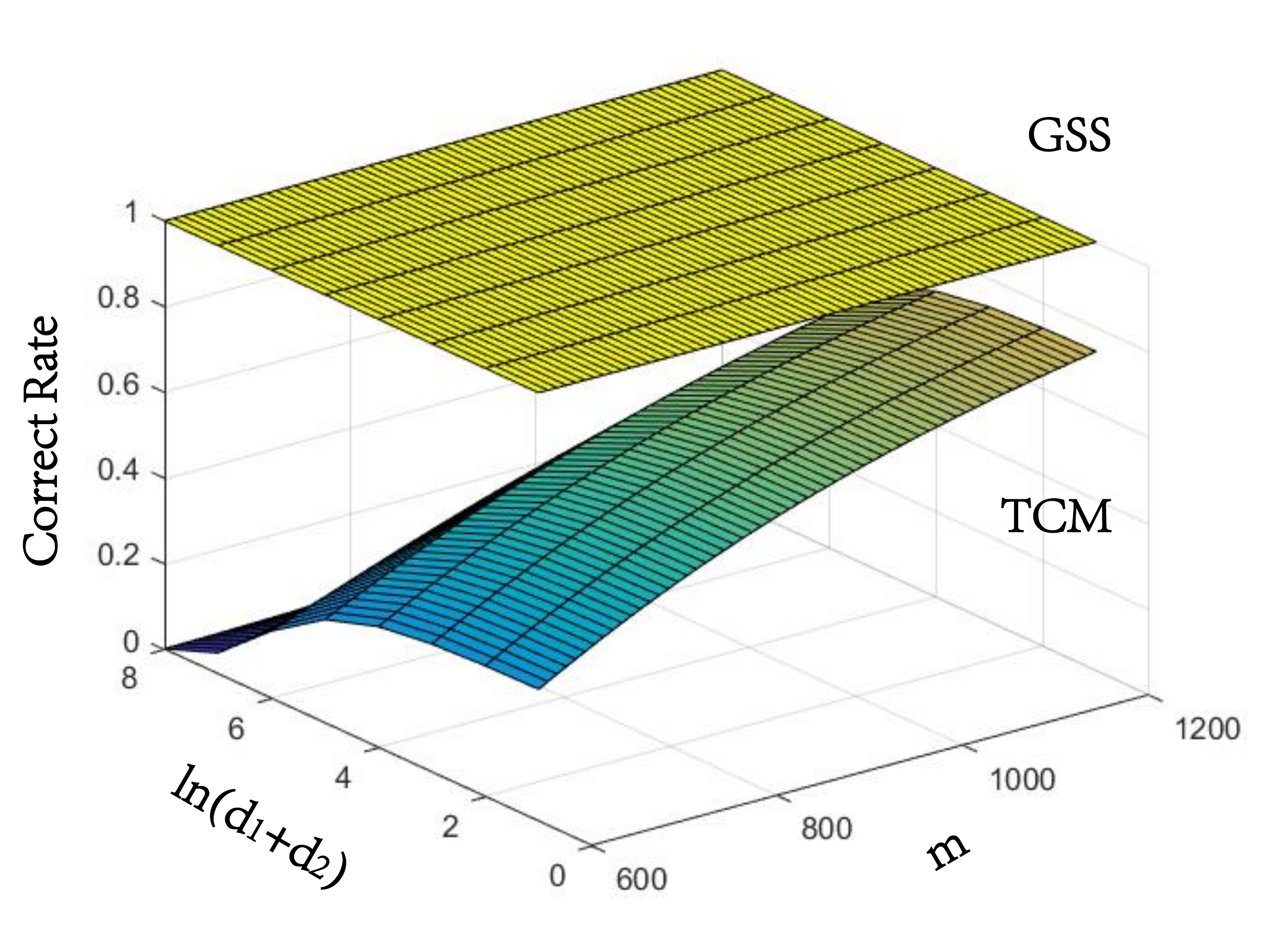}
\end{minipage}
}
\subfigure[1-hop Successor Query]{
\begin{minipage}[b]{0.3\textwidth}
\centering
\includegraphics[width=1\textwidth]{Theo-successor.pdf}
\end{minipage}
}
\subfigure[1-hop Precursor Query]{
\begin{minipage}[b]{0.3\textwidth}
\centering
\includegraphics[width=1\textwidth]{Theo-precursor.pdf}
\end{minipage}
}
\caption{Theoretical Accuracy}
\label{TA}
\end{figure*}
In graph stream summarization, updates and queries to $G$ are all mapped to operations over the compressed graph $G_h$ and then are conducted using the data structure $DS$. }

In the context of streaming graphs, $G$ changes with every data item in the graph stream $S$, which is mapped to updating the graph sketch $G_h$, and conducted in data structure $DS$. For every new item $(\overrightarrow{s,d};t;w)$ in $S$, we map edge $\overrightarrow{s,d}$ in $G$ to edge $\overrightarrow{H(s),H(d)}$ in $G_h$ with weight $w$ and then insert it into $G_h$. Similarly, queries over $G$ are also mapped to the same kind of queries over the graph sketch $G_h$. In order to support any kind of graph queries, we first define three graph query primitives as follows, since many kinds of graph queries can be answered using these primitives.
\nop{There are three cases for inserting edge $\overrightarrow{H(s),H(d)}$ into $G_h$. First, edge $\overrightarrow{H(s),H(d)}$ has been in the graph sketch $G_h$ already, we only update the edge weight. Second, if both nodes $H(s)$ and $H(d)$ have been in the graph sketch $G_h$ but edge $\overrightarrow{H(s),H(d)}$ does not occur, we introduce edge $\overrightarrow{H(s),H(d)}$ with the weight $w$ to $G_h$ directly. Third, if either of $H(s)$ or $H(d)$ does not occur in $V(G_h)$, we introduce the node and then add the edge $\overrightarrow{H(s),H(d)}$ with the weight $w$ into $G_h$. }

\nop{
\begin{Def}
\textbf{Edge Insertion}: the insertion of an edge $(s, d)$ with weight $w$ to a graph $G=(V, E)$ includes the following procedures. First, check if $s$ and $d$ are in $V$, and insert them to $V$ if not. Second, check if edge $(s, d)$ is already in $E$. Insert it to $E$ if not, and initialize its weight to $0$. Third, add $w$ to the weight of $(s, d)$.
\end{Def}
}

\nop{
can also be mapped to the same kind of queries upon the corresponding nodes and edges in $G_h$. For example, when querying for the reachability of node $s$ and $d$ in $G$, we return the reachability of $H(s)$ and $H(d)$ in $G_h$ as result. There may be errors in queries because of collisions, which means different nodes in $G$ are mapped to the same node in $G_h$, or different edges are mapped to one in $G_h$. The error rate is determined by the compression rate. The higher the compression rate is, the easier we get a wrong answer.To answer different queries for $G$, we also need to be able to answer various kinds of queries for $G_h$ with $DS$. Here we define $3$ basic query operators for a graph:
}

\begin{Def}
\label{GQP}
\textbf{Graph Query Primitives}: Given a graph $G(V,E)$, the three graph query primitives are:
\begin{itemize}
\item \textbf{Edge Query}: given an edge $e=\overrightarrow{s,d}$, return its weight $w(e)$ if it exists in the graph and return $-1$ if not.
\item \textbf{$1$-hop Successor Query}: given a node $v$, return a set of nodes that are 1-hop reachable from $v$, and return $\{-1\}$ if there is no such node;
\item \textbf{$1$-hop Precursor Query}: given a node $v$, return a set of nodes that can reach node $v$ in 1-hop, and return $\{-1\}$ if there is no such node.
\end{itemize}
\end{Def}

With these primitives, we can re-construct the entire graph. We can find all the node IDs in the hash table. Then by carrying out 1-hop successor queries or 1-hop precursor queries for each node, we can find all the edges in the graph. The weight of the edges can be retrieved by the edge queries. As the graph is reconstructed, all kinds of queries and algorithms can be supported. In fact, in many situations, it is not necessary to re-construct the entire graph. We can just follow the specific algorithm and use the primitives to get the information when needed. Therefore, The data structure $DS$ needs to support these $3$ query primitives.

\nop{When answering queries over the streaming graph $G$, we return the answer of the same queries in the graph sketch $G_h$ as an approximate answer. Therefore, the data structure $DS$ should be able to conduct all three query primitives over $G_h$ to support various kinds of queries.}

\nop{
Most queries in the graph is composed of these $3$ basic operators. In this paper, we use the following $3$ queries as examples:

\textbf{Edge Queries}: the same as the edge query operator. It is the most basic queries in applications. For example, query for the number of packers one host set to another in a data center, or query for the interactions between two users in a social network.

\textbf{Node Queries}: a node query for node $v$ is to compute the sum of the weights of all edges $(v, v_s)$ where $v_s$ is an 1-hop successor of $v$. It is a combination of 1-hop successor queries and edge queries, and can be used to check the behavior of a node in the graph.

\textbf{Reachability Queries}: a reachability query means given two nodes $s$ and $d$, we want to find if in the graph there is a path
$\{p_0, p_1, p_2, p_3,....p_{n-1}, p_n\}$ 
where $p_0=s$, $p_n=d$, and $\forall i \in [1, n]$, there is an edge $(p_{i-1}, p_i)$. There have been many algorithms for reachability query, like \cite{}. In this paper, we use the most basic algorithm, breadth first search (BFS) as an example. It is composed of a series of 1-hop successor queries. First we initialize a set $S =\{s\}$. Then for every node in $S$ we add its 1-hop successors to $S$. We carry out this procedure recursively until $d$ is added to $S$ or $S$ does not change any more. If after the algorithm ends $d$ is still not in $S$, we report false for the reachability query, otherwise we report yes.
}


\nop{Although TCM \cite{Tang2016Graph} and gMatrix \cite{Khan2016Query} can answer these query primitives, they suffer from inaccuracy of query results. Figure \ref{TA} illustrates the theoretical accuracy of different approaches for these graph query primitives. The accuracy is calculated according to the accuracy analysis in Section \ref{analysis}. For the simplicity of the presentation, we fix the graph size $|E|=421578$ and $|V|=34546$ in the figure. $m$ is the size of matrix in these data structures. In the edge query primitive, $d_1$ and $d_2$ represent the out-degree of the source node and the in-degree of the destination node, respectively. In the 1-hop successor query primitive, $d_{out}$ represents the out degree of the node we query, and in the 1-hop precursor query primitive, $d_{in}$ represents the in degree of the node we query. which is the size of a graph dataset we use in the experiment. From Figure \ref{TA}, we know that our method outperforms TCM and gMatrix by orders of magnitudes in the accuracy ratio, especially in the 1-hop successor/precursor queries. Note that the theoretical results are consistent with our experimental results in section \ref{sec:experiment}, which confirms the superiority of our approach in the query accuracy. }

\nop{
\begin{center}
\begin{table}
\caption{Notation Table}
\label{NT}
\begin{tabular}{|p{3cm}|p{5cm}|}
\hline 
Notation&Meaning\\
\hline  
$S=\{e_1, e_2, ....e_n\}$&The Graph stream $S$\\
\hline  
$G=(V, E)$&Streaming graph with edge set $E$ ,node set $V$\\
\hline 
$G_h=(V_h, E_h)$&Graph sketch with edge set $E_h$, node set $V_h$\\
\hline 
$e=(s, d)$&The edge $e$ from node $s$ to node $d$\\
\hline 
$v$&A node in the streaming graph\\
\hline 
$w(e)$&The weight of edge $e$ \\
\hline 
$H(\cdot)$&The map function from $G$ to $G_h$\\
\hline 
$M$&The size of value range of function $H(.)$\\
\hline 
$H(v)$&ID of the node mapped by node $v$ \\
\hline 
$H(e)=(H(s), H(d))$&The edge mapped by edge $e=(s, d)$ \\
\hline 
$f(v)$&Fingerprint of node $v$\\
\hline 
$F$&Size of the value range of fingerprints\\
\hline 
$h(v)$&Hash value of node $v$\\
\hline 
$m$&Length of the matrix\\
\hline
$r$&Length of the hash address sequence\\
\hline 
$\{h_i(v)|1\leqslant i\leqslant r\}$&Hash address sequence of node $v$\\
\hline 
$\{q_i(v)|1\leqslant i\leqslant r\}$&$LR$ sequence of node $v$\\
\hline
$k$ & Number of candidate buckets for each edge.\\
\hline 
$\{q_i(e)|1\leqslant i\leqslant k\}$&$LR$ sequence of edge $e$\\
\hline 
$l$ & The number of rooms in each bucket\\
\hline
\end{tabular}
\end{table}
\end{center}
}

\nop{The large gap in the accuracy ratio between TCM and gMatrix with ours is because that they use an $m \times m$ traditional 
adjacency matrix as the corresponding data structure $DS$.
With such an adjacency matrix we can only represent a graph sketch with $|V_h|=m$ nodes. To meet the linear space complexity, they requires that $m \leqslant \sqrt[]{|E|}$. However, as in a sparse streaming graph $|V|\gg \sqrt[]{|E|}  \geqslant |V_h| $, that means that a lot of nodes in the original streaming graph $G$ are mapped into the same node in the graph sketch $G_h$, which leads to a very low accuracy. Figure \ref{TA} illustrates the accuracy ratio is quite low when $m \leqslant \sqrt[]{|E|}$. The theoretical analysis of accuracy ratio in TCM and gMatrix has been given in Section \ref{analysis}.
In \fname, we use fingerprints to further distinguish the nodes in the matrix to gain more accurate results. 
}



\nop{
TCM is the state of the art of graph stream summarization. It satisfies the former requirements. However, it simply uses an $m \times m$ adjacency matrix as $DS$, where $m$ is the number of nodes in the compressed graph $G_h$. This makes it suffer the same shortcoming as adjacency matrix: inefficiency in memory usage. If $m$ is very small, the compression rate will be high and the accuracy will be poor, but if $m$ is large, the matrix will be large and most parts will be empty, wasting a lot of memory.  In this paper, we propose a novel graph stream summarization algorithm \fname, which makes best use of the memory, and achieves much higher accuracy with the same memory consumption.  
}
\presec
\section{GSS: Basic Version}\label{sec:idea}
\postsec

In this section, we describe a conceptually simple scheme to help illustrate intuition and benefit of our approach. The full approach, presented in Section \ref{GSS_AA}, is designed with more optimizations. As stated above, to produce a graph stream summarization, we first need to design a graph sketch $G_h=(V_h, E_h)$ for the streaming graph $G$. Initially, we use the same strategy as TCM to generate the graph sketch. We choose a hash function $H(\cdot)$ with value range $[0, M)$, then $G_h$ is generated as following:
\begin{enumerate}
\item \textbf{Initialization}: Initially, $V_h=\varnothing$, and $E_h=\varnothing$.
\item \textbf{Edge Insertion}: For each edge $e=(s, d)$ in $E$ with weight $w$, we compute hash values $H(s)$ and $H(d)$. If either node with ID $H(s)$ or $H(d)$ is not in $V_h$ yet, we insert it into $V_h$. Then we set $H(e)=\overrightarrow{H(s),H(d)}$. If $H(e)$ is not in $E_h$, we insert $H(e)$ into $E_h$ and set its weight $w(H(e))=w$. If $H(e)$ is in $E_h$ already, we add $w$ to the weight.
\end{enumerate}
$G_h$ is empty at the beginning and expands with every data item in the graph stream. We can store $\langle H(v), v \rangle$ pairs with hash tables to make this mapping procedure reversible. This needs $O|V|$ additional memory, as $|V|\leqslant |E|$, the overall memory requirement is still within $O(|E|)$.

\prefig
\begin{figure}[htbp]
\centering
\includegraphics[width=0.4\textwidth]{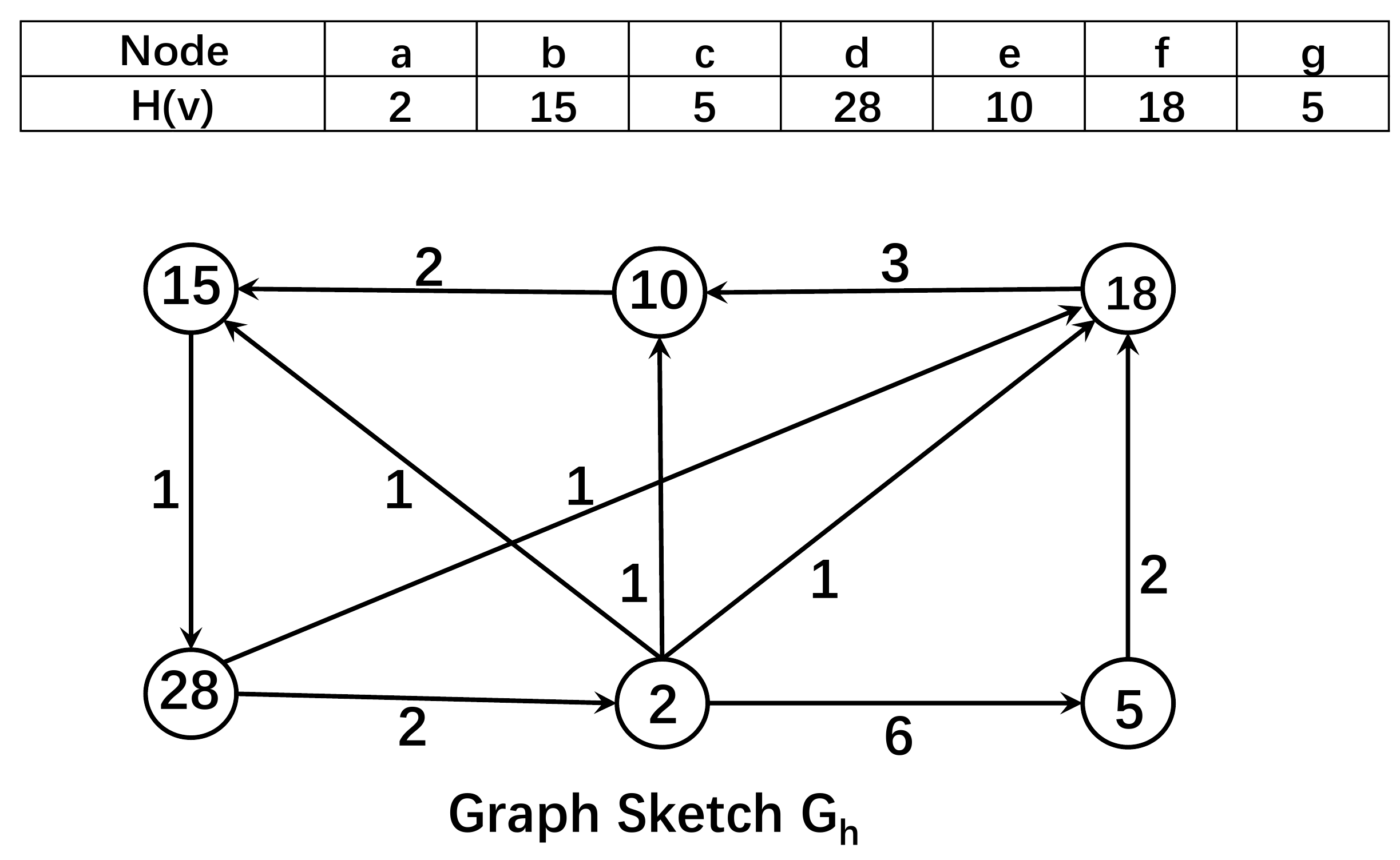}
\caption{A sample map function}
\label{compress}
\end{figure}
\postfig

\begin{Exp}
A graph sketch $G_h$ for the streaming graph $G$ in Figure \ref{sample} is shown in Figure \ref{compress}. The value range of the hash function $H(\cdot)$ is $[0,32)$. In the example, nodes $c$ and $g$ are mapped to the same node with ID $5$ in $G_h$. In $G_h$, the weight of edge $(2, 5)$ is $6$, which is the summary of the weight of edge $(a, c)$ and edge $(a, g)$ in $G$.
\end{Exp}

\begin{figure*}[htbp]
\subfigure[Edge Query]
{
\begin{minipage}[b]{0.3\textwidth}
\centering
\includegraphics[width=1\textwidth]{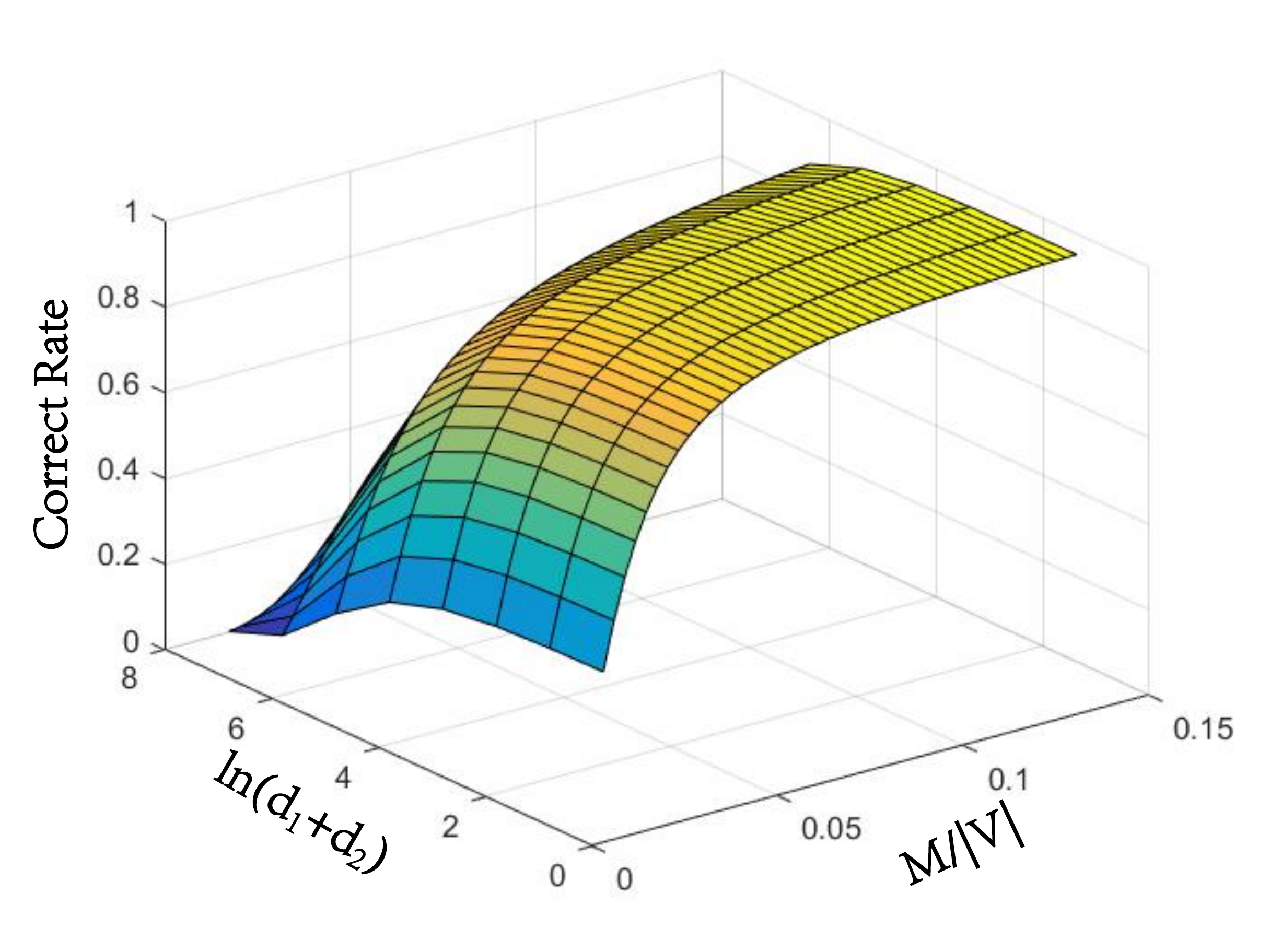}
\end{minipage}
}
\subfigure[1-hop Successor Query]{
\begin{minipage}[b]{0.3\textwidth}
\centering
\includegraphics[width=1\textwidth]{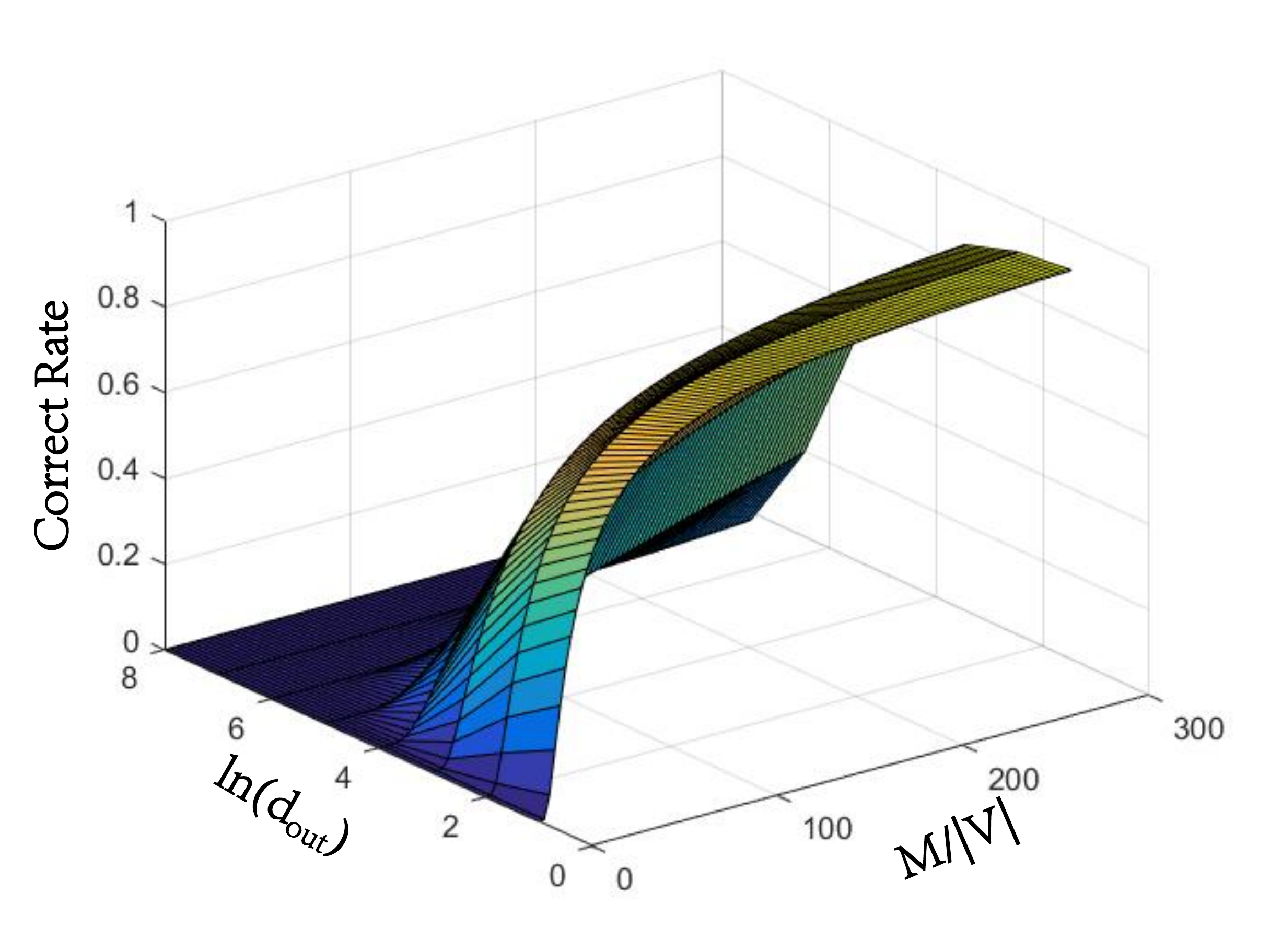}
\end{minipage}
}
\subfigure[1-hop Precursor Query]{
\begin{minipage}[b]{0.3\textwidth}
\centering
\includegraphics[width=1\textwidth]{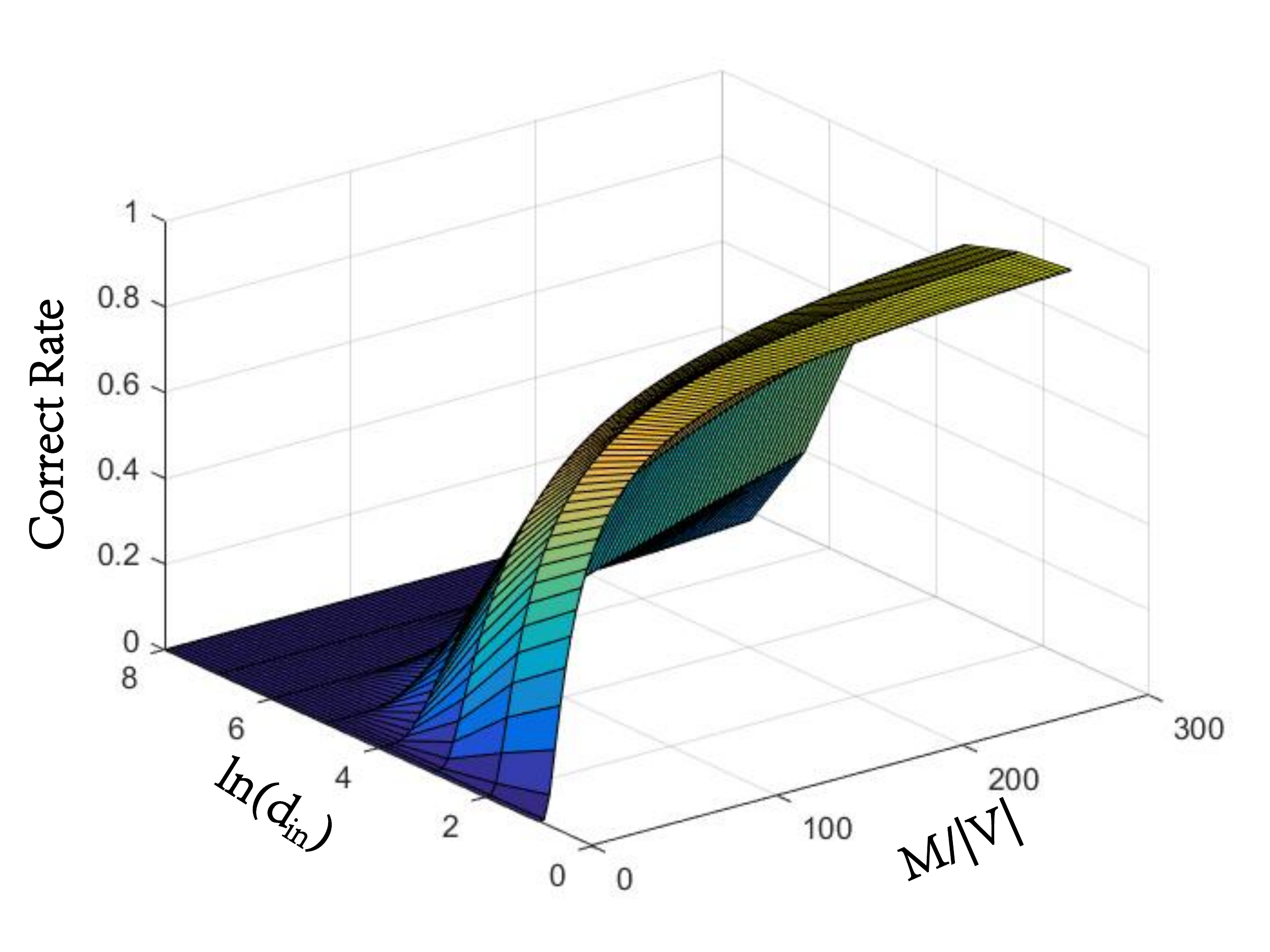}
\end{minipage}
}
\caption{Influence of M on Accuracy}
\label{MA}
\end{figure*}

Obviously, the size of the value range of the map function $H(\cdot)$, which we represent with $M$, will significantly influence the accuracy of the summarization, especially in the 1-hop successor / precursor query primitives. 
\textbf{
In a uniform mapping with the hash function, each node in $G$ has the probability $\frac{1}{M}$ to collide with another, which means they are mapped to the same node in $G_h$. When there are $|V|$ nodes, the probability that a node $v$ does not collide with any other nodes is ${(1-\frac{1}{M})}^{|V|-1} \approx {e}^{\frac{|V|-1}{M}}$. In the 1-hop successor / precursor queries, if $v$ collides with others, the query result about it will definitely have errors. Therefore we have to use a large $M$ to maximize this probability. }

\textbf{
Figure 3 shows the theoretical results of the relationship between $M$ and the accuracy of the query primitives . The results are computed according to analysis in Section VI-B. }
(In the figure of the edge query, $d_1$ and $d_2$ means the in-degree of the source node and the out-degree of the destination node of the queried edge. In the figure of the 1-hop successor / precursor query, $d_{in}$ and $d_{out}$ means the in-degree and the out-degree of the queried node, respectively). 
\textbf{
The figure shows that we have to use a large $M$ to achieve high accuracy in the query primitives, which is not possible in the prior works.}
According to Figure 3, only when  $\frac{M}{|V|}>200$, the accuracy ratio is larger than {$80\%$} in 1-hop successor / precursor queries. When  $\frac{M}{|V|} \leq 1$, the accuracy ratio falls down to nearly {$0$}, which is totally unacceptable.

Both TCM and the gMatrix resort to an adjacency matrix to represent $G_h$. In this case, the matrix rank $m$ equals to $M$, i.e, the value range of the map function. To keep the memory usage of the graph sketch within $O(|E|)$ (Condition 2 in Definition \ref{def:problem}), $m$ must be less than $\sqrt[]{E}$, that means $m=M < \sqrt[]{E} \ll |V|$ for a sparse streaming graph. According to our theoretical analysis\footnote{The detailed analyses are given in Section \ref{AccuracyAnalysis}} in Figure \ref{MA}, the query results' accuracy is quite low in them. Our experiments in Section \ref{sec:experiment} also confirm the theoretical analysis.

\prefig
\begin{figure}[htbp]
\includegraphics[width=0.5\textwidth]{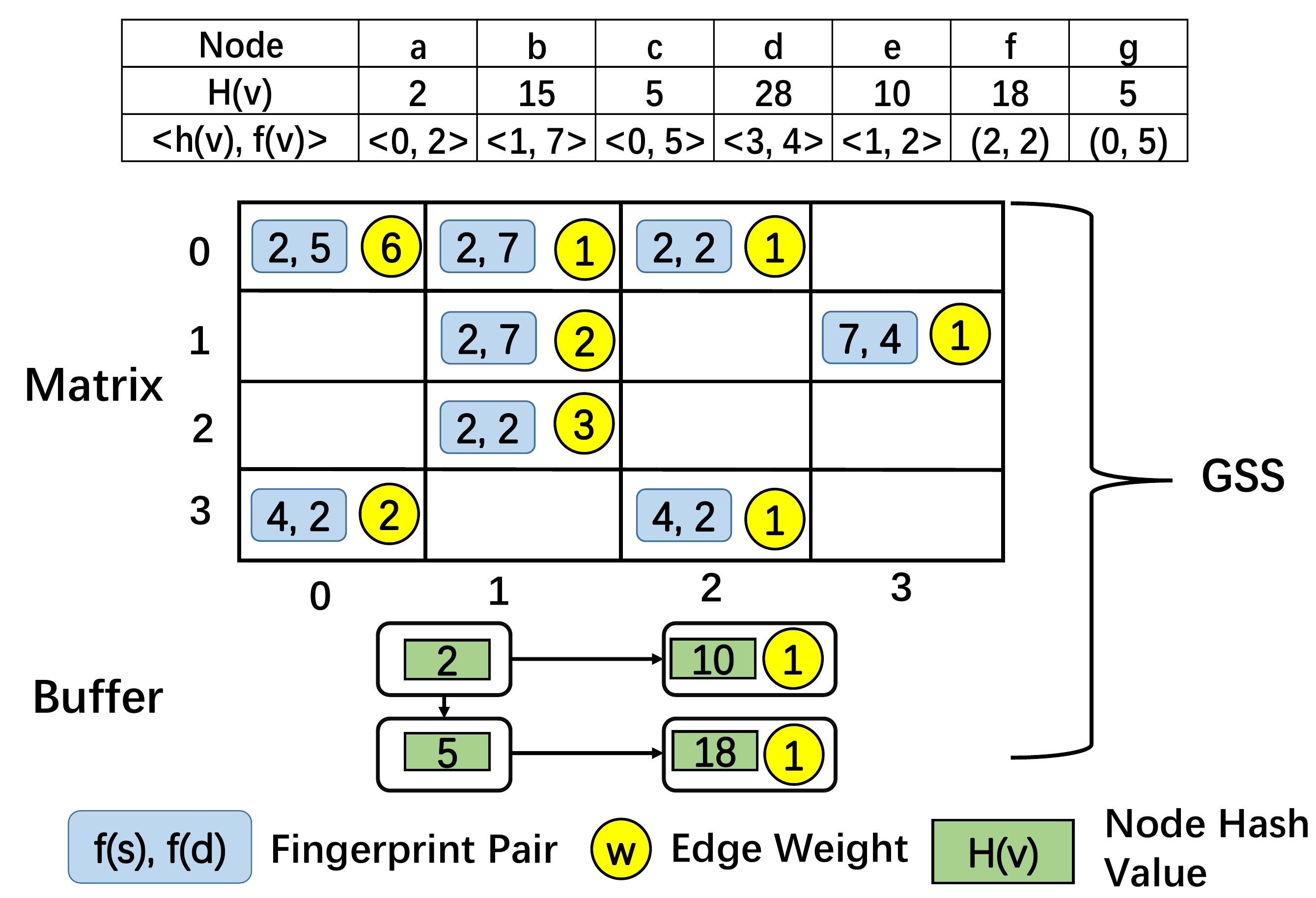}
\caption{A sample of the basic version of data structure}
\label{v1}
\end{figure}
\postfig

Considering the above limitations, we design a novel data structure for \emph{g}raph \emph{s}tream \emph{s}ummarization, called GSS.

\begin{Def} \textbf{GSS}:\label{def:gss} Given a streaming graph $G=(V,E)$, we have a hash function $H(\cdot)$ with value range $[0,M)$ to map each node $v$ in graph $G$ to node $H(v)$ in graph sketch $G_h$. Then we use the following  data structure to represent the graph sketch $G_h$:
\begin{enumerate}
\item $GSS$ consists of a size $m \times m$ adjacency matrix $X$ and an adjacency list buffer $B$ for left-over edges.
\item For each node $H(v)$ in sketch graph $G_h$, we define an address $h(v)(0 \leqslant h(v) \leqslant m)$ and a fingerprint $f(v)(0 \leqslant f(v) \leqslant F)$ where $M=m\times F$ and $h(v)=\lfloor \frac{H(v)}{F}\rfloor$, $f(v)=H(v)\%F$.
\item Each edge $\overrightarrow{H(s),H(d)}$ in the graph sketch $G_h$ is mapped to a bucket in the row $h(s)$, column $h(d)$ of the matrix $X$. We record $[\langle f(s), f(d)\rangle, w]$ in the corresponding bucket of the matrix, where $w$ is the edge weight and $f(s)$, $f(d)$ are fingerprints of the two endpoints.
\item Adjacency list buffer $B$ records all left-over edges in $G_h$, whose expected positions in the matrix $X$ have been occupied by other previous inserted edges already.
\end{enumerate}
\end{Def}
When implementing a \fname\ for a graph stream, in order to satisfy the $O(|E|)$ memory cost requirement, we usually set $m=\alpha \times \sqrt[]{|E|}$, where $\alpha$ should be a constant approximate to $1$. To achieve high accuracy, we set $M \gg |V|$. This can be achieved by setting large $F$, in other words, using long fingerprints. When the memory is not sufficient, we can also set smaller $M$ with smaller $m$ and $F$, but this will decrease the accuracy. 
\begin{Exp}
The basic version of \fname\ to store $G_h$ in Figure \ref{compress} is shown in Figure \ref{v1}. Here we set $F=8$. The nodes in the original streaming graph and their corresponding $H(v)$, $h(v)$ and $f(v)$ are shown in the table. In this example, edge $\overrightarrow{2, 10}$ and edge $\overrightarrow{5,18}$ in $G_h$ are stored in the buffer because of collisions with other edges.
\end{Exp}

We discuss the insertion and primitive query operations over GSS as follows:

\textbf{Edge Updating}: When a new item $(s, d; t; w)$ comes in the graph stream $S$, we map it to an edge $\overrightarrow{H(s),H(d)}$ with weight $w$ in graph sketch $G_h$.
Then we find the bucket in row $h(s)$, column $h(d)$. If the bucket is empty, we store the fingerprints pair $\langle f(s), f(d)\rangle$ together with the edge weight $w$ in the bucket. If it is not empty, we compare the fingerprint pair of this edge with the fingerprint pair $\langle f(s^{\prime}), f(d^{\prime})\rangle$ that is in the bucket already. If they are same, we add the weight $w$ to the existing one; otherwise, it means this bucket has been occupied by other edges, and we store edge $\overrightarrow{H(s), H(d)}$ in the adjacency list in the buffer $B$. We call this kind of edges as \emph{left-over} edges.


\nop{Obviously, in this data structure the probability that an edge is stored in the matrix is the same as the correct rate of edge query in a TCM with the same matrix size. From Fig .\ref{TA} we can see that this probability is easy to guarantee. Moreover, in section \ref{DS}, we will further improve the data structure to minimize the size of buffer. Therefore, in \fname\ most edges are stored in the matrix rather than the buffer, to insert them only need $O(1)$ memory accesses. Even if the edge has to be stored in the buffer, because the adjacency list is short, visit to it will be fast. The edge insertion operator has small time cost, which is necessary for updating in the graph stream.
}

\textbf{Graph Query Primitives}: The three primitives (defined in Definition \ref{GQP}) are all supported with our proposed data structure GSS.

\vspace{-0.01cm}
\emph{Edge Query}: Given an edge query $e=\overrightarrow{s,d}$, we work as follows. We check the bucket in row $h(s)$, column $h(d)$ in the matrix. Let $\langle f(s^{\prime}), f(d^{\prime})\rangle$ be the fingerprint pair stored at the bucket. If $\langle f(s^{\prime}), f(d^{\prime})\rangle$ equals to the the fingerprint pair $\langle f(s), f(d)\rangle$ of edge $\overrightarrow{s,d}$, we return the weight in the bucket. Otherwise we search the buffer $B$ for edge $\overrightarrow{H(s),H(d)}$ using the adjacency list. If we cannot find it in the matrix $X$ or in the buffer $B$, we return $-1$, i.e. reporting that the edge $e=\overrightarrow{s,d}$ does not exists.

\vspace{-0.01cm}
\emph{$1$-hop Successor Query}: To find the 1-hop successors of node $v$, we work as follows. First, we search all buckets in row $h(v)$ of the matrix $X$. If a bucket in row $h(v)$ and column $c$ has a fingerprint pair $\langle f(v), f(v_s)\rangle$, we add node $H(v_s)=c\times F+ f(v_s)$ to the 1-hop successors set $SS$. After that, we also need to search the buffer area to find all edges with source node $H(v)$, and add its destination node to the 1-hop successors set $S$. We return $-1$ if we find no result, i.e., $|SS|=0$. Otherwise, for each $H(s)$ in successors set $SS$, we obtain the original node IDs by accessing the hash table.

\vspace{-0.01cm}
\emph{$1$-hop Precursor Query}: To find the 1-hop precursors of node $v$, we have the analogue operations with $1$-hop Successor Query if we switch the columns and the rows in the matrix $X$. The details are omitted due to space limit.
\nop{
In GSS, we use a combination of fingerprints and hash addresses, and both of them help us to distinguish different nodes and edges in the graph sketch. For example, in Figure  \ref{compress}, node $a$ and node $f$ are mapped to different nodes in the graph sketch, and they have the same fingerprint but different addresses. In the data structure the edges connected to them are stored in different areas, thus we will not mix them up. On the other hand, node $a$ and node $c$ have the same address but different fingerprints. Though the edges connected to them are stored in the same area, we can use fingerprints to distinguish them.}

In \fname, we store edges with different source nodes in $G_h$ in one row of the matrix, because the graph is sparse and each node is usually connected to very few edges. We can use fingerprints to distinguish them. For example, edge $\overrightarrow{15,28}$ and edge $\overrightarrow{10,15}$ are all stored in row $1$, but they have different source node fingerprints, namely $2$ and $7$, thus we know exactly which nodes they are from. It is similar in columns. Fingerprints also help us to distinguish edges when they are mapped into the same bucket. This enables us to apply a map function with a much larger value range, and generate a much larger graph sketch with the same size of matrix as TCM. With a $4\times 4$ matrix as in Figure \ref{compress}, TCM can only support a map function with $M=4$ , and the number of nodes in the graph sketch will be no more than $4$, thus the accuracy will he much poorer.

\nop{This technique makes the memory more efficiently used. In TCM, we need to build a ${M}^2=32 \times 32$ matrix to store the graph sketch $G_h$ in Fig \ref{compress}. While in \fname, because it is actually a 2-dimension hash table to store the edges in the compressed graph $G_h$, we only need $O(|E_h|)$ memory. Even if each bucket is larger than TCM, as $O(|E_h|)<<O(|V_h|)^2$ in large graphs, \fname\ still need much smaller memory to get the same accuracy.}

\presec
\section{GSS: Augmented Algorithm}
\postsec
\label{GSS_AA}
As we know, GSS has two parts: a size $m \times m$ matrix $X$ and an adjacency list buffer $B$  for left-over edges. Obviously, we only need $O(1)$ time to insert an edge into $X$, but the linear time $O(|B|)$ if the edge must goto the buffer $B$, where $|B|$ represents the number of all left-over edges. Therefore $|B|$ both influences the memory and time cost. In this section, we design several solutions to reduce the size of buffer $B$. 
\presub
\subsection{Square Hashing}
\postsub
In the basic version, an edge is pushed into buffer $B$ if and only if its mapped position in the matrix $X$ has been occupied. The most intuitive solution is to find another bucket for it. Then where to find an empty bucket? We further notice the skewness in node degrees. In the real-world graphs, node degrees usually follow the power law distribution. In other words, a few nodes have very high degrees, while most nodes have small degrees. Consider a node $v$ that has $A$ out-going edges in the graph sketch $G_h$. For a $m\times m$ adjacency matrix $X$ in GSS (see Definition \ref{def:gss}), there are at least $A-m$ edges that should be inserted into buffer $B$, as these $A$ edges must be mapped to the same row (in $X$) due to the same source vertex $v$. These high degree nodes lead to crowed rows and result in most left-over edges in buffer $B$. On the other hand, many other rows are uncrowded. We have the same observation for columns of matrix $X$. \emph{Is it possible to make use of the unoccupied positions in uncrowded rows/columns?} It is the motivation of our first technique, called \emph{\hashname}.

\nop{
If we can map each edge into multiple buckets, the probability that it find an empty one will be much higher. Moreover, \nop{we notice that in equation .\ref{}, the probability that an edge is stored in the buffer is influenced by a parameter $D$, which is the summary of the current in-degree of its source node and the out-degree of its destination node in the moment this edge is inserted.} in most real-world streaming graphs, the distribution of the node degrees is usually highly skewed. Some nodes have very high degrees, and the rows or columns corresponding to them are crowed, thus the edges connected with them has a high probability to be stored in the buffer. They are usually the main component of the buffer. We hope to distribute these edges into the less crowed area, which will further decrease the buffer.
\nop{The average degree, in other others, $\frac{|E|}{|V|}$, is within $10$, while some nodes may be connected to hundreds or even thousands of edges. This skewness will be maintained in the graph sketch, and bring problems to the storage. For a node $v$ with high degree in the streaming graph $G$, node $H(v)$ in the graph sketch $G_h$ will have higher degrees, and row $h(v)$ and column $h(v)$ will be very crowded. For example, in graph $G$ in Fig.\ref{sample}, the out degree of node $a$, in other words, the number of edges with source node $a$, is $4$, much higher than other nodes. In $G_h$ in Fig. \ref{compress}, node $2$ which $a$ is mapped to also has out degree $4$. Therefore, in the matrix in Fig. \ref{v1}, row $0$ is nearly full and edge $(2,10)$ and $(5, 18)$ which are mapped to this row have to be stored in the buffer. If $a$ has a higher degree, there will be more edges in the buffer. On the other hand, other rows in the matrix still have many empty buckets. This leads to both a waste of memory and a decrease in speed, as when the buffer grows it takes more time to search for an edge in it.}
}
\nop{
To achieve these goals, in the modified version we introduce a technique called \hashname.
}

For each node with ID $H(v)=\langle h(v), f(v) \rangle$ in $G_h$, we compute a sequence of hash addresses $\{h_i(v)| 1\leqslant i \leqslant r\}(0\leqslant h_i(v) <m)$ for it. Edge $\overrightarrow{H(s), H(d)}$ is stored in the first empty bucket among the $r\times r$ buckets with addresses 
\vspace{-0.08in}
\begin{center}
$\{(h_i(s), h_j(d)) |(1\leqslant i \leqslant r, 1\leqslant j \leqslant r)\}$ 
\end{center}
where $h_i(s)$ is the row index and $h_j(d)$ is the column index. We call these buckets \emph{mapped buckets} for convenience. Note that we consider row-first layout when selecting the first empty bucket. 

\begin{figure}[htbp]
\includegraphics[width=0.5\textwidth]{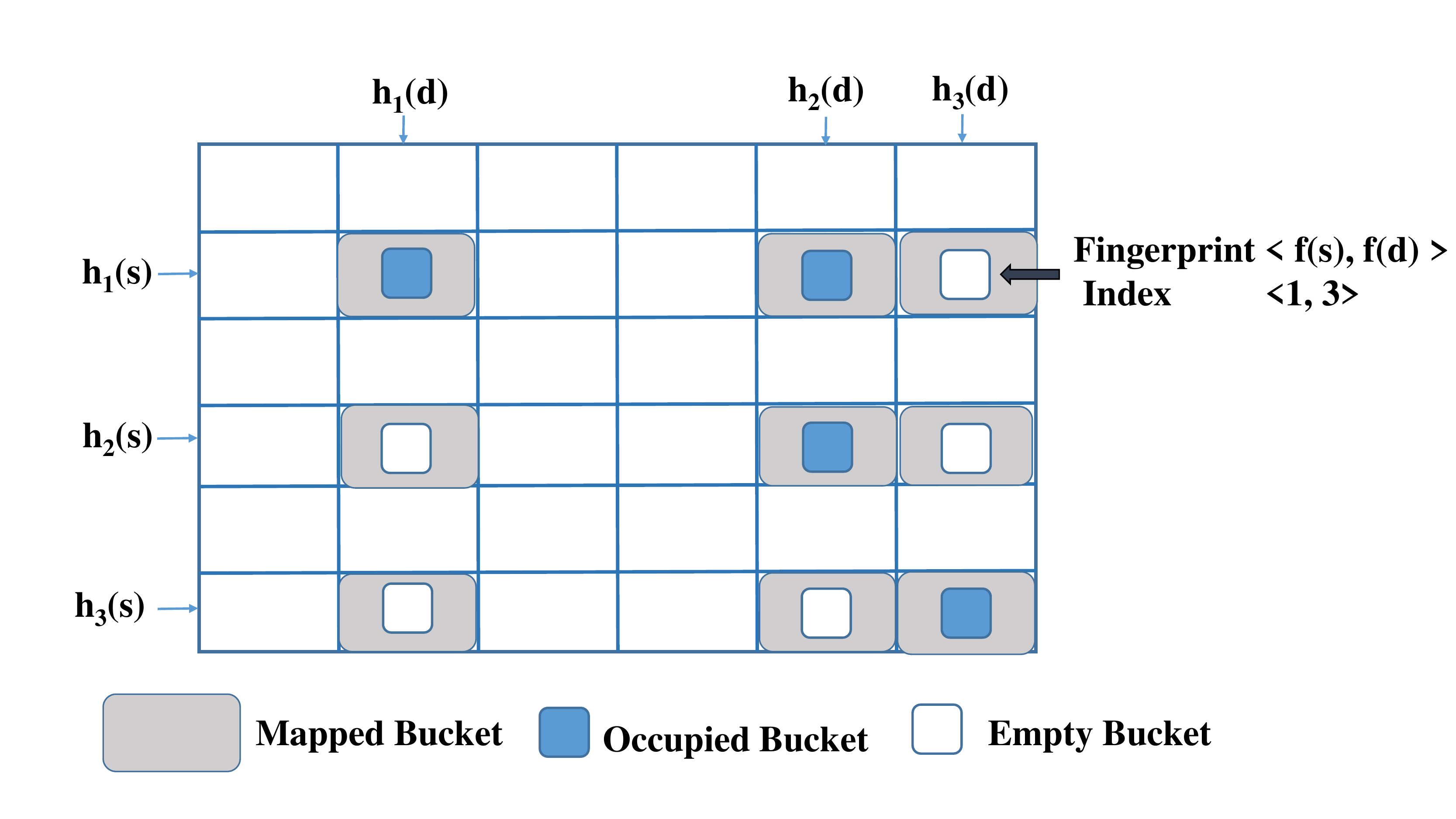}
\caption{The square hashing}
\label{DS}
\end{figure}

\begin{Exp}
An example of \hashname\ is shown in Figure \ref{DS}. The inserted edge is mapped to $9$ buckets, and the first 2 with address $(h_1(s), h_1(d))$ and $(h_1(s), h_2(d))$ have been already occupied. Therefore the edge is inserted in the third mapped bucket. In the bucket, we store the weight, the fingerprint pair, together with an index pair $\langle 1,3 \rangle$ which indicates the position of this bucket in the mapped buckets sequence. We will talk about the use of index pair later.  
\end{Exp}
\prefig
\postfig

The following issue is how to generate a \emph{good} hash address sequence $\{h_i(v)|1\leq i \leq r \}$ for a vertex $v$. There are two requirements:

\Paragraph{Independent:} For two nodes $v_1$ and $v_2$, we use $P$ to represent the probability that $\forall 1\leq i \leq r, h_i(v_1)=h_i(v_2)$. Then we have $P=\prod_{i=1}^{r}Pr(h_i(v_1)=h_i(v_2))$.
In other words, the randomness of each address in the sequence will not be influenced by others. This requirement will help to maximize the chance that an edge finds an empty bucket among the $r\times r$ mapped buckets.

\Paragraph{Reversible:} 
Given a bucket in row $R$ and column $C$ and the content in it, we are able to recover the representation of the edge $e$ in the graph sketch $G_h$: $\overrightarrow{H(s), H(d)}$, where $e$ is the edge in that bucket. This property of indexing is needed in the 1-hop successor query and the 1-hop precursor query. As in these queries, we need to check the potential buckets to see if they contain edges connected to the queried node $v$ and retrieve the other end point in each qualified bucket.

\nop{(The following two paraphrases can be removed, something can be moved above..)

The method to generate the hash address sequence is not easy to find. On one hand, we hope that the addresses in the sequence are independent with each other, thus each edge has higher probability to find an empty bucket. On the other hand, the addresses must be \emph{reversible}, which we define as follows:
\begin{Def}
Reversible addressing: A method to generate addresses for edges so that given a bucket in row $R$ and column $C$ and the content in it, we are able to recover the representation of the edge $e$ in the graph sketch $G_h$, where $e$ is the edge in that bucket.
\end{Def}
This property of indexing is needed in the 1-hop successor query and the 1-hop precursor query.
It is a little difficult to meet the above 2 requirements at the same time. For example, if the sequence of each node is generated by several independent hash functions, it is not reversible unless using additional memory, and if the sequence is an ascending array, the addresses are not independent and the congestion can not be solved well. In \hashname,  this hash address sequence of $H(v)$ is generated with a pseudo random algorithm using both $h(v)$ and $f(v)$ as input. The procedure is as following:}

To meet the above requirements, we propose to use \emph{linear congruence method}\cite{L1999Tables} to generate a sequence of $r$ random values $\left \{q_i(v)|1 \leqslant i \leqslant r \right \}$ with $f(v)$ as seeds. We call this sequence the linear congruential (LR) sequence for convenience. The linear congruence method is as following: select a timer $a$, small prime $b$ and a module $p$, then
\vspace{-0.08in}
\begin{equation}
\label{eq1}
\left\{
\begin{aligned}
q_1(v)& = (a\times f(v)+b)\%p \\
q_i(v)& =(a \times q_{i-1}(v)+b)\% p, (2 \leqslant i \leqslant r) \\
\end{aligned}
\right.
\end{equation}
\vspace{-0.08in}

By choosing $a$, $b$ and $p$ carefully, we can make sure the cycle of the sequence we generate is much larger than $r$, and there will be no repetitive numbers in the sequence \cite{L1999Tables}. Then  we generate a sequence of hash addresses as following:
\vspace{-0.08in}
\begin{equation}
\left \{h_i(v)| h_i(v) = (h(v)+q_i(v))\%m, 1\leqslant i \leqslant r \right \}
\end{equation}

When storing edge $\overrightarrow{H(s), H(d)}$ in the matrix, besides storing the pair of fingerprints and the edge weight, we also store an index pair $(i_s, i_d)$, supposing that the bucket that contains this room has an address $(h_{i_s}(s), h_{i_d}(d))$. As the length of the sequence, $r$, is small, the length of each index will be less than $4$ bits. Therefore storing such a pair will cost little.

Note that the hash sequence $\left \{q_i(v)|1 \leqslant i \leqslant r \right \}$ generated by the linear congruence method are both \emph{independent} and \emph{reversible}. The independence property has been proved in \cite{Khan2016Query}. We show how to recover the original hash value $H(v)$ based on the $f(v)$, $h_i(v)$ and the index $i$ as follows. First, we compute the LR sequence $\{q_i(v)\}$ with $f(v)$ following equation \ref{eq1}. Second we use the equation 
$(h(v)+q_i(v))\%m=h_i(v)$
to compute the original hash address $h(v)$. As $h(v)<m$, the equation has unique solution. At last we use $H(v)=h(v)\times F+f(v)$ to compute $H(v)$. Given a bucket in the matrix, the fingerprint pair $(f(s), f(d))$ and the index pair $(i_s, i_d)$ are all stored in it, and we have $h_{i_s}(s)=R$, $h_{i_d}(d)=C$, where $R$ and $C$ are the row index and the column index of the bucket in the matrix, respectively. Therefore we can retrieve both $H(s)$ and $H(d)$ as above.

\prefig
\begin{figure}[htbp]
\includegraphics[width=0.5\textwidth]{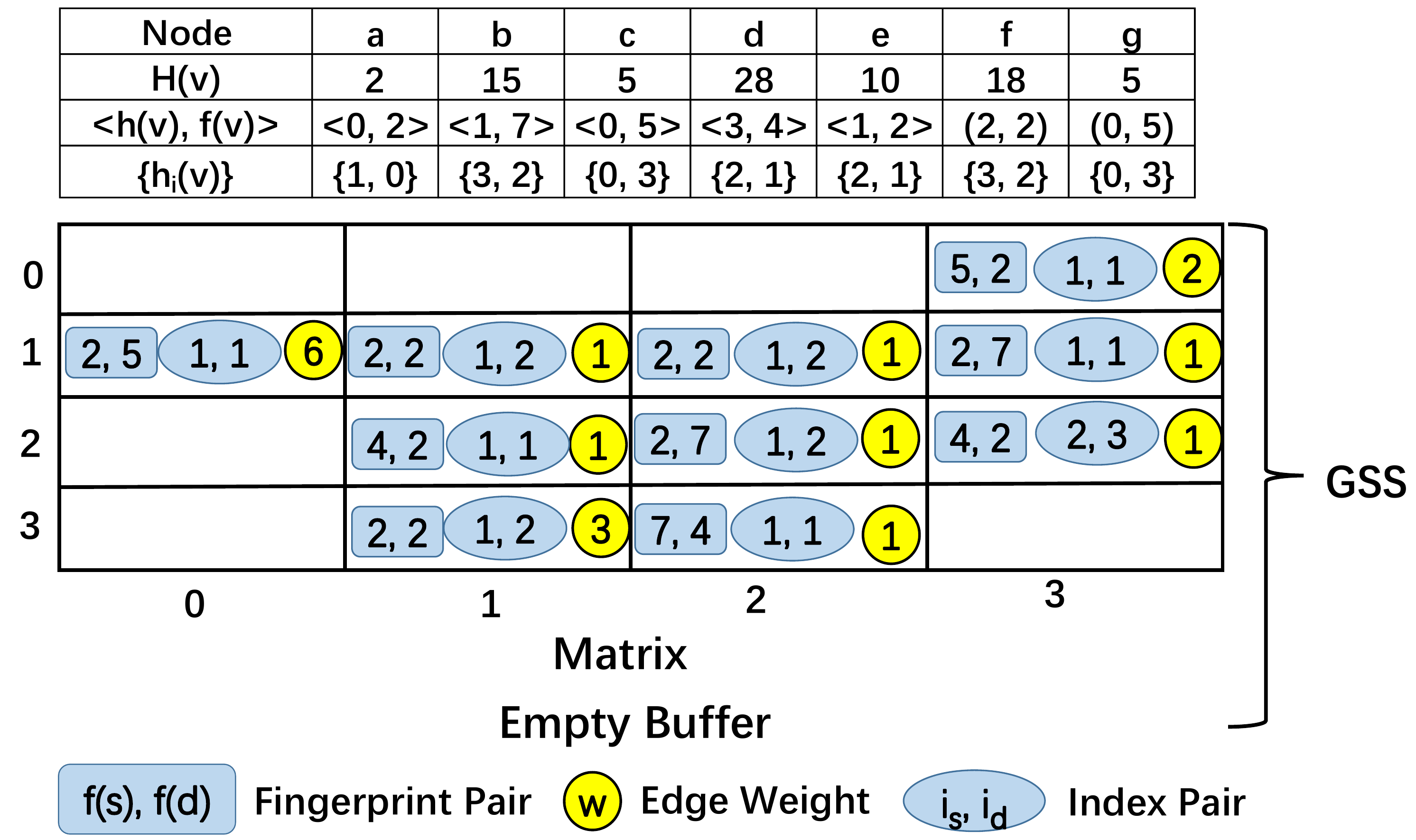}
\label{v2}
\caption{A sample of the modified version of data structure}
\end{figure}
\postfig

\begin{Exp}
An example of the modified version is shown in Fig. \ref{v2}. In the matrix we stored $G_h$ in Fig. \ref{compress}, which is a compressed graph of $G$ in Fig.\ref{sample}. In this example we set $F=8$, $m=4$, $r=2$, and the equation in the linger congruence method is
\vspace{-0.08in}
\begin{equation}
\left\{
\begin{aligned}
q_1(v)& = (5\times f(v)+3)\%8 \\
q_i(v)& =(5 \times q_{i-1}(v)+3)\% 8, (2 \leqslant i \leqslant r) \\
\end{aligned}
\right.
\end{equation}
\end{Exp}

Compared to the basic version, in the modified version all edges are stored in the matrix, and the number of memory accesses we need to find an edge in the matrix is within $2^2=4$. In fact in the example we only need one memory access to find most edges, and $2$ for a few ones.

In the following, we illustrate the four basic operators in this data structure GSS. 

\textbf{Edge Updating}: When a new item $(s,d,t;w)$ comes in the graph stream $S$, we map it to edge $\overrightarrow{H(s), H(d)}$ in the graph sketch $G_h$ with weight $w$. Then we compute two hash address sequences $\{h_i(s)\}$ and $\{h_i(d)\}$ and check the $r^2$ mapped buckets with addresses $\{(h_i(s),h_j(d))|1\leqslant i\leqslant r, 1\leqslant j \leqslant r\}$ one by one. For a bucket in row $h_{i_s}(s)$ and column $h_{i_d}(d)$, if it is empty, we store the fingerprint pair $(f(s), f(d))$ and the index pair $(i_s, i_d)$ and weights $w$ in it, and end the procedure. If it is not empty, we check the fingerprint pair $(f(s^{\prime}), f(d^{\prime}))$ and the index pair $(i_s^{\prime}, i_d^{\prime})$ stored in the bucket. If the fingerprint pair and the index pair are all equal to the corresponding pairs of the new inserted edge $\overrightarrow{H(s), H(d)}$, we add $w$ to the weights in it, and end the procedure. Otherwise it means this bucket has been occupied by other edges and we consider other hash addresses following the hash sequence. If all $r^2$ buckets have been occupied, we store edge $\overrightarrow{H(s), H(d)}$ with weight $w$ in the buffer $B$, like the basic version of GSS. 

\nop{For example, if a new data item $(a, f; 1; t_{13})$ comes in the graph stream in Fig.\ref{sample}. After computation, we find edge $(a, f)$ is mapped to edge $(2, 18)$ in the graph sketch $G_h$, and insert it into the matrix. The hash address sequence for node $2$ is ${1, 0}$, and the hash address sequence for node $18$ is ${3, 2}$. Therefore  we first check bucket in row $1$, column $3$. We find the fingerprint pair in it is $(2, 7)$, which is not equal to the fingerprint pair we need $(2, 2)$, thus we proceed on. For similar reasons the second bucket in address $(1, 2)$ and the third in $(0, 3)$ are all eliminated. At last we find that the forth bucket in address $(0, 2)$ is empty, and we store the fingerprint pair $(2, 2)$, the index pair $(2, 2)$ and the weight $1$ in it.}

\textbf{Graph Query Primitives:} The three graph query primitives are supported in the modified data structure as follows:

\emph{Edge Query}: When querying an edge $e=\overrightarrow{s, d}$, we map it to edge $\overrightarrow{H(s),H(d)}$ in the graph sketch, and use the same \hashname method to find the $r^2$ mapped buckets and check them one by one.  Once we find a bucket in row $h_{i_s}(s)$ and column $h_{i_d}(d)$ which contains the fingerprint pair $(f(s), f(d))$ and the index pair $(i_s, i_d)$, we return its weight as the result. If we find no results in the $r^2$ buckets, we search the buffer for  edge $\overrightarrow{H(s),H(d)}$ and return its weights. If we still can not find it, we return $-1$.

\nop{For example, when querying edge $(e, b)$ in Fig.\ref{v2}, which is mapped to $(10, 15)$ in the graph sketch. We find $4$ mapped buckets in address $(2, 3)$, $(2, 2)$, $(1, 3)$, $(1, 2)$ and check them one by one. \nop{In the bucket with address $(2, 3)$, we find the fingerprint pair stored in it is $(4, 2)$, not equal to the one we need, and continue to check next bucket.} In bucket with address $(2, 2)$, we find that the fingerprint pair is $(2, 7)$, equal to$(f(e), f(b))$, and the index pair is $(1, 2)$, also what we need. Therefore we return its weight 1 as the result.}

\emph{1-hop Successor Query}: to find the 1-hop successors of node $v$, we map it to node $H(v)$ in $G_h$. Then we compute its hash address sequence ${h_i(v)}$ according to $H(v)$, and check the $r$ rows with index $h_i(v),(1 \leqslant i \leqslant r)$. If a bucket in row $h_{i_s}(v)$, column $C$ contains fingerprint pair $((f(v), f(x))$ and index pair $(i_s, i_d)$ where $f(x)$ is any integer in range $[0, F)$ and $i_d$ is any integer in range $[1,r]$, 
\nop{we use $f(x)$ to compute the random value sequence ${q_i(x)}$, and use equation
\begin{equation}
\centering
(h(x)+q_{c_2}(x))\% m=C
\end{equation}
}
we use $f(x)$, $i_d$ and $C$ to compute $H(x)$ as stated above. Then we add $H(x)$ to the 1-hop successor set $SS$. After searching the $r$ rows, we also need to check the buffer to see if there are any edges with source node $H(v)$ and add their destination node to $SS$. We return $-1$ if we find no result, otherwise we obtain the original node IDs from $SS$ by accessing the hash table $\langle H(v), v \rangle$.

\emph{1-hop Precursor Query}: to answer an 1-hop precursor query, we have the analogue operations with $1$-hop Successor Query if we switch the columns and the rows in the matrix $X$. The details are omitted due to space limit.

After applying \hashname, the edges with source node $H(v)$ in $G_h$ are on longer stored in a single row, but spread over $r$ rows with addresses $h_i(v) (1\leqslant i \leqslant r)$. Similarly, edges with destination node $H(v)$ are stored in the $r$ different columns. These rows or columns are shared by the edges with different source nodes or destination nodes. The higher degree a node has, the more buckets its edges may take. This eases the congestion brought by skewness in node degrees. Moreover, as each bucket has multiple mapped buckets, it has higher probability to find an empty one. Obviously, \hashname\ will reduce the number of \emph{left-over edges}.
\nop{for node $H(v)$ in $G_h$, we first get its hash address sequence ${h_i(v)}$, and check the $r$ columns with index $h_i(v),(1 \leqslant i \leqslant r)$. If a bucket in row $C$, column $h_{c_1}(v)$ contains fingerprint pair $((f(x), f(v))$ and index pair $(c_2, c_1)$ where $f(x)$ is any integer in range $[0, F)$ and $c_2$ is any integer in range $[1,r]$, we use $f(x)$ to compute the random value sequence ${q_i(x)}$, and use equation
\begin{equation}
\centering
h(x)+q_{c_2}(x))\% m=C
\end{equation}
to compute $h(x)$. Then we add node ID $H(x)=h(x)\times F+f(x)$ to the 1-hop precursor set. After searching the $r$ columns, we also need to check the buffer to see if there are any edges with destination node $H(v)$ and add their source node to the 1-hop precursor set.}

\nop{As the 1-hop successor query and the 1-hop precursor query has similar procedure, we use one example to illustrate them. In Fig. \ref{v2}, when we query for the 1-hop precursor set of node $e$, we check column $2$ and column $1$ of the matrix. In each column, we search the buckets one by one, skipping the unqualified ones and collecting the source nodes of the qualified edges. For example, in row $2$ column $2$, the second fingerprint in the bucket is not equal to the fingerprint of node $e$, thus we skip this bucket. In row $1$ column $2$, the bucket contain index pair $(1, 2)$. In other words, for the destination of this edge, the column index $2$ is its second hash address. For node $e$, $2$ is its first hash address. Therefore the destination node of the edge is not $e$.   In row $1$ column $1$, we find the second fingerprint and the second index both match. Therefore, we determine that the destination node in this bucket is node $e$. Then we use the first fingerprint in this bucket, $2$, to compute the LR sequence, which is $\{5,4\}$. As the first index is $1$ in this bucket and the row index is $1$, we compute the equation
\begin{equation}
\centering
(h(x)+5)\%4=1, (h(x)<4)
\end{equation}
and get the hash value of the source node $h(x)=0$. At last we compute the ID of the source $0\times 8+2=2$, and add it to the 1-hop precursor set. Similarly we check the other rows in this column and another get 1-hop precursor $18$. As the buffer is empty, we end the procedure with these two 1-hop precursors and retrieve the original node id $a, f$ with the hash table.}

\presub
\subsection{Further Improvements}
\postsub
There are some other improvements which can be implemented to \fname. 
\subsubsection{Mapped Buckets Sampling}
In the modified version of \fname, each edge has $r^2$ mapped buckets. We usually set $r$ to integers from $4$ to $16$. When the skewness of node degrees is serious, $r$ can be larger. If we check all the $r^2$ buckets when inserting an edge, it will be time consuming. To improve the updating speed, which is very important for graph stream summarization, we can use a sampling technique to decrease the time cost. Instead of check all the $r^2$ buckets, we select $k$ buckets as a sample from the mapped buckets, we call these buckets \emph{candidate buckets} for short. For each edge we only check these $k$ buckets in updating and query, and the operations are the same as above. \nop{ In this procedure, once we find the edge to insert, we add the new weight to it and end the procedure, or once we find an empty bucket, we insert the edge into this bucket and end the procedure.} The method to select these $k$ buckets for an edge $e$ is also a linear congruence method. We add the fingerprint of the source node and the destination node of $e$ to get a seed $seed(e)$, then we compute a $k$ length sequence as

\vspace{-0.08in}
\begin{equation}
\label{eq1}
\left\{
\begin{aligned}
q_1(e)& = (a\times seed(e)+b)\%p \\
q_i(e)& =(a \times q_{i-1}(e)+b)\% p, (2 \leqslant i \leqslant k) \\
\end{aligned}
\right.
\end{equation}
where $a$, $b$ and $p$ are the same integers used above. We choose the $k$ buckets with address
\begin{equation}
\centering
\left\{ (h_{\left \lfloor \frac{q_i(e)}{r}\right \rfloor \%r   }(s), h_{(q_i(e)\%r)}(d))| 1 \leqslant i \leqslant k \right\}
\end{equation}
$\{h_i(s)\}$ and $\{h_i(d)\}$ are the hash address sequence of the source node and the destination node, respectively.

\subsubsection{Multiple Rooms}
When the memory is sufficient, we do not need to use multiple matrices to increase accuracy as TCM, as the accuracy is already very high. Instead, in order to further decrease the buffer size, we can separate each bucket in the matrix into $l$ segments, and each segments contains an edge, including the weight, the fingerprint pair and the index pair. We call each segment a \textit{room} for convenience.
When performing the basic operators, we use the same process as above the find the buckets we need to check, and search all the rooms in them to find qualified edges or empty rooms. \nop{This modification enables each bucket to contain multiple edges, and decreases the size of the buffer further.}

\begin{figure}[htbp]
\centering
\includegraphics[width=0.4\textwidth]{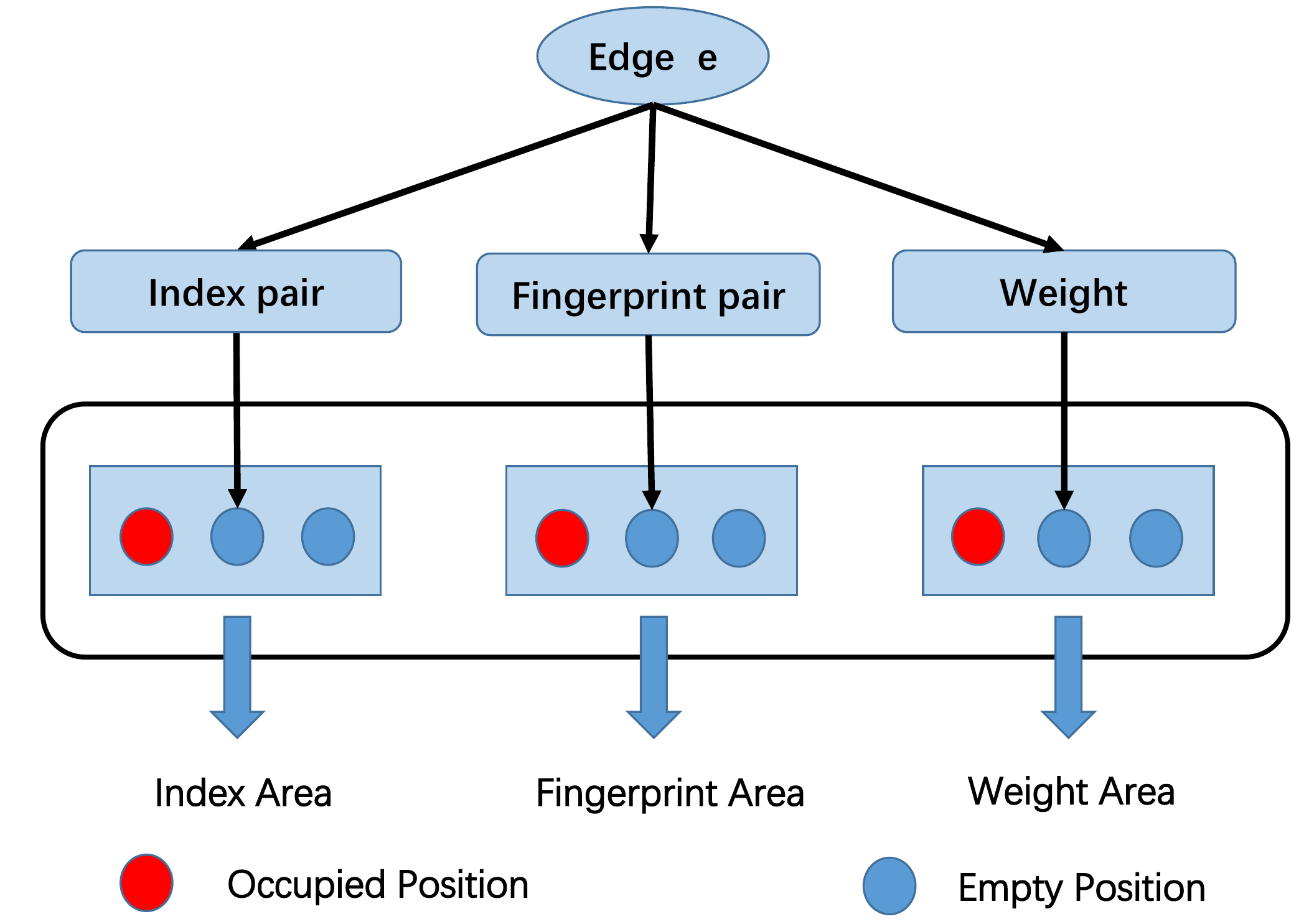}
\label{split}
\caption{Bucket Separation}
\end{figure}

However, when the rooms in each bucket are stored separately, the speed will probably decrease, as we can not fetch the $l$ rooms in one memory access in most cases, and multiple memory accesses increase the time cost. As shown in Fig. \ref{split}, we separate the bucket into $3$ area: the index area, the fingerprint area, and the weight area. Each area contains the corresponding parts of the $l$ rooms. When we check this bucket to find certain edges, we can first check all the index pairs. If we find a matched index pair, we check the corresponding fingerprint pair, and if the fingerprint pair is also matched, we fetch the corresponding weights. If we do not find any matched index pair, we can just move on and do not need to check the fingerprint pairs any more. As the index pairs are very small, usually no more than $1$ byte, we can fetch all the index pairs in one memory access. This will omit a lot of unnecessary memory accesses.

\prefig
\postfig
\presub
\presec
\section{Analysis}
\postsec
\label{analysis}

\subsection{Memory and Time Cost Analysis}
\label{CostAnalysis}
\postsub
As stated above, GSS has $O(|E|)$ memory cost and constant update speed. The memory cost of GSS is $O(|E_h|+|B|)$, to be precise, where $|E_h|$ is the number of edges in the graph sketch $G_h$ and $|B|$ is the size of buffer. When we use hash table to store the original ID, additional $O(|V|)$ memory is needed, but the overall memory cost is still $O(|E|)$. 
The update time cost is $O(k+\frac{|B|}{|E_h|}|B|)$, where $k$ is the number of sampled buckets and is a small constant. When an edge is stored in the matrix, we only need to check at most $k$ candidate buckets, which takes $O(k)$ time. Each edge has probability $\frac{|B|}{|E_h|}$ to be stored in the buffer. When it is stored in the buffer, the update takes additional $O(|B|)$ time, as the buffer is an adjacency list.  
In implementations the buffer stores $0$ edges in most cases, which will be shown in section \ref{buffersize} and \ref{buffesize}. Therefore $\frac{|B|}{|E_h|}|B|$ is also a small constant. When it is necessary to store the ID of nodes in applications, one insertion to the hash table is needed, which also takes constant time. Overall, the update time cost is $O(1)$.

The time cost of queries is based on the algorithms we use. We consider the time cost of the operators as an evaluation. The time cost of the edge query operator is the same as the update, and the time cost of the 1-hop successor query and 1-hop precursor query is $O(rm+|B|)$, where $m$ is the side length of the matrix and $r$ is the length of the hash address sequence.

\nop{The average time cost in update in $O(1)$. In the matrix, the update need constant time. As very few (less than $0.01\%$), even none edges are stored in the buffer, we do not need to access the buffer in most cases. Even when we need to check it, the speed is still high as there are few edges in it. Therefore the average time cost of inserting an edge is still $O(1)$. As for the time cost of query, it depends on the kind of query and the algorithm we use. We consider the time cost of the operators as an evaluation of the query speed. For edge query operator, as the process is the similar to edge insertion, the time cost is also $O(k)$. For the 1-hop successor query and the 1-hop precursor query, the time cost is $O(r\times m)$, where $r$ is the length of the hash address sequence, and $m$ is the side length of the matrix.}

\presub
\subsection{Accuracy Analysis}
\label{AccuracyAnalysis}
\postsub

In this part we evaluate the accuracy of \fname. Before we analyze the probability of errors, we first propose the following theorem:
\begin{thm}\label{thm:1}
The storage of the graph sketch $G_h$ in the data structure of \fname\ is accurate. Which means for any edge $e_1=\overrightarrow{H(s_1), H(d_1)}$ and $e_2=\overrightarrow{H(s_2), H(d_2)}$ in $G_h$, the weights of them will be added up if and only if $H(s_1)=H(s_2), H(d_1)=H(d_2)$.
\end{thm}

As the buffer is an adjacency list that stores edges in $G_h$ accurately, we only need to check the matrix.
If we want to prove the storage of the graph sketch is accurate, we need to prove that if such collision happens to $e_1$ and $e_2$, we have $e_1=e_2$ in $G_h$, in other words, $H(s_1)=H(s_2), H(d_1)=H(d_2)$ .We assume that the bucket contains the wights of $e_1$ and $e_2$ is in row $R$ and column $C$ in the matrix. Obviously $e_1$ and $e_2$ must have the same fingerprint pair, otherwise it will be easy for us to differentiate them. With the same fingerprints, these two edges will produce the same LR sequences $\{q(s)\}$ and $\{q(d)\}$. Moreover, this bucket must have the same index pair for these two edges, and we represent this pair with $(c_1, c_2)$. Then we have
 \begin{equation}
 \left\{
 \begin{aligned}
  &(h(s_1)+q_{c_1}(s_1))\%m=R\\
  &((h(s_2)+q_{c_1}(s_2)))\%m=R \\
  &q_{c_1}(s_1)=q_{c_1}(s_2)\\
  \end{aligned}
  \right.
  \end{equation}
 and
 \begin{equation}
 \left\{
 \begin{aligned}
  &(h(d_1)+q_{c_2}(d_1))\%m=C\\
  &((h(d_2)+q_{c_2}(d_2)))\%m=C \\
  &q_{c_2}(d_1)=q_{c_2}(d_2)\\
  \end{aligned}
  \right.
 \end{equation}
With these equations, we can get that $h(d_1)=h(d_2)$, $h(s_1)=h(s_2)$ when $ 0 \leqslant h(s_1), h(s_2), h(d_1), h(d_2)< m$. With the same fingerprint pair and hash values, we have $H(s_1)=H(s_2)$, $H(d_1)=H(d_2)$. $e_1$ and $e_2$ are the same edge in $G_h$. Therefore the storage of $G_h$ is accurate.

 This theorem means we only need to consider the procedure of mapping $G$ to $G_h$, as all errors happen in this procedure.
We use $\hat{P}$ to represent the probability of the following event:
\begin{Def}
\textbf{Edge Collision}: An edge collision means that given an edge $e$, there is at least one $e'$ in $G$ and $e' \neq e$ which satisfies $H(e)=H(e')$ in the compressed graph $G_h$.
\end{Def}

We set $P=1-\hat{P}$, and $P$ is the main component of the error rate of all the $3$ graph query primitives.

In the edge query, $P$ is just the correct rate. 
In the 1-hop successor query for a node $v$, the correct rate is ${P}^{|V|-d}$, where $|V|$ is the number of nodes in $G$, and $d$ is the out-degree of the queried node. Because we will get a correct answer if and only if for each $v'$ in $G$ which is not a 1-hop successor of $v$, $(v, v')$ does not collide with any existing edges, and the probability of such an event is $P$. It is the same in the 1-hop precursor query. 
Therefore we need to compute $P$ to evaluate the accuracy of \fname.

\presub
\subsection{Collision Rate}
\label{CollisionAnalysis}
\postsub
Now we show the probability that an edge $e$ suffers from edge collision, $\hat{P}$.
For $e=\overrightarrow{s, d}$ in $G$, we assume there are $D$ edges with source node $s$ or destination node $d$ in $G$ besides $e$, and there are totally $|E|$ edges in $G$. We represent the size of the value range of the map function $H(\cdot)$ with $M$.

For an edge share no common endpoints with $e$ , it will collide with $e$ when both its source node and destination node collide with the corresponding node of $e$. The probability that it collides with $e$ in map function $H(\cdot)$ is:
\begin{equation}
p_1 = \frac{1}{M^2}
\end{equation}
The probability that all the $|E|-D$ edges have no collisions with $e$ is
\begin{equation}
Pr_1={(1-p_1)}^{|E|-D}
\end{equation}
For those $D$ edges connected to $e$, as one of the two end points is the same, the probability that such an edge has a collision with $e$ is
\begin{equation}
p_2 = \frac{1}{M}
\end{equation}
The probability that all the $D$ edges have no collisions with $e$ is
\begin{equation}
Pr_2={(1-p_2)}^{D}
\end{equation}
Therefore the correct rate of $e$, in other words, all the $|E|$ edges do not have collisions with $e$ in mapping is
\begin{equation}
\begin{aligned}
P&=Pr_1 \times Pr_2\\
&=({(1-p_1)}^{|E|-D})\times ({(1-p_2)}^{D})\\
&=e^{-p_1 \times (|E|-D)} \times e^{-p_2 \times D}\\
&=e^{-\frac{|E|-D}{M^2}} \times e^{-\frac{D}{M}}\\
&=e^{-\frac{|E|+(M-1)\times D}{M^2}}\\
\end{aligned}
\end{equation}
And $\hat{P}=1-P$. In \fname\ we have $M=m\times F$, where $m$ is the length of the matrix, and $F$ is the maximum size of the fingerprints. 
The above correct rate is usually very high in applications. For example, suppose that the fingerprints we use are $8$-bit, in other words, $F=256$, and when querying an edge $e$, we have $|E|=5 \times 10^5$,$D=200$. We use a matrix with side length $m=1000$. Then the correct rate of this edge query is $e^{-0.00078}=0.9992$. 
On the other hand, in TCM the accuracy analysis is the same as \fname\, but we have $M=m$. This lead to the difference on accuracy with the same size of matrix.
With the same matrix size, TCM only has a probability of $0.497$ to get a correct weight for $e$.

\presub
\subsection{Buffer Size Analysis}
\label{buffersize}
\postsub

After all the improvements, the buffer in \fname is very small. The mathematical expression of the buffer size is very complicated and is influenced by many details of the graph. Therefore we give an expression of the probability that a new edge $e$ becomes a \emph{left-over edge}, which means inserted into the buffer, as a measurement. Assuming that there are already $N$ different edges in the graph stream, and among them $D$ edges have common source node or common destination node with $e$. \nop{After mapping to the graph sketch $G_h$, there will be $Cr\times N$ different edges, where $Cr$ is the average correct rate of edge query. This is the accurate number of edges that inserted into the data structure, but as shown above, $Cr$ is very close to $1$ in \fname. Therefore we ignore it for simplicity.} The length of the matrix is $m$, and each bucket in the matrix has $l$ rooms. For each node we compute a hash address sequence with length $r$. For each edge we choose $k$ \emph{candidate buckets} among the $r^2$ mapped buckets. Then the probability that $e$ becomes a left-over edge is:
For each candidate bucket of $e$, as the $N-D$ non-adjacent edges are randomly inserted into the matrix with area $m^2$, the probability that there are $a_1$ non-adjacent edges inserted into it is:
\begin{equation}
\begin{aligned}
p_1(a_1)&=\binom{N-D}{a_1}\times {(\frac{1}{m^2})}^{a_1} \times {(1-\frac{1}{m^2})}^{N-D-a_1}\\
&=\binom{N-D}{a_1}\times {(\frac{1}{m^2})}^{a_1} \times {e}^{-\frac{N-D-a_1}{m^2}}
\end{aligned}
\end{equation}
As the $D$ adjacent edges are randomly inserted in an area of $r \times m$ ($r$ rows or $r$ columns in the matrix), the probability that there are $a_2$ adjacent edges inserted into this bucket is:
\begin{equation}
\begin{aligned}
p_2(a_2)&=\binom{D}{a_2}\times {(\frac{1}{r \times m})}^{a_2} \times {(1-\frac{1}{r\times m})}^{D-a_2}\\
&=\binom{D}{a_2}\times {(\frac{1}{r \times m})}^{a_2} \times {e}^{-\frac{D-a_2}{r\times m}}
\end{aligned}
\end{equation}
The probability that there are already $n$ edges inserted into this bucket is:
\begin{equation}
\begin{aligned}
p(n)&=\sum_{a=0}^{n}p_1(a)\times p_2(n-a)\\
\end{aligned}
\end{equation}
The probability that there are less than $l$ edges inserted into this bucket is:
\begin{equation}
\begin{aligned}
&Pr=\sum_{n=0}^{l-1}p(n)\\
&=\sum_{n=0}^{l-1}\sum_{a=0}^{n}p_1(a)\times p_2(n-a)\\
&=\sum_{n=0}^{l-1}\sum_{a=0}^{n}\binom{N\!-\!D}{a}\binom{D}{n\!-\!a}{(\frac{1}{m^2})}^{a}{(\frac{1}{rm})}^{n\!-\!a}{e}^{-(\frac{N\!-\!D\!-\!a}{m^2}+\frac{D\!-\!n\!+\!a}{rm})}\\
\end{aligned}
\end{equation}
This is also the lower bound that the bucket is still available for $e$.  The probability that $e$ can not be inserted into the matrix is the probability that all the $k$ candidate buckets are not available, which is:
 \begin{equation}
\begin{aligned}
P=(1-Pr)^k
\end{aligned}
\end{equation}
where
\begin{small}
\begin{equation}
\begin{aligned}
&Pr
=\sum_{n=0}^{l-1}\sum_{a=0}^{n}\binom{N\!-\!D}{a}\binom{D}{n\!-\!a}{(\frac{1}{m^2})}^{a}{(\frac{1}{rm})}^{n\!-\!a}\!\!{e}^{-(\frac{N\!-\!D\!-\!a}{m^2}+\frac{D\!-\!n\!+\!a}{rm})}\\
\end{aligned}
\end{equation}
\end{small}
Notice that this is an upper bound as we ignore collisions in the map procedure from $G$ to $G_h$. This probability is rather small. 
For example if $N=1\times 10^6$, $D=10^4$, we still set the side length of the matrix to $w=1000$, and set $r=8$, $l=3$, $k=8$, the upper bound probability of insertion failure is only $0.002$. Experiments show that when the size of matrix is nearly equal to the number of edges, there will be almost no edges inserted into the buffer.  
\presec
\section {Experimental Evaluation}
\postsec
\label{sec:experiment}

\newcommand{\dataseta}{email-EuAll}
\newcommand{\datasetb}{cit-HepPh}
\newcommand{\datasetc}{web-NotreDame}
\newcommand{\datasetd}{lkml-reply}
\newcommand{\datasete}{Caida-networkflow}

In this section, we show our experimental studies of \fname. 
\nop{Among the prior arts, only TCM and the gMatrix support all kinds of queries, and despite of its versatility, the gMatrix has the same accuracy as TCM. Therefore we regard TCM as state of the art and compare \fname\ with it in experiments. We first explain the data sets in experiments (\ref{dataset}) and evaluation metrics (\ref{metrics}). Next,} We compare \fname\ with TCM on the three graph query primitives: edge query , 1-hop successor query, 1-hop precursor query (\ref{edgequery}) and two compound queries, node queries (\ref{nodequery}) and reachability queries (\ref{reachabilityquery}). We also evaluate the size of buffer (\ref{buffersize})and update speed of \fname\ (\ref{EUS}). Then we further compare \fname\ with the state-of-the-art graph processing algorithms on triangle counting and subgraph matching \ref{ECQ}. 

All experiments are performed on a server with dual 6-core CPUs (Intel Xeon CPU E5-2620 @2.0 GHz, 24 threads) and 62 GB DRAM memory, running Ubuntu. All algorithms including \fname\ and TCM are implemented in C++.
\presub
\subsection{Data Sets}\label{dataset}
\postsub
We choose three real world data sets. Details of three data sets are described as follows:

1)\textbf{\dataseta}\label{dataset1}
\footnote{http://snap.stanford.edu/data/email-EuAll.html}.This data set is communication network data generated using email data from a large European research institution for a period of 18 months. Each node in the directed graph corresponds to an email address. Each edge between node \emph{src} and \emph{dst} represents \emph{src} sent at least one email to \emph{dst}. The directed graph contains 265214 nodes and 420045 edges. We use the Zipfian distribution to add the weight to each edge and the edge weight represents the appearance times in the stream.

2)\textbf{\datasetb}\label{dataset2}\footnote{http://snap.stanford.edu/data/email-EuAll.html}.It is the Arxiv HEP-PH (high energy physics phenomenology) citation graph. If a paper \emph{src} cites paper \emph{dst}, the data set contains a directed edge from  \emph{src} to \emph{dst}. The data set covers  34,546 papers as nodes with 421,578 edges. The edge weights are also added using Zipfian distribution.

3)\textbf{\datasetc}\label{dataset3}
\footnote{http://konect.uni-koblenz.de/networks/lkml-reply}.It is a web graph collected from the University of Notre Dame. Nodes represent web pages and directed edges represent hyperlinks between pages. The data set contains 325729 nodes and 1497134 edges.  We use the Zipfian distribution to generate weights for the edges in the data set, and insert the edges into the data structure one by one to simulate the procedure of real-world incremental updating. 

4)\textbf{\datasetd}\label{dataset4}\footnote{http://konect.uni-koblenz.de/networks/lkml-reply}.It is a collection of communication records in the network of the Linux kernel mailing list. It contains 63399 email addresses (nodes) and 1096440 communication records(edges). Each edge is weighted by its frequency in the data set, and has a timestamp indicating the communication time. We feed the data items to the data structure according to their timestamps to simulate a graph stream.

5)\textbf{\datasete}\label{dataset5} \footnote{www.caida. org} It is the “CAIDA Internet Anonymized Traces 2015 Dataset”. It contains 445440480 communication records (edges) concerning 2601005 different IP addresses (nodes). Each edge is weighted by its frequency in the data set, and has a timestamp indicating the communication time. We feed the data items to the data structure according to their timestamps to simulate a graph stream.

The function we use to cumulate the edge weights is addition. In this case, TCM and \fname\ only have over-estimations. The codes are open sourced\footnote{https://github.com/Puppy95/Graph-Stream-Sketch}
\presub
\subsection{Metrics}\label{metrics}
\postsub
In this part we give a definition of the metrics we use in experiments.

\textbf{Average Relative Error (ARE):} ARE measures the accuracy of the reported weights in edge queries and node queries. Given a query $q$, the \emph{relative error} is defined as
\[
\begin{aligned}
RE(q) = \frac{\hat{f(q)}}{f(q)}-1
\end{aligned}
\]
where $f(q)$ and $\hat{f(q)}$ are the real answer and the estimated value of $q$. When giving a query set, the \emph{average relative error} ($ARE$) is measured by averaging the relative errors over all queries int it. A more accuracy data structure will have smaller $ARE$.


\textbf{Average Precision} We use average precision as the evaluation metric in 1-hop successor queries, 1-hop precursor queries and graph pattern matching. Given such a query $q$, we use $SS$ to represent the accurate set of 1-hop successors / precursors of the queried node $v$, and $\hat{SS}$ to represent the set we get by $q$. As TCM and \fname、 have only false positives, which means $SS \subseteq \hat{SS}$, we define the precision of $q$ as:
\vspace{-0.2cm}
\[
\begin{aligned}
Precision(q) = \frac{|SS|}{|\hat{SS}|}
\end{aligned}
\]
Average precision of a query set is the average value of the precision of all queries in it.
A more accuracy data structure will have higher \emph{Average Precision} 

\textbf{True Negative Recall:} It measures the accuracy of the reachability query. Because connectives of all edges are kept, there is no false negatives in TCM and \fname, which means if we can travel to $d$ from $s$ in the streaming graph, the query result of these data structures will be definitely yes. Therefore in experiments we use reachability query sets $Q=\{q_1,q_2,...,q_k\}$ where $\forall q_i\in Q$, source node $s$ and destination node $d$ in $q_i$ are unreachable. \emph{True negative recall} is defined as  the number of queries reported as unreachable divided by the number of all queries in $Q$.

\textbf{Buffer Percentage:} It measures buffer size of \fname.  \emph{Buffer percentage} is defined as the number of edges that the buffer contains divided by the total number of edges in the graph stream.
\presub
\subsection{Experiments settings}\label{setting}
\postsub
In experiments, we implement two kinds of GSS with different fingerprint sizes: 12 bits and 16 bits, and vary the matrix size. We use \emph{fsize} to represent the fingerprint size by short. We apply all improvements to GSS, and the parameters are as follows. Each bucket in the matrix contains $l=2$ rooms. The length of the address sequences is $r=16$, and the number of \emph{candidate buckets} for each edge is $k=16$ ($r=8$,$k=8$ for the small data set \dataseta and \datasetb). As for $TCM$, we apply $4$ graph sketches to improve its accuracy, and allow it to use larger memory, because otherwise the gap between it and \fname\ will be so huge that we can hardly compare them in one figure. 
In edge query primitives, we allow TCM to use $8$ times memory, and in other queries we implement it with $256$ times memory, as its accuracy is too poor in these queries (in \datasetc, we implement it with $16$ times memory because of the limitation of the memory of the server). This ratio is the memory used by all the 4 sketches in TCM divided by the memory used by \fname\ with 16 bit fingerprint. When the size of \fname\ varies, the size of matrix in TCM also varies correspondingly to keep the ratio unchanged.
\presub
\subsection{Experiments on query primitives}\label{edgequery}
\postsub
\nop{In this section, we evaluate the performance of \fname \ in the basic primitives.   To make the two data structure comparable in accuracy, we compare \fname\  with TCM by fixing the ratio of the memory used by TCM and \fname\  8 times in edge queries, and 256 times in 1-hop successor/precursor queries. In all experiments , we implement TCM and \fname \  using square matrices. Because without a prior knowledge of graph stream distribution, it's difficult to decide a reasonable size of the non-square matrix.}

\begin{figure*}[htb]
\centering
\subfigure[\dataseta]{
\begin{minipage}[b]{0.28\textwidth}
\begin{center}
\includegraphics[width=1\textwidth]{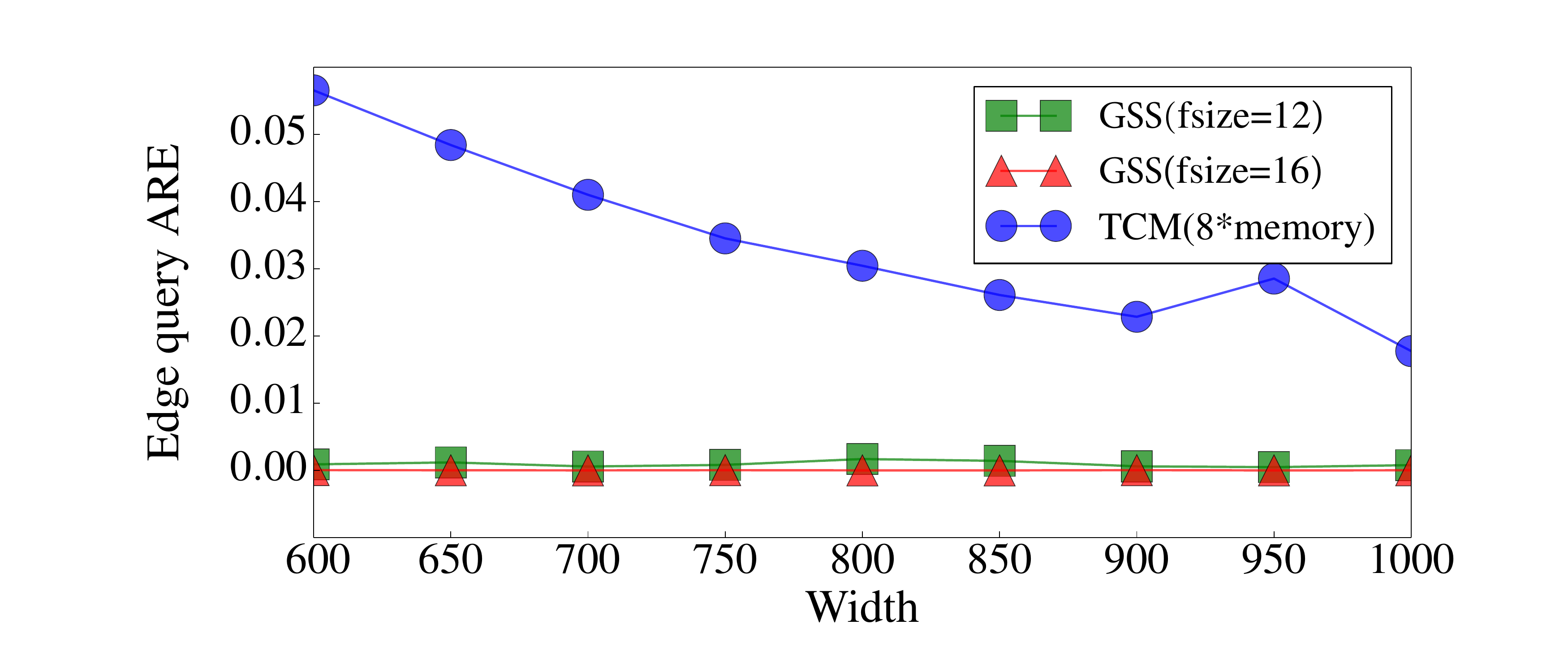}\end{center}
\label{EEedge-ARE}
\vspace{-0.4cm}
\end{minipage}
}
\subfigure[\datasetb]{
\begin{minipage}[b]{0.28\textwidth}
\begin{center}
\includegraphics[width=1\textwidth]{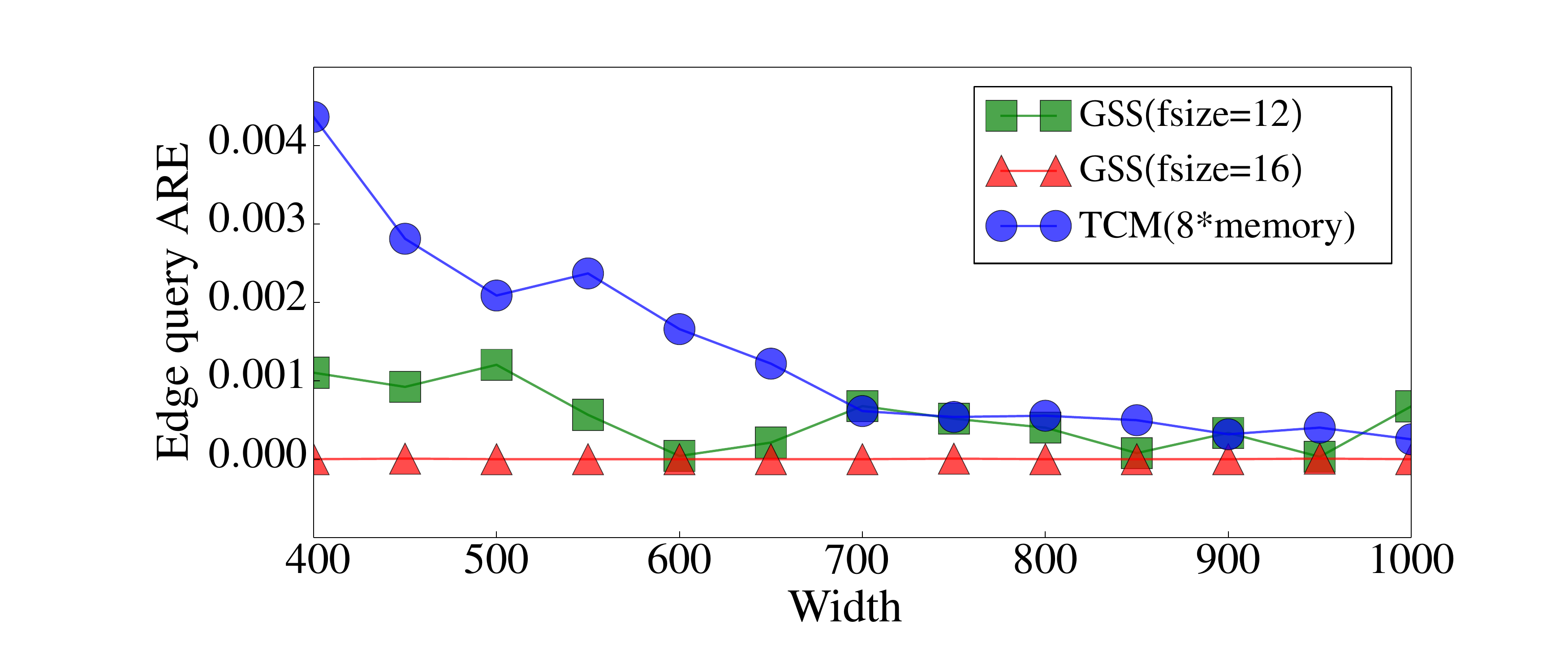}\end{center}
\label{citedge-ARE}
\vspace{-0.4cm}
\end{minipage}
}
\subfigure[\datasetc]{
\begin{minipage}[b]{0.28\textwidth}
\begin{center}
\prefig
\includegraphics[width=1\textwidth]{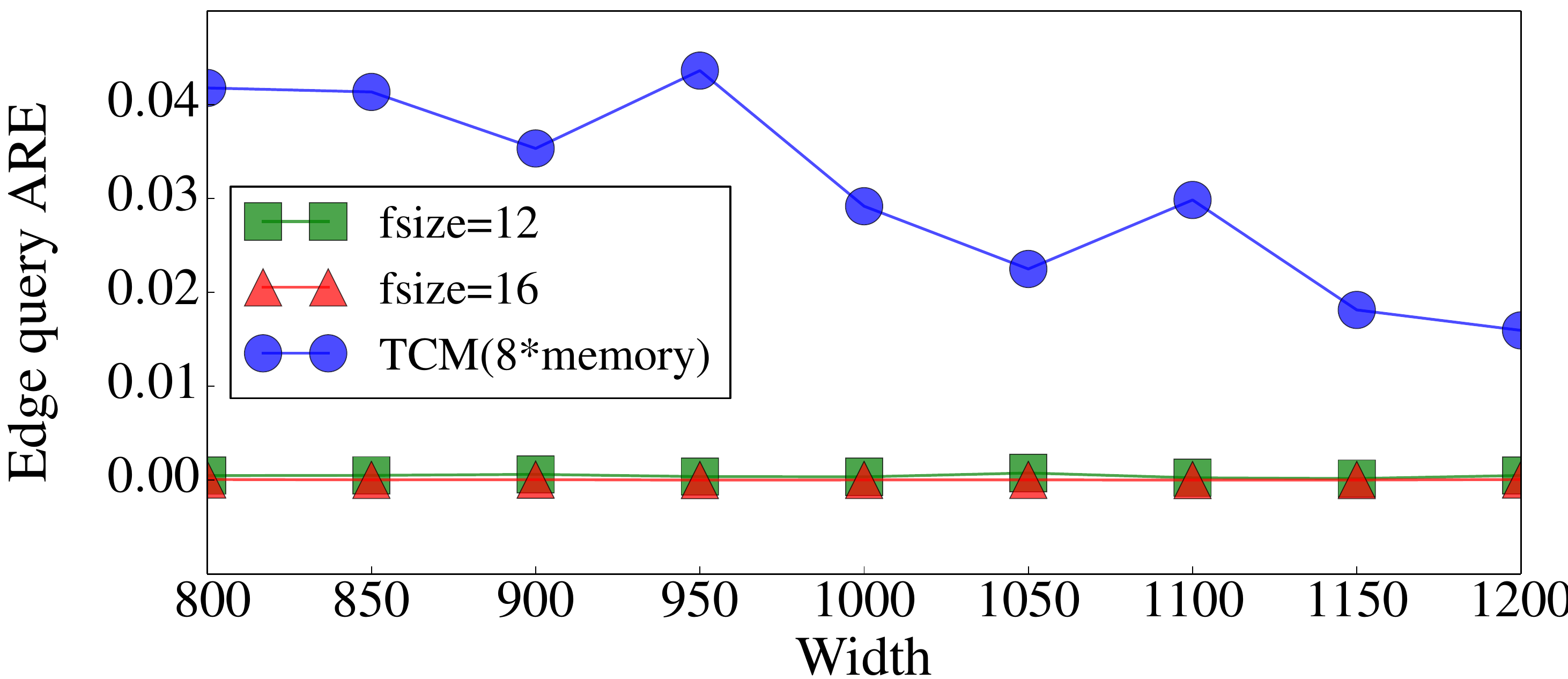}\end{center}
\postfig
\label{EEedge-ARE}
\vspace{-0.4cm}
\end{minipage}
}
\subfigure[\datasetd]{
\begin{minipage}[b]{0.28\textwidth}
\begin{center}
\prefig
\includegraphics[width=1\textwidth]{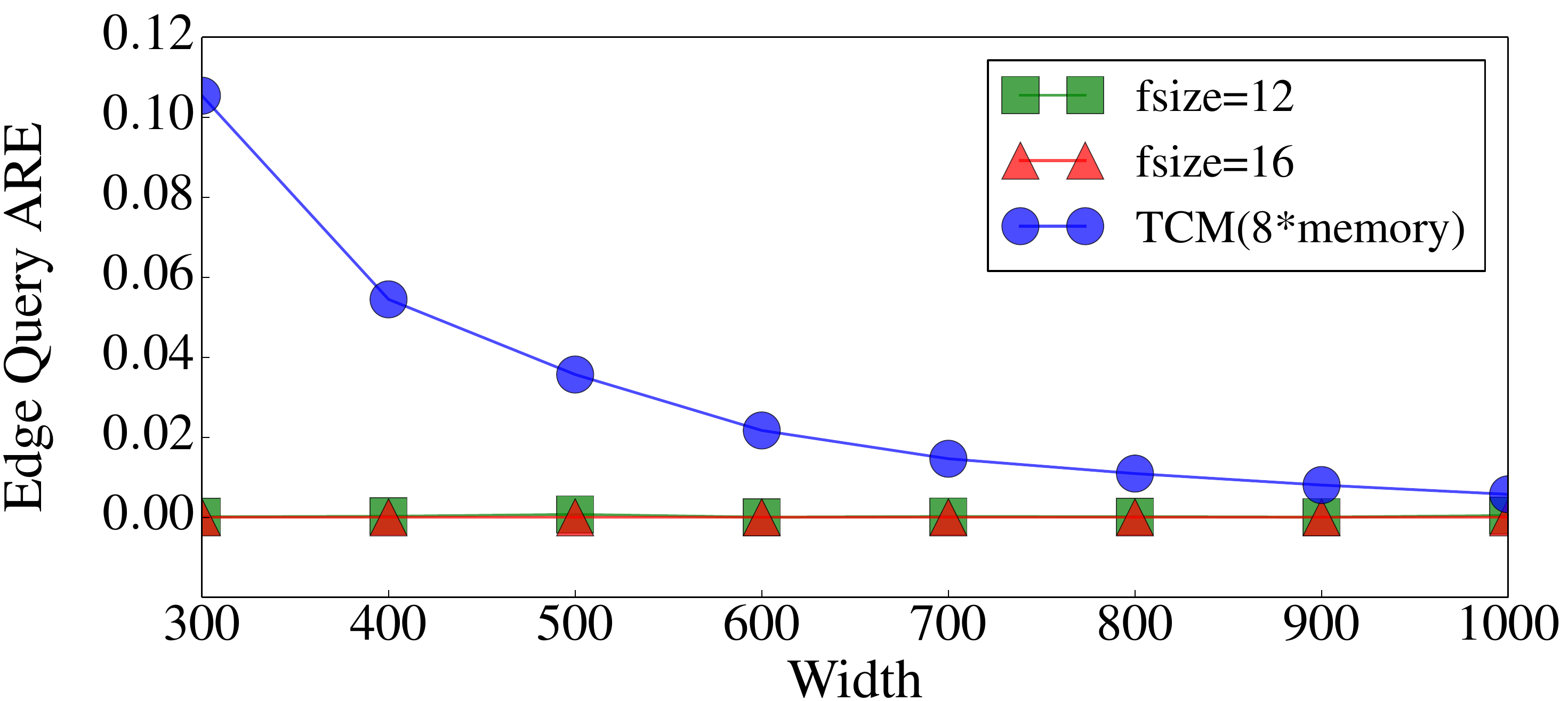}
\end{center}
\postfig
\label{WNedge-ARE}
\vspace{-0.4cm}
\end{minipage}
}
\subfigure[\datasete]{
\begin{minipage}[b]{0.28\textwidth}
\begin{center}
\prefig
\includegraphics[width=1\textwidth]{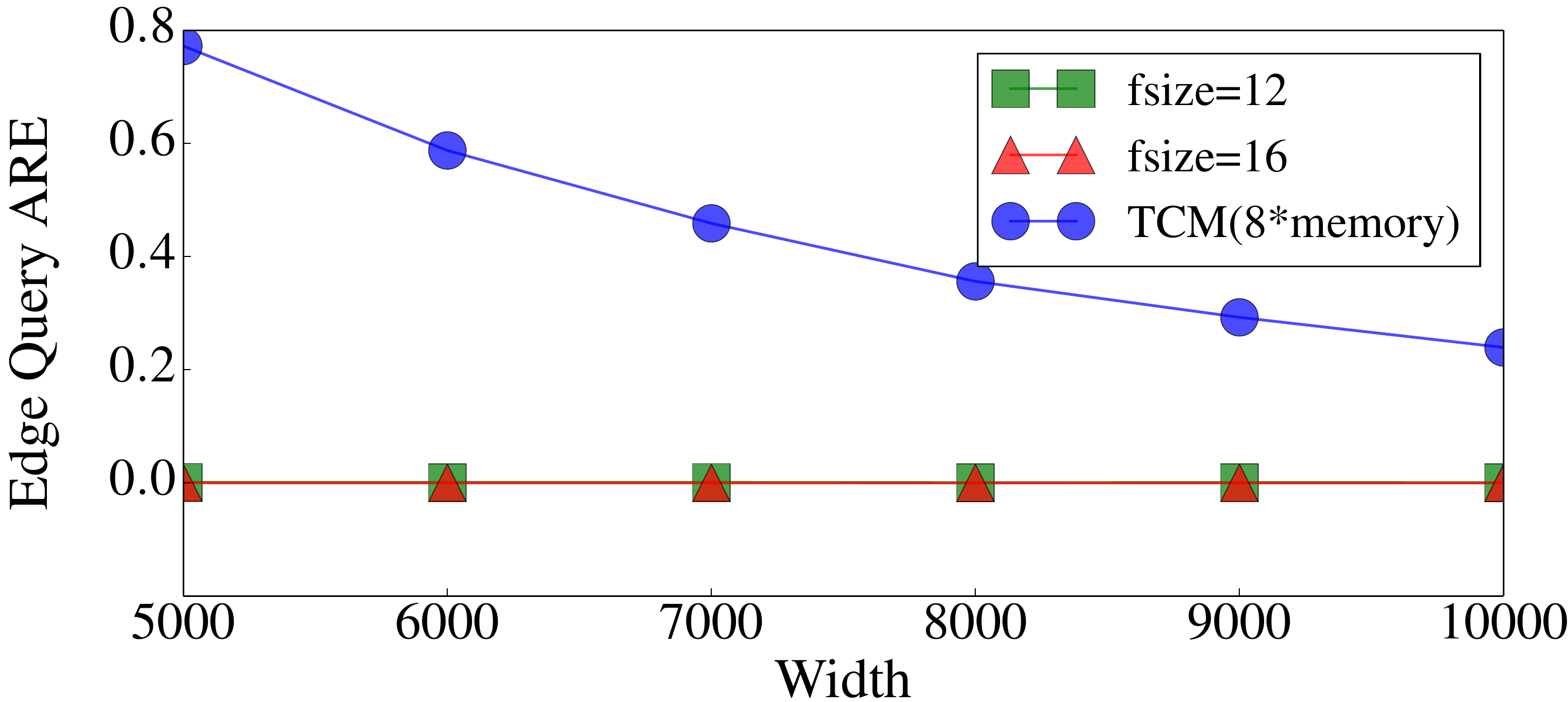}
\end{center}
\postfig
\label{citedge-ARE}
\vspace{-0.4cm}
\end{minipage}
}
\prefigcaption
\caption{Average Relative Error of Edge Queries}\label{edgeARE}
\end{figure*}
\postfig

\prefig
\begin{figure*}[htb]
\centering
\subfigure[\dataseta]{
\begin{minipage}[b]{0.28\textwidth}
\centering
\includegraphics[width=1\textwidth]{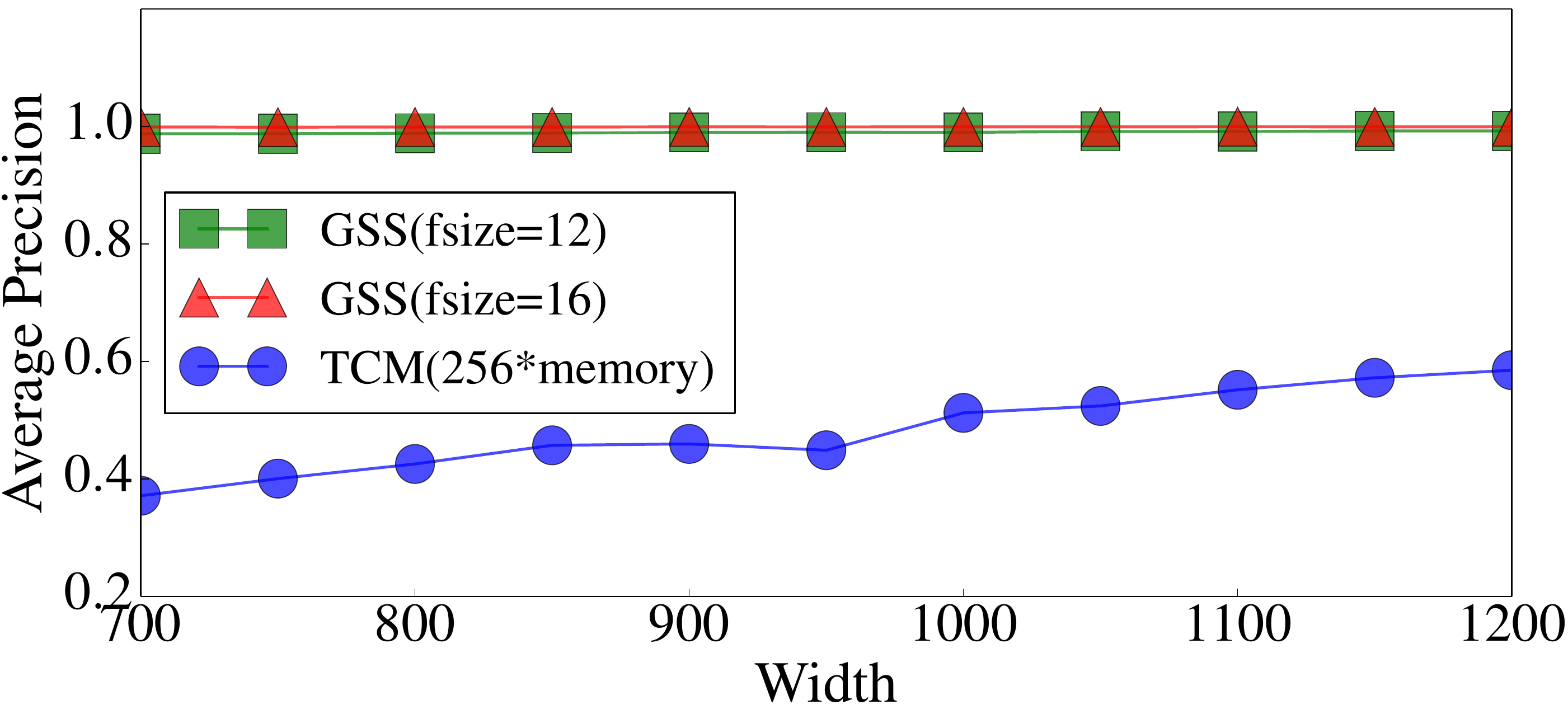}
\label{eeprecursor}
\vspace{-0.4cm}
\end{minipage}
}
\subfigure[\datasetb]{
\begin{minipage}[b]{0.28\textwidth}
\centering
\includegraphics[width=1\textwidth]{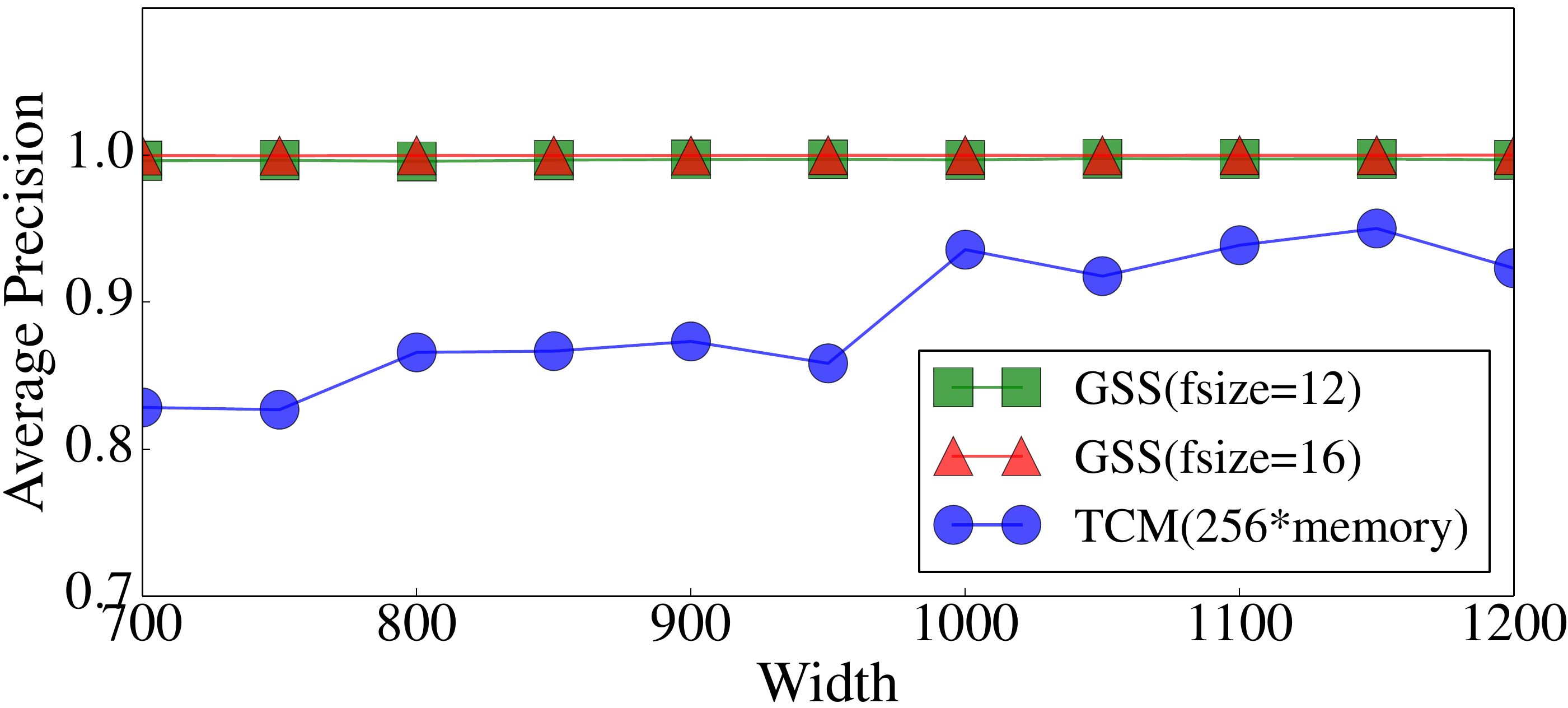}
\label{citprecursor}
\vspace{-0.4cm}
\end{minipage}
}
\subfigure[\datasetc]{
\begin{minipage}[b]{0.28\textwidth}
\centering
\includegraphics[width=1\textwidth]{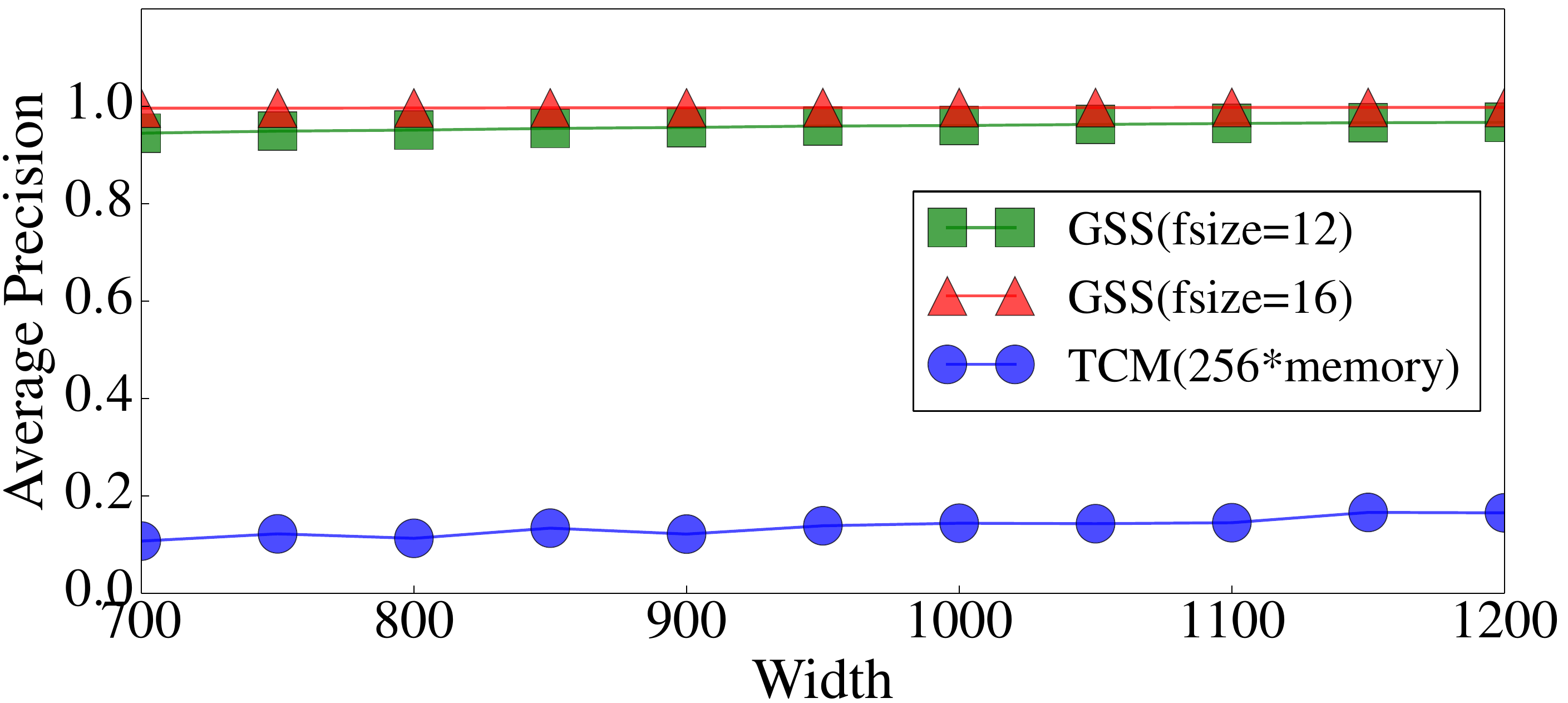}
\label{wnprecursor}
\vspace{-0.4cm}
\end{minipage}
}
\subfigure[\datasetd]{
\begin{minipage}[b]{0.28\textwidth}
\centering
\includegraphics[width=1\textwidth]{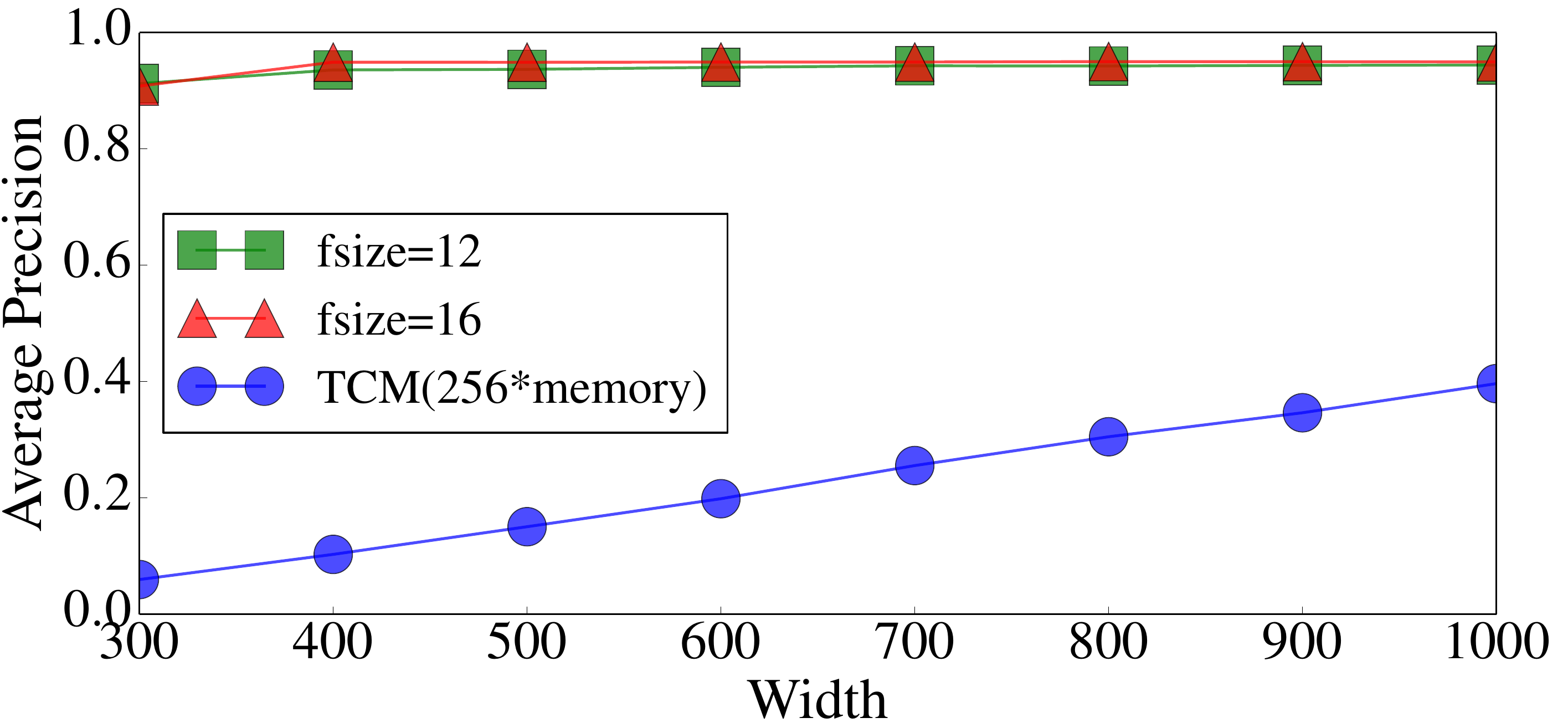}
\label{lkmlprecursor}
\vspace{-0.4cm}
\end{minipage}
}
\subfigure[\datasete]{
\begin{minipage}[b]{0.28\textwidth}
\centering
\includegraphics[width=1\textwidth]{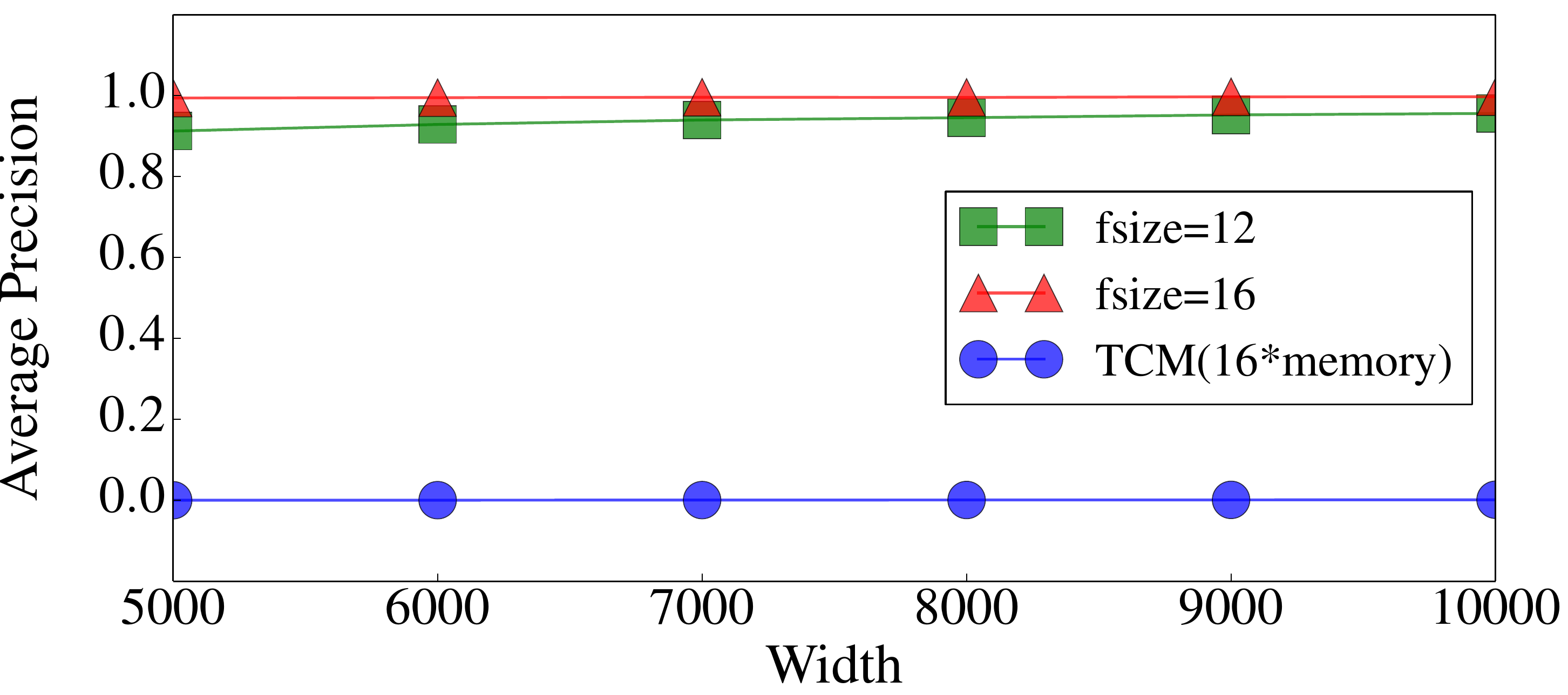}
\label{caidaprecursor}
\vspace{-0.4cm}
\end{minipage}
}
\prefigcaption
\caption{Average Precision of 1-hop Precursor Queries}\label{precursor}
\end{figure*}
\postfig

\prefig
\begin{figure*}[htb]
\centering
\subfigure[\dataseta]{
\begin{minipage}[b]{0.28\textwidth}
\centering
\includegraphics[width=1\textwidth]{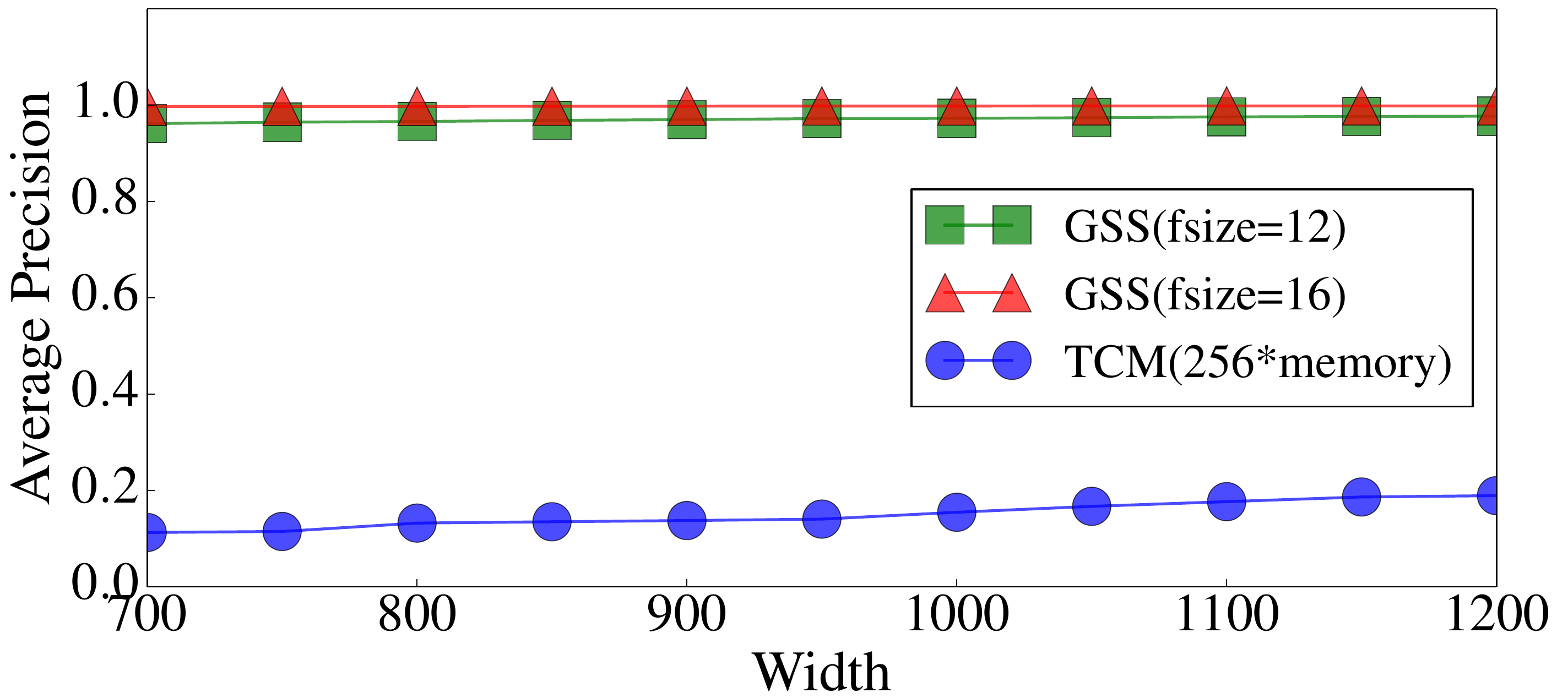}
\label{eesuccessor}
\vspace{-0.4cm}
\end{minipage}
}
\subfigure[\datasetb]{
\begin{minipage}[b]{0.28\textwidth}
\centering
\includegraphics[width=1\textwidth]{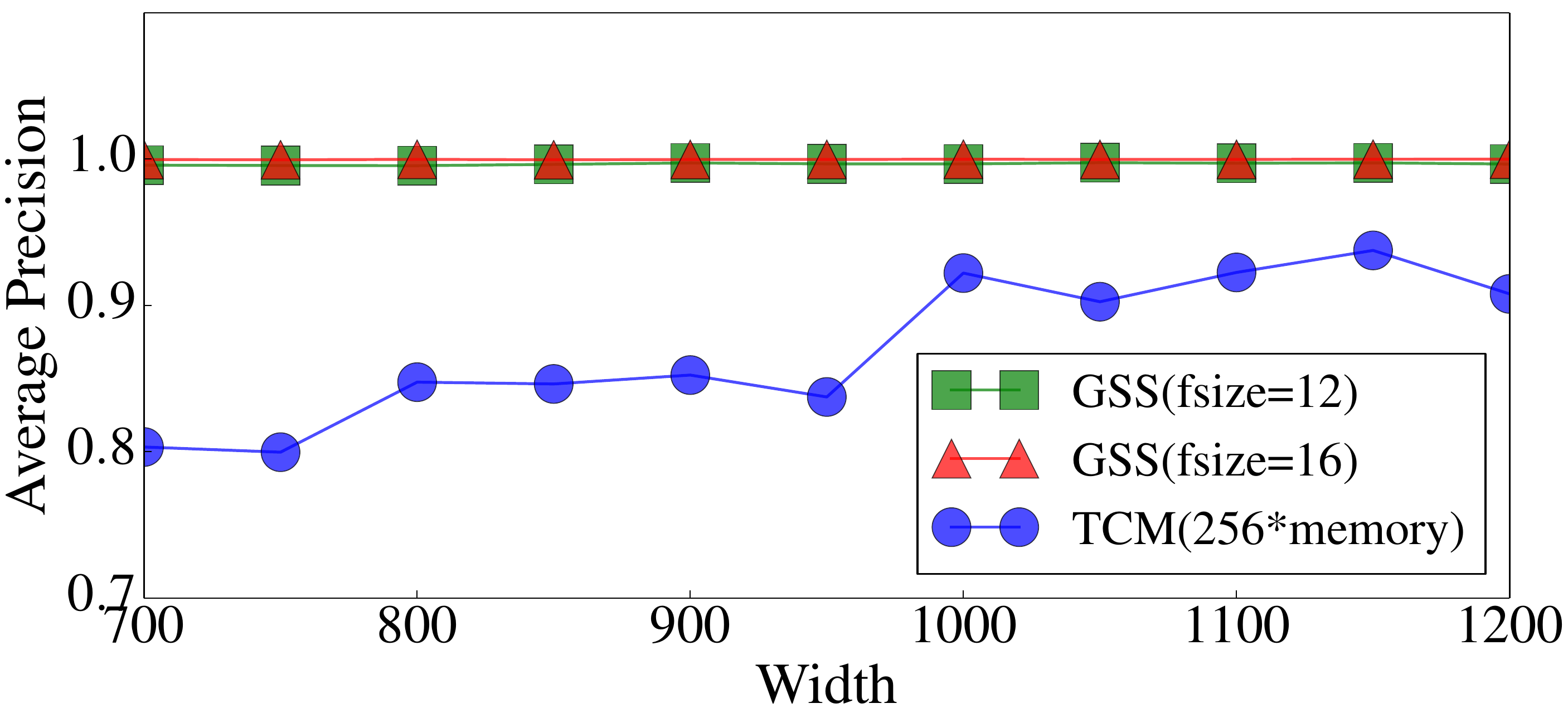}
\label{citsuccessor}
\vspace{-0.4cm}
\end{minipage}
}
\subfigure[\datasetc]{
\begin{minipage}[b]{0.28\textwidth}
\centering
\includegraphics[width=1\textwidth]{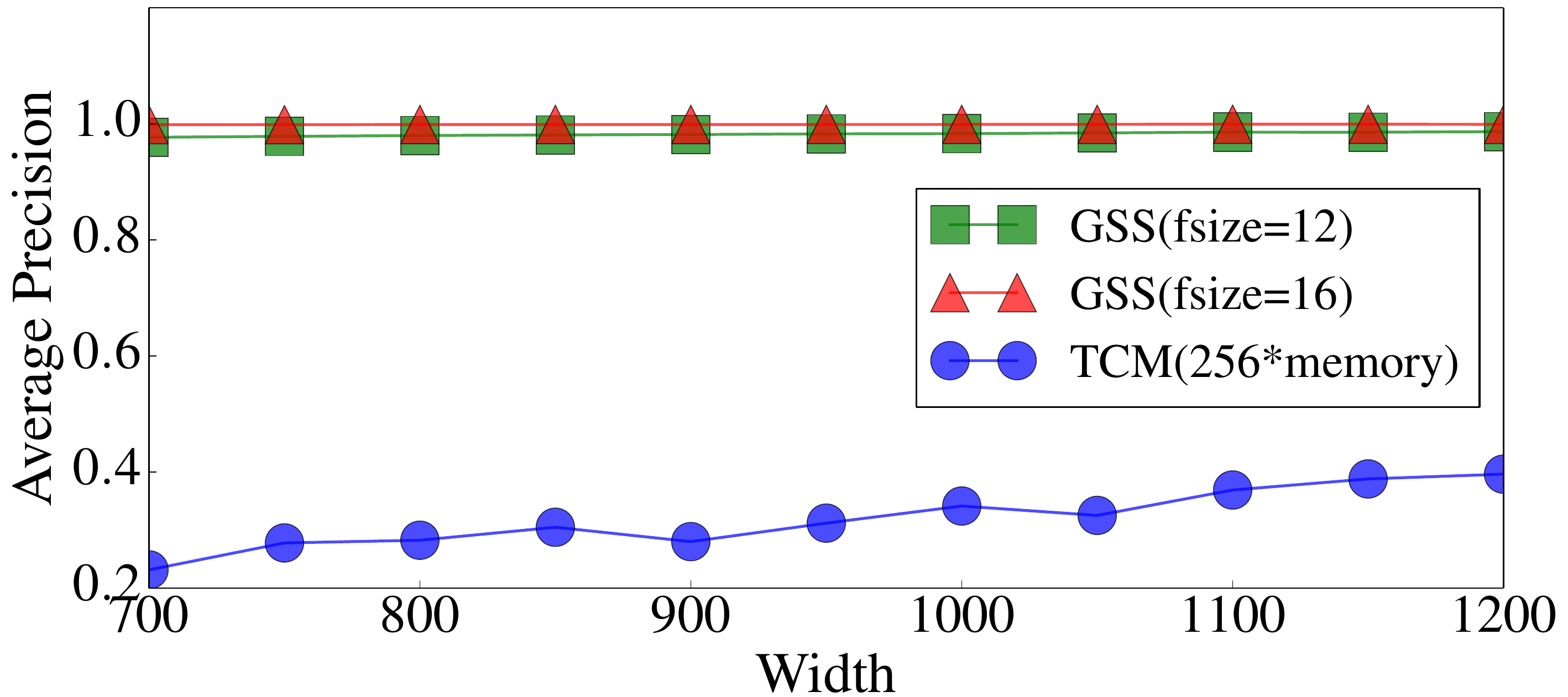}
\label{wnsuccessor}
\vspace{-0.4cm}
\end{minipage}
}
\subfigure[\datasetd]{
\begin{minipage}[b]{0.28\textwidth}
\centering
\includegraphics[width=1\textwidth]{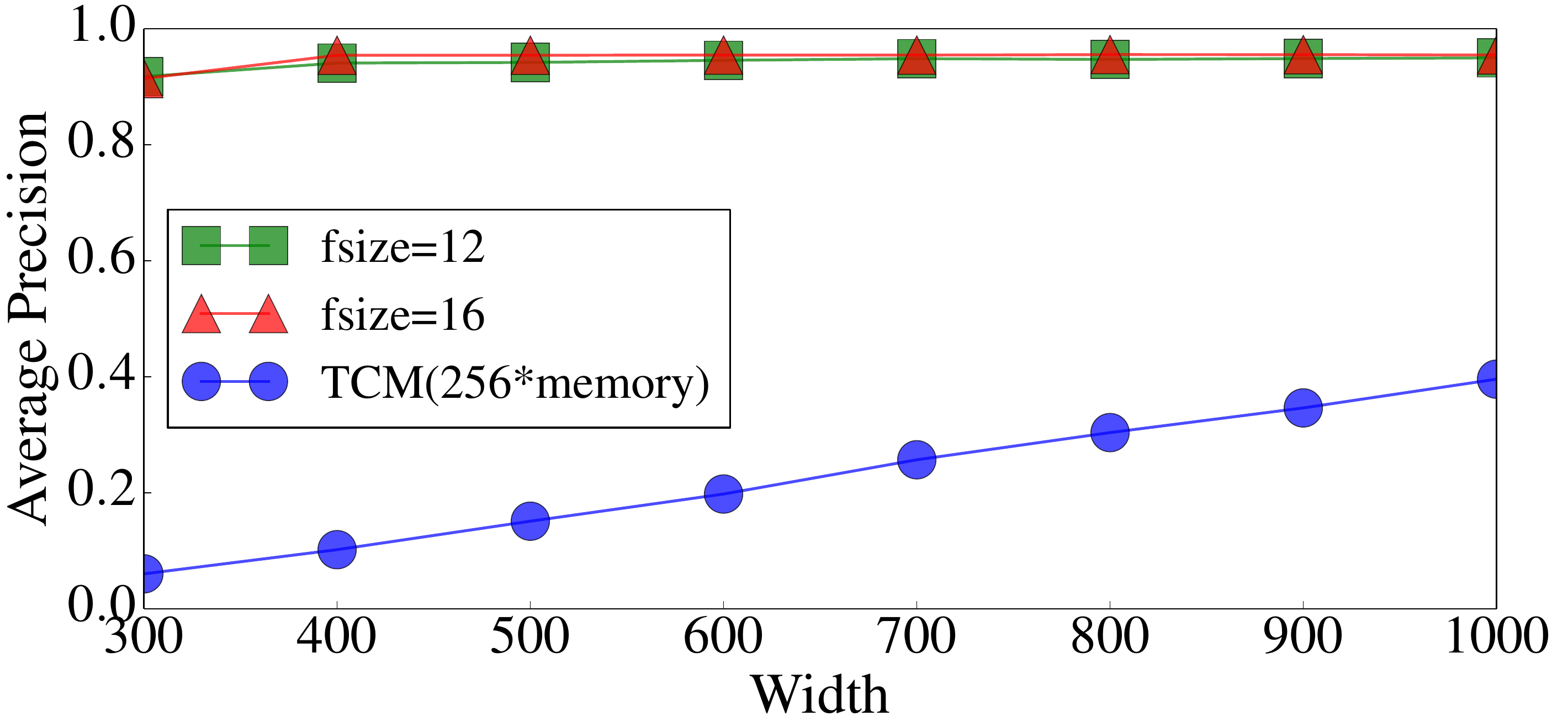}
\label{lkmlsuccessor}
\vspace{-0.4cm}
\end{minipage}
}
\subfigure[\datasete]{
\begin{minipage}[b]{0.28\textwidth}
\centering
\includegraphics[width=1\textwidth]{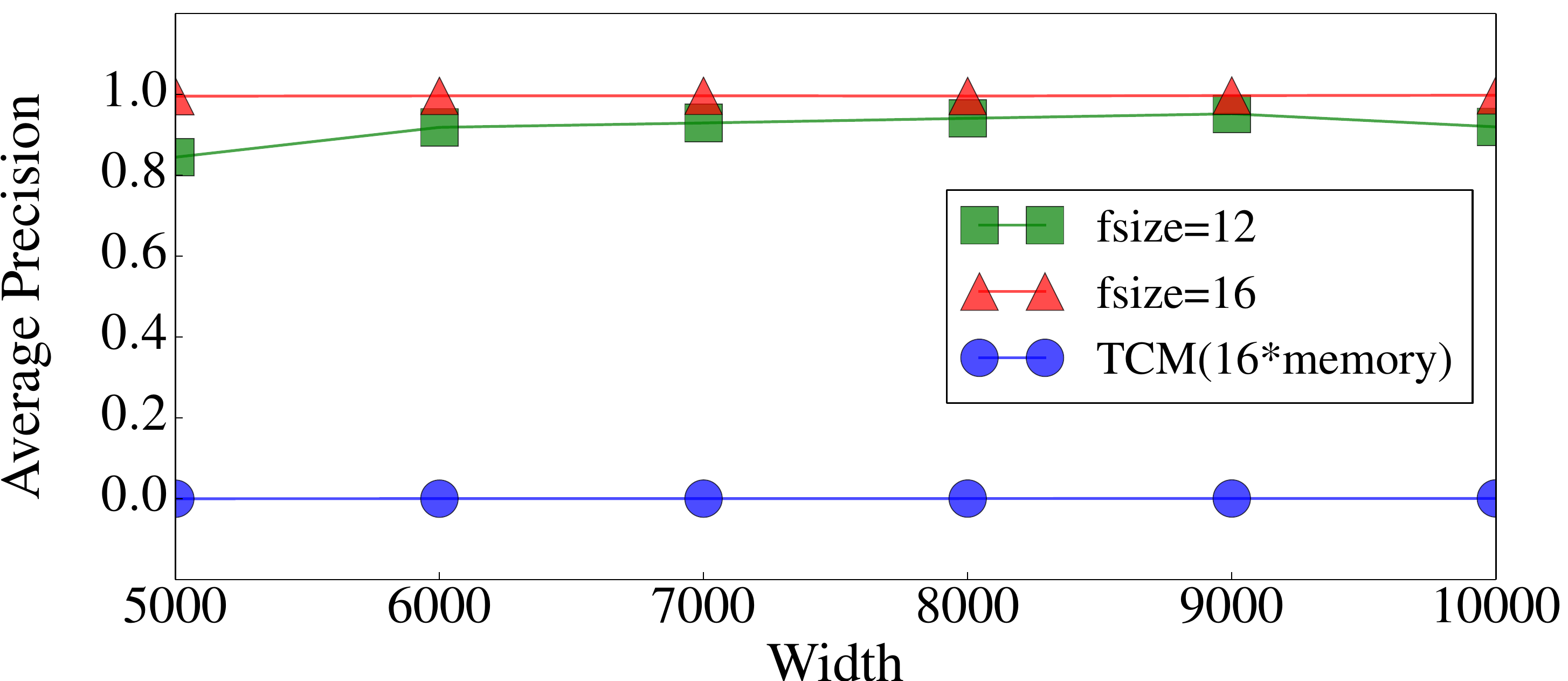}
\label{caidasuccessor}
\vspace{-0.4cm}
\end{minipage}
}
\prefigcaption
\caption{Average Precision of 1-hop Successor Queries}\label{successor}
\end{figure*}
\postfig
%

In this section, we evaluate the performance of \fname \ in the 3 basic graph query primitives: the edge query, the 1-hop precursor query and the 1-hop successor query. \nop{We implement TCM with 8 times of memory in edge queries, and 256 times of memory in other 2 query primitives to make the result comparable.( In \datasete, we implement TCM with 16 times of memory rather than 256 times in the 1-hop successor/precursor query because of the limitation of the host memory).} Figure \ref{edgeARE}, Figure \ref{precursor}, and Figure \ref{successor} show that ARE of edge queries and average precision of 1-hop precursor / successor queries for the data sets, respectively. 
To reduce random error introduced by the selection of the data sample, the edge query set contains all edges in the graph stream, and the 1-hop precursor / successor query set contains all nodes in the graph stream.
The results tell us that \fname\  performs much better in supporting these query primitives than TCM, especially in the 1-hop precursor / successor query primitives.
In both \fname\ and TCM, the ARE decreases, and the precision increases with the growth of the width of the matrix. This trend is not very significant in \fname\ as the accuracy is high and there are no errors in most experiments. Also, when the length of fingerprint becomes longer, the accuracy of \fname\ increases. 
\presub
\subsection{Experiments on Node Query}\label{nodequery}
\postsub
\begin{figure*}[htb]
\centering
\subfigure[\dataseta]{
\begin{minipage}[b]{0.28\textwidth}
\centering
\includegraphics[width=1\textwidth]{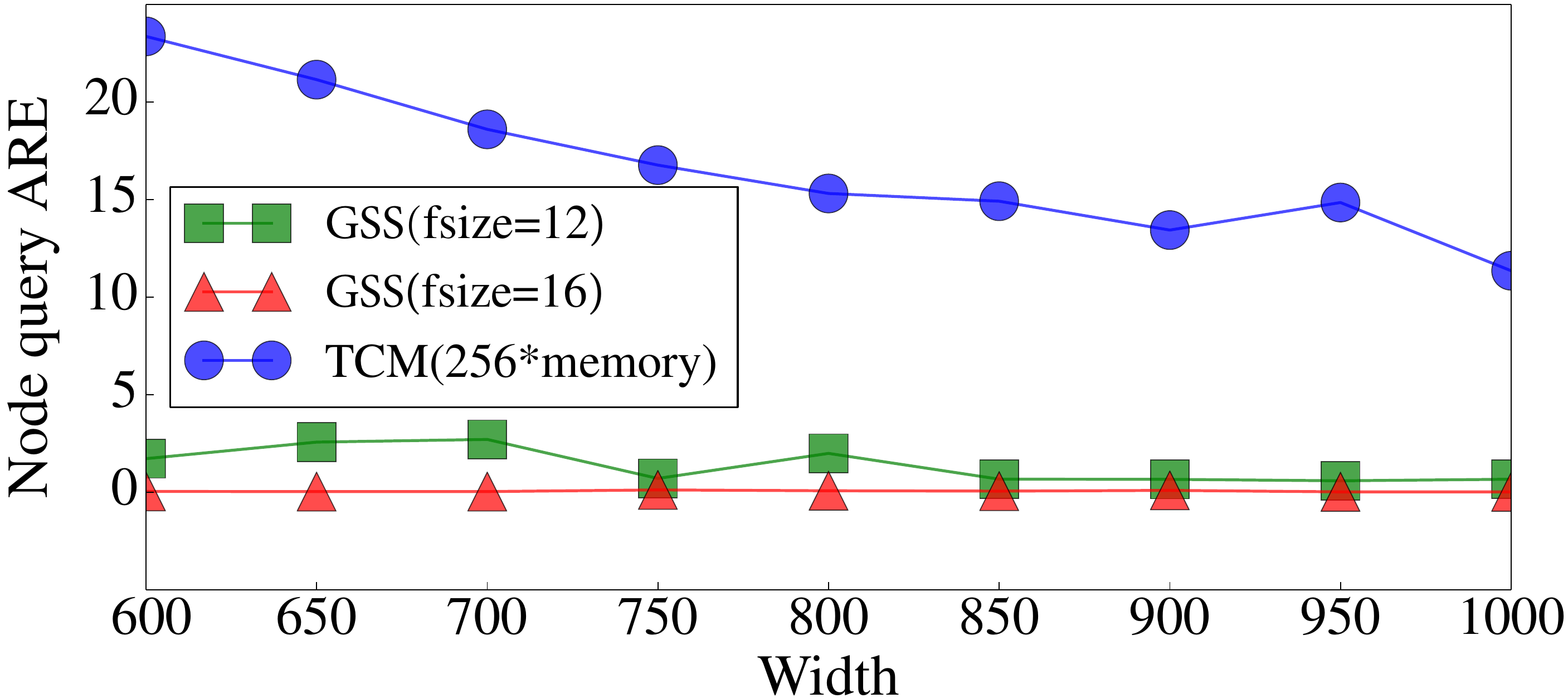}
\label{eenode}
\vspace{-0.4cm}
\end{minipage}
}
\subfigure[\datasetb]{
\begin{minipage}[b]{0.28\textwidth}
\centering
\includegraphics[width=1\textwidth]{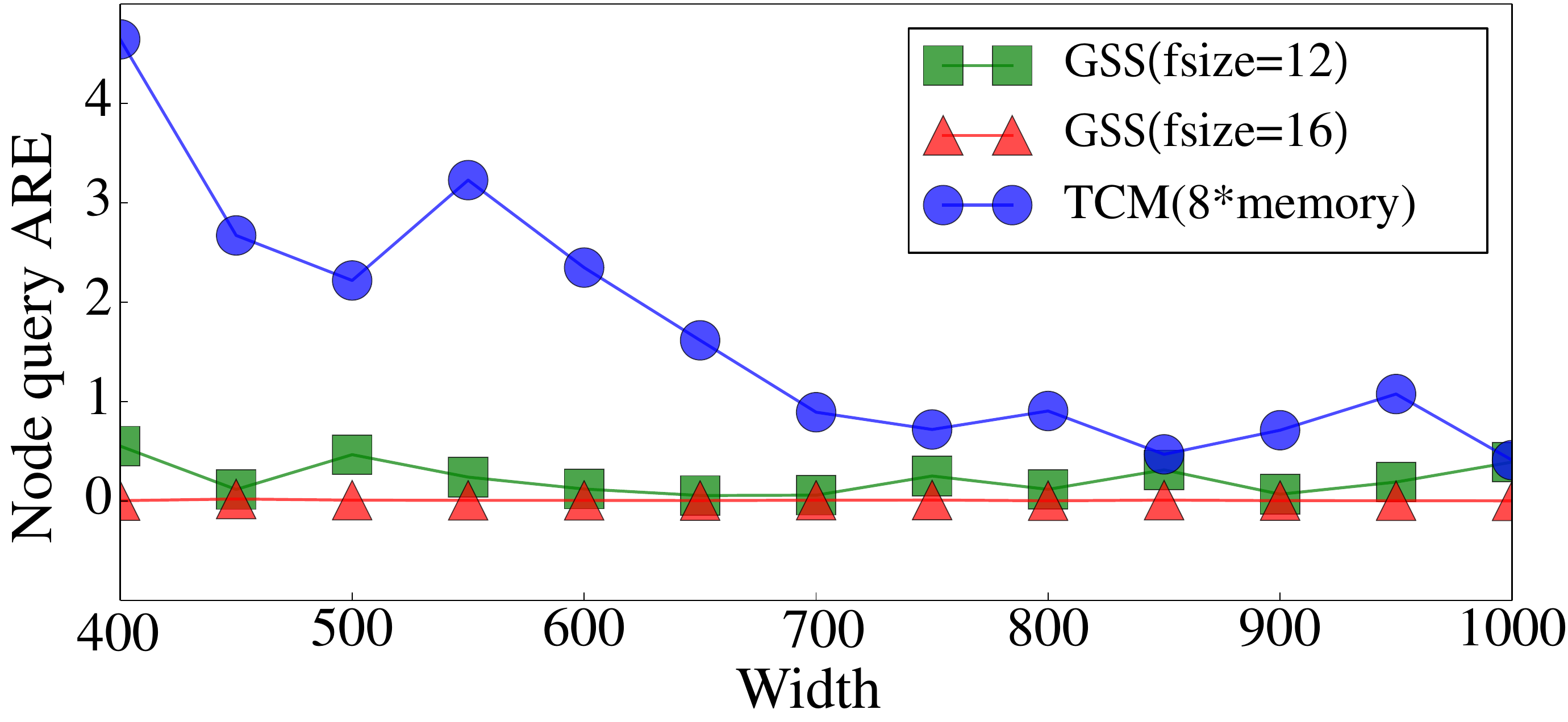}
\label{citnode}
\vspace{-0.4cm}
\end{minipage}
}
\subfigure[\datasetc]
{
\begin{minipage}[b]{0.28\textwidth}
\centering
\includegraphics[width=1\textwidth]{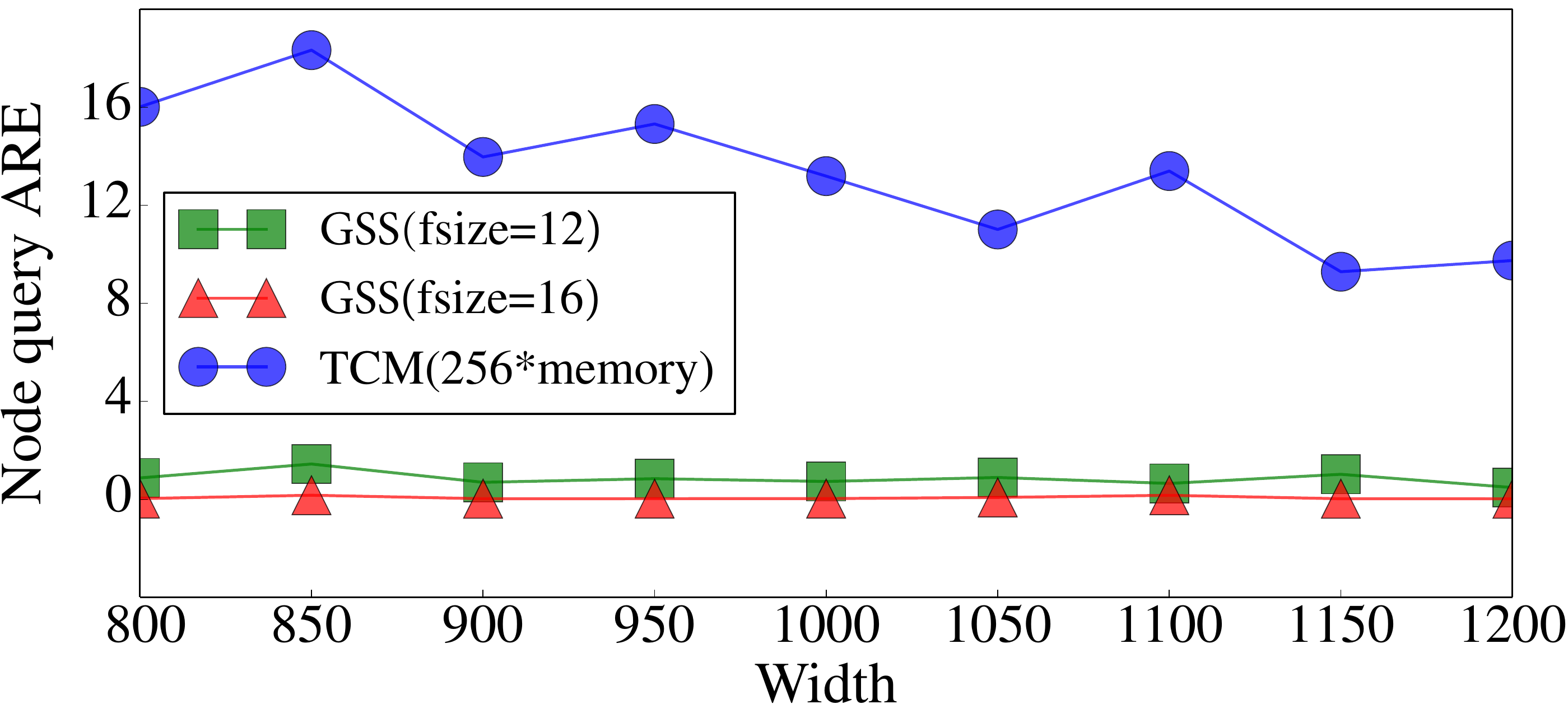}
\label{wnnode}
\vspace{-0.4cm}
\end{minipage}
}
\subfigure[\datasetd]{
\begin{minipage}[b]{0.28\textwidth}
\centering
\includegraphics[width=1\textwidth]{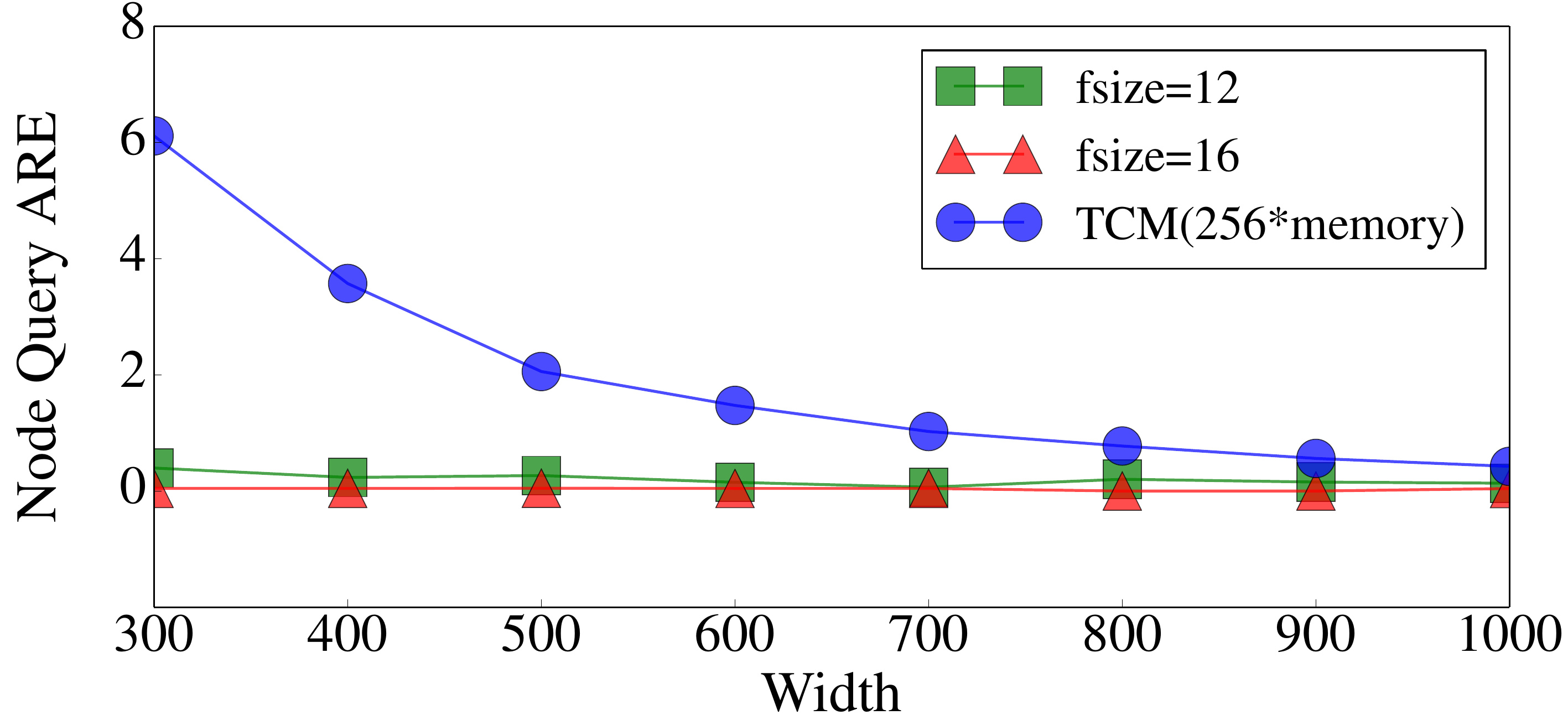}
\label{lkml-node}
\vspace{-0.4cm}
\end{minipage}
}
\subfigure[\datasete]{
\begin{minipage}[b]{0.28\textwidth}
\centering
\includegraphics[width=1\textwidth]{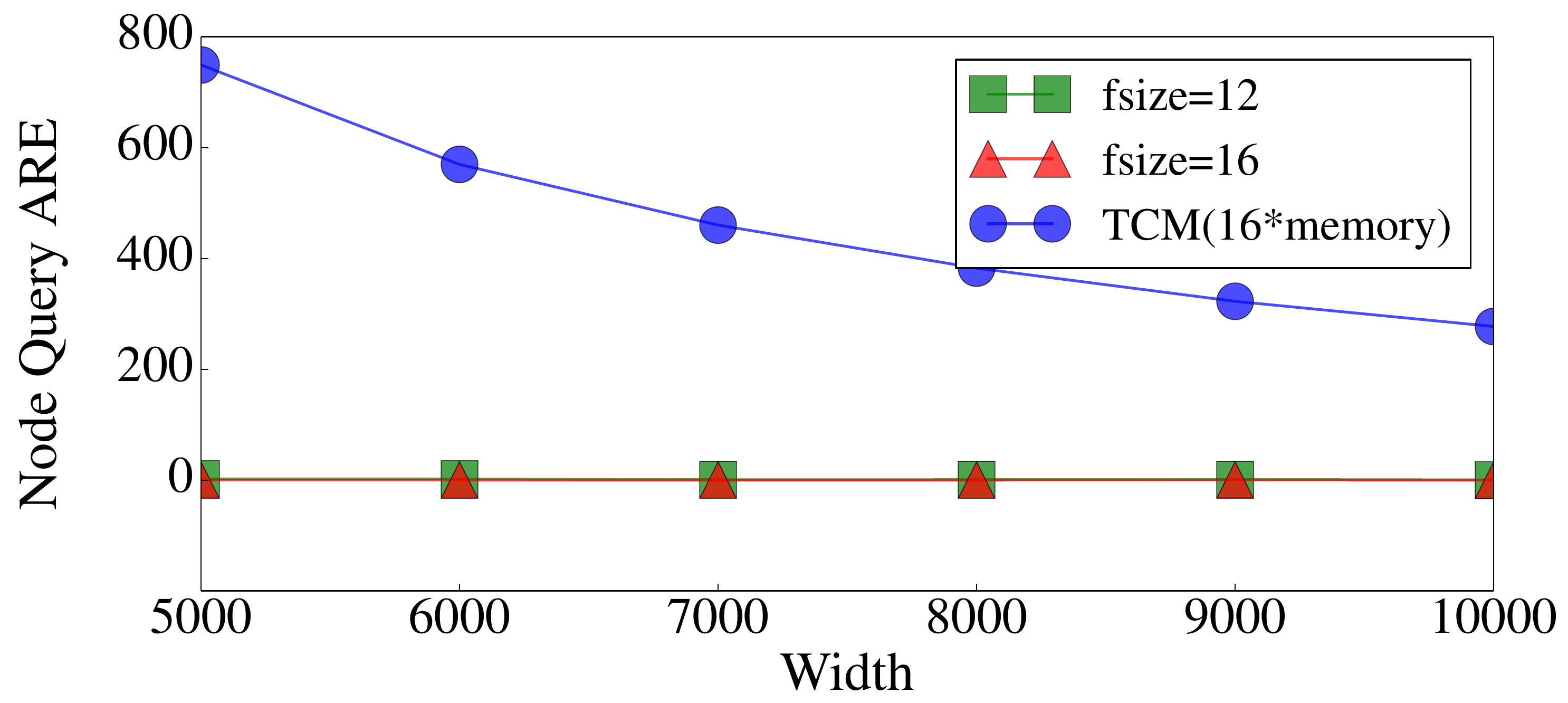}
\label{caidanode}
\vspace{-0.4cm}
\end{minipage}
}
\prefigcaption
\caption{Average Relative Error of Node Queries}\label{nodeARE}
\end{figure*}
\postfig

In this section, we evaluate the performance of \fname\ in estimating the accuracy of node query. A node query for a node $v$ is to compute the summary of the weights of all edges with source node $v$. For each dataset, node query set contains all nodes in the graph stream.
Figure \ref{nodeARE} shows the ARE of node queries in data sets \dataseta, \datasetb \ and \datasetc, respectively.
\nop{As TCM has very poor accuracy in topology queries, to make TCM comparable with \fname, we fixed the memory of TCM 256 times as large as \fname\ in the first 2 datasets. In the third dataset we fixed it to 16 times because of the limitation of the memory of the server.} The figure shows that although we unfairly fix the ratio of memory used by TCM and \fname, \fname\ still can achieve better performance than TCM.

\presub
\subsection{Experiments on Reachability Query}\label{reachabilityquery}
\postsub
In this section, we evaluate the performance of \fname\  in supporting reachability queries. 
Each reachability query set $Q$ contains 100 unreachable pairs of nodes which are randomly generated from the graph. 
\begin{figure*}[htb]
\prefig
\centering
\subfigure[\dataseta]{
\begin{minipage}[b]{0.28\textwidth}
\centering
\includegraphics[width=1\textwidth]{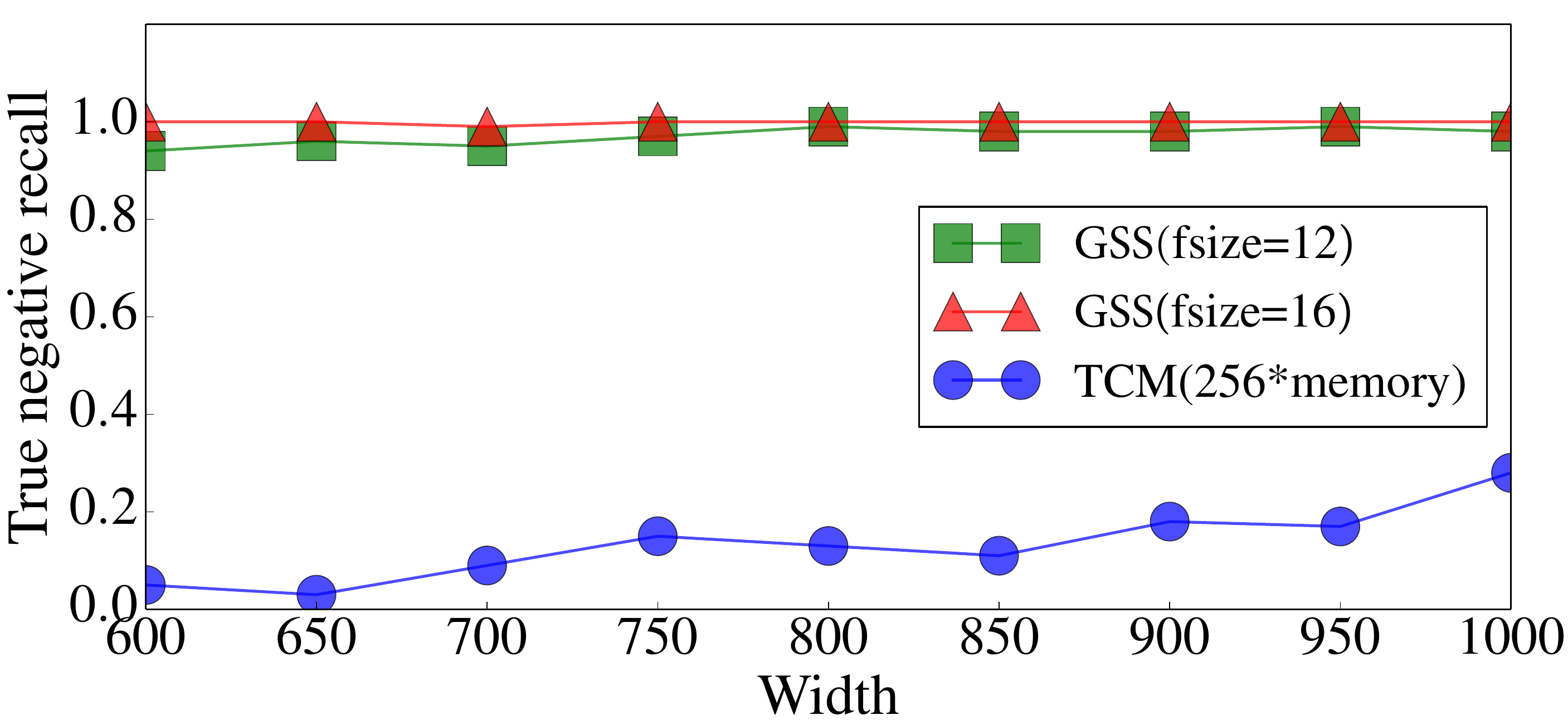}
\label{eetrans}
\vspace{-0.4cm}
\end{minipage}
}
\subfigure[\datasetb]{
\begin{minipage}[b]{0.28\textwidth}
\centering
\includegraphics[width=1\textwidth]{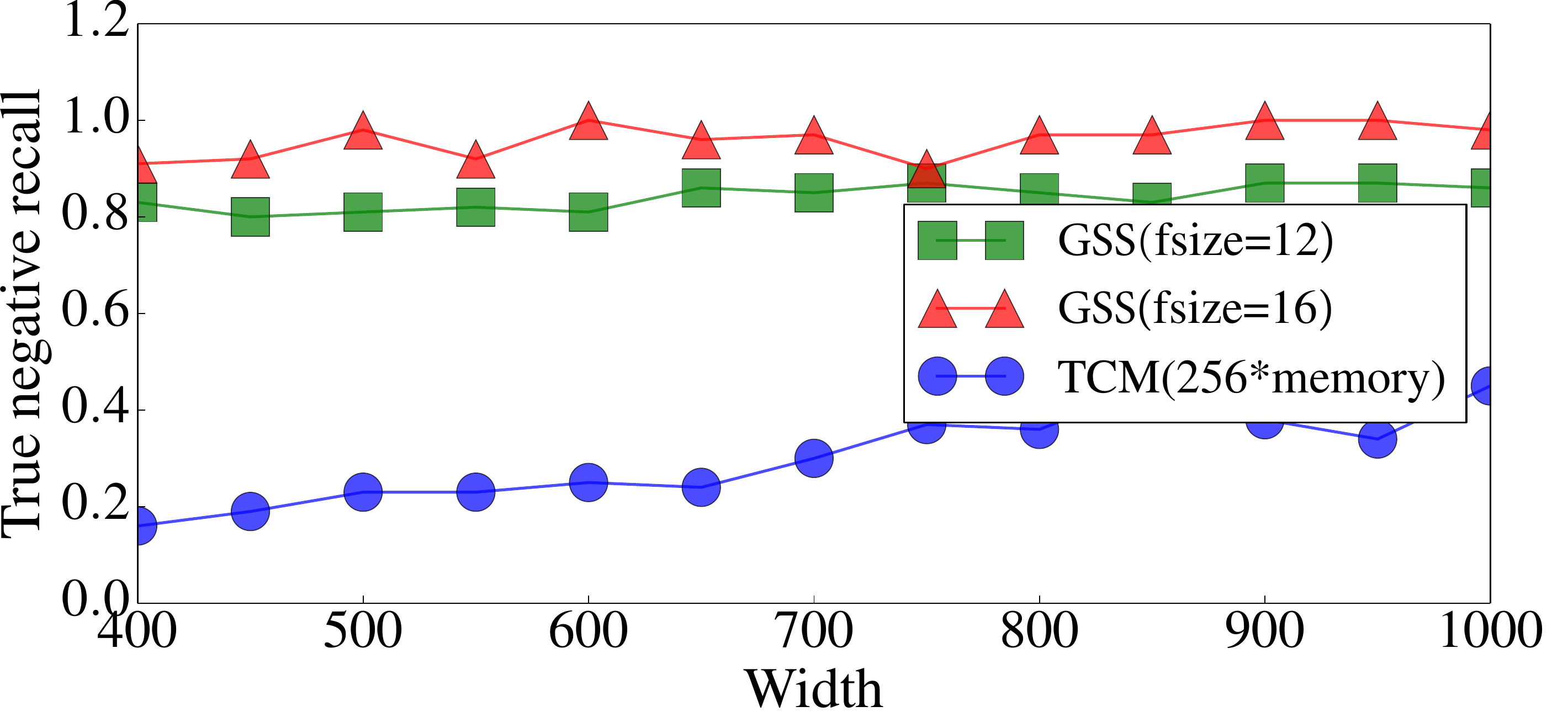}
\label{cittrans}
\vspace{-0.4cm}
\end{minipage}
}
\subfigure[\dataseta]{
\begin{minipage}[b]{0.28\textwidth}
\centering
\includegraphics[width=1\textwidth]{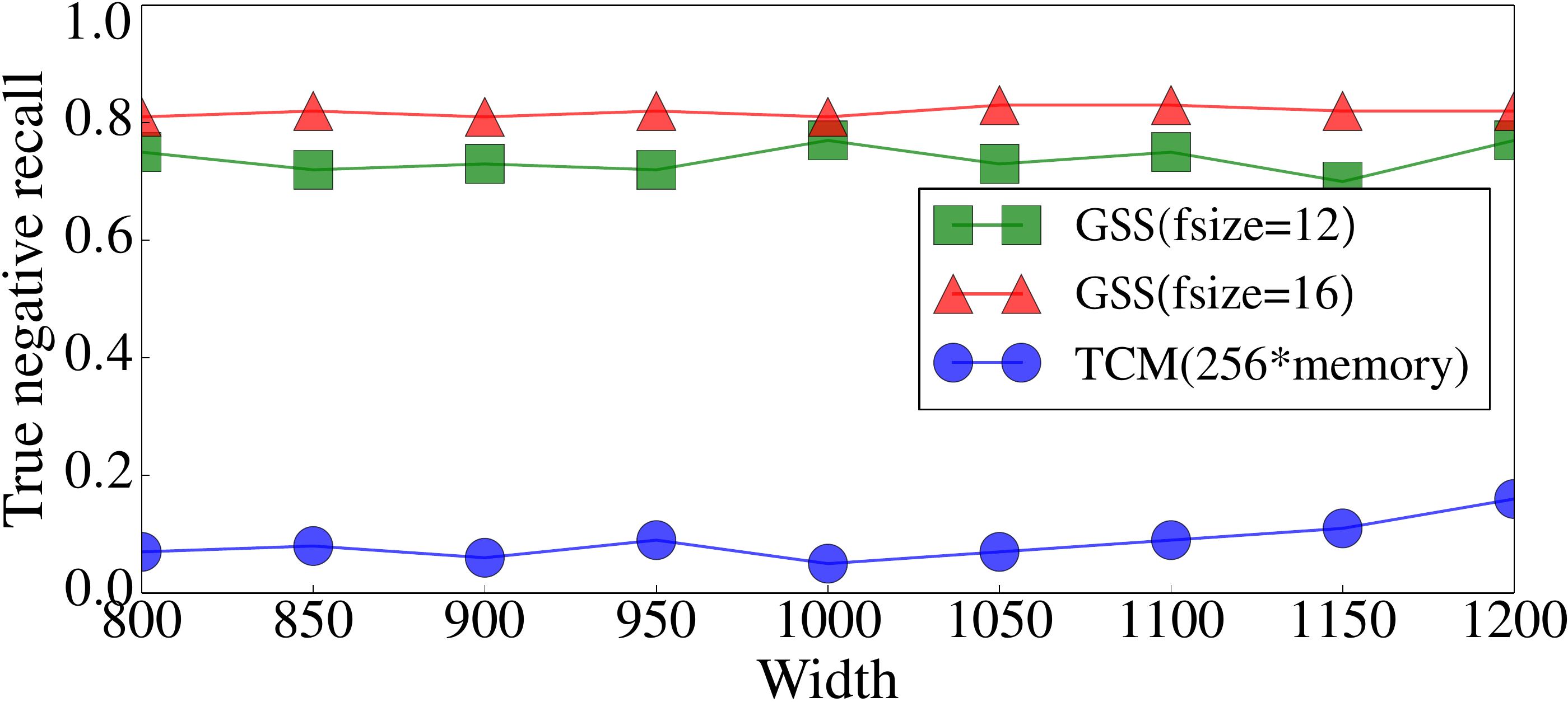}
\label{wntrans}
\vspace{-0.4cm}
\end{minipage}
}
\subfigure[\datasetb]{
\begin{minipage}[b]{0.28\textwidth}
\centering
\includegraphics[width=1\textwidth]{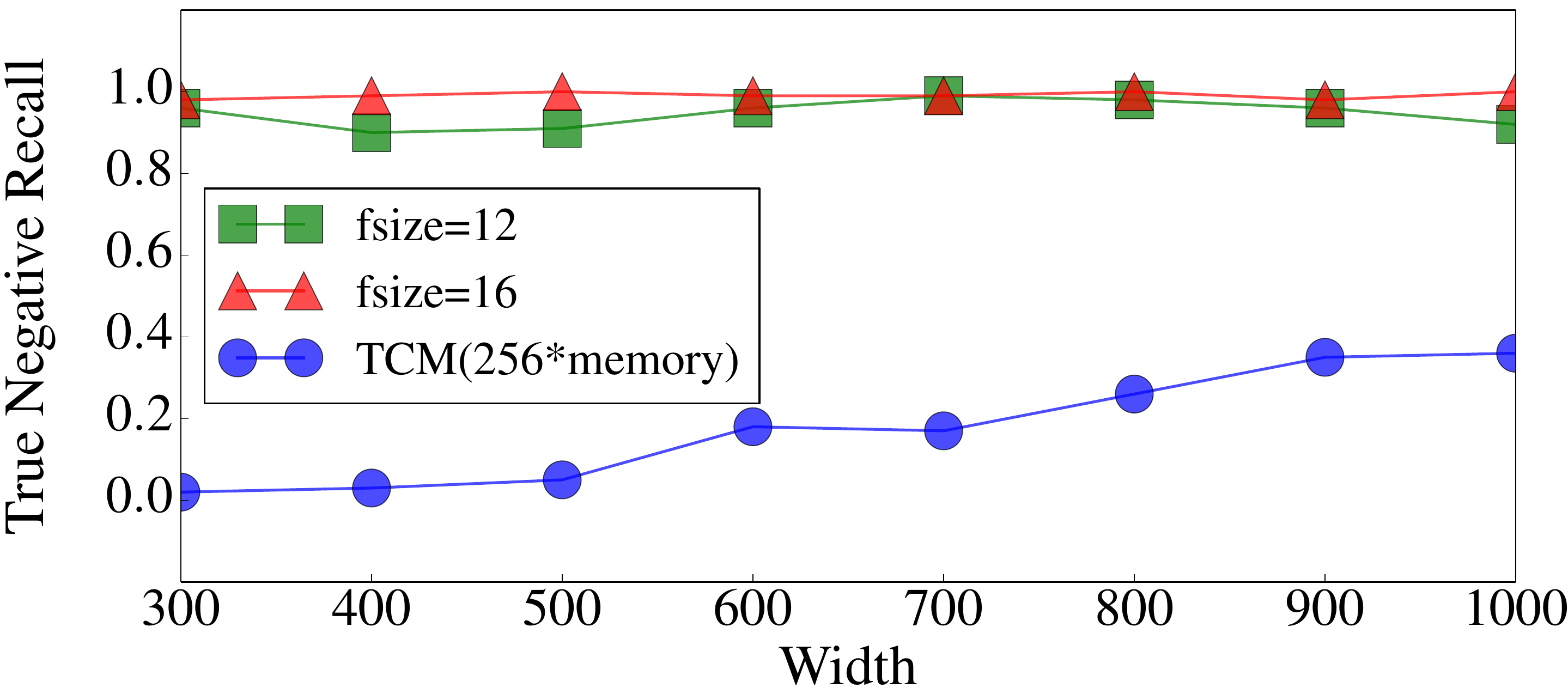}
\label{lkmltrans}
\vspace{-0.4cm}
\end{minipage}
}
\subfigure[\datasetc]{
\begin{minipage}[b]{0.28\textwidth}
\centering
\includegraphics[width=1\textwidth]{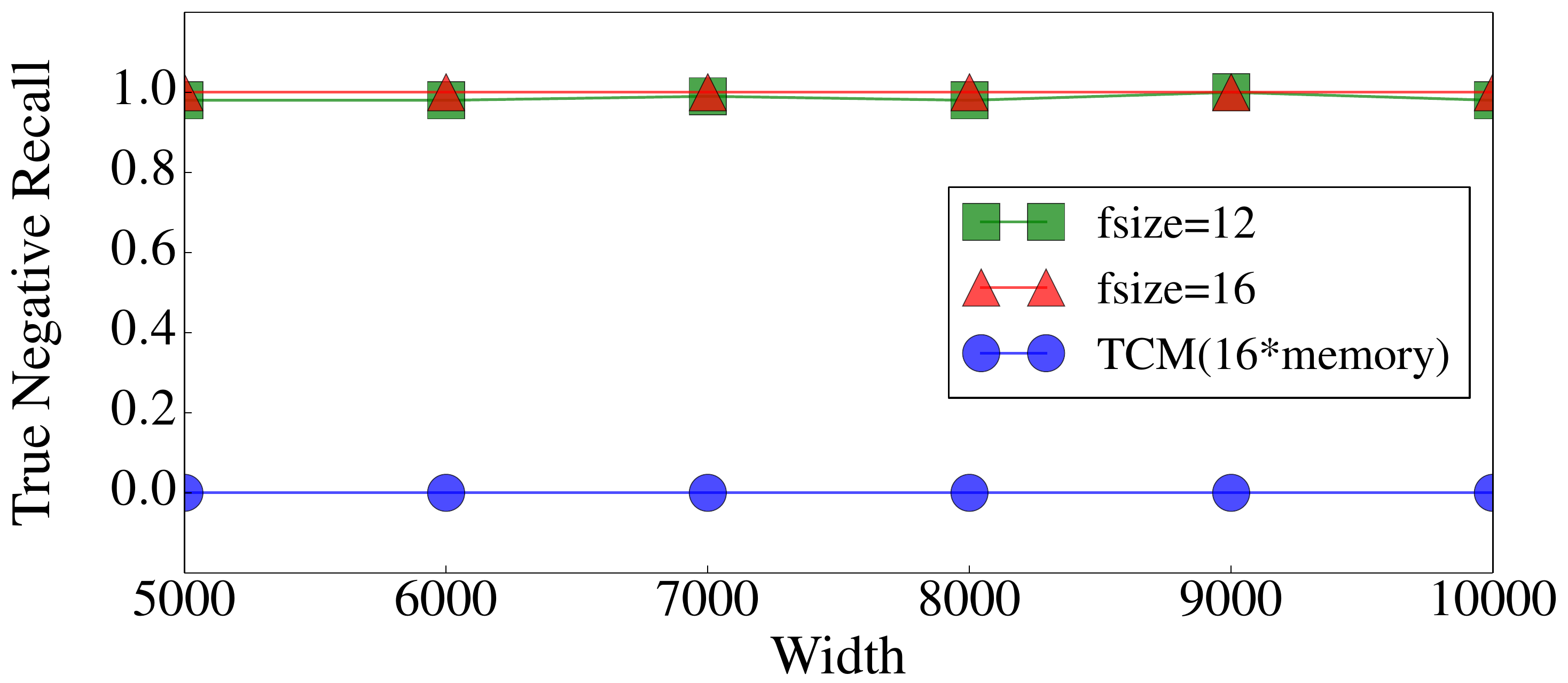}
\vspace{-0.4cm}
\label{caidatrans}
\end{minipage}
}
\prefigcaption
\caption{True Negative Recall of Reachability Queries}\label{trans}
\end{figure*}
\postfig
Figure \ref{trans} shows the true negative recall of reachability query for the data sets \dataseta, \datasetb \ and \datasetc, respectively. From the figures we can see that the accuracy of \fname\ is much higher than TCM even when TCM uses much larger memory. The gap varies with the size of the graph. Along with increasing the memory and the length of the fingerprint, \fname\  can achieve better performance. We can also see that the accuracy of TCM is so poor that it can barely support this query.
\begin{figure*}[htb]
\centering
\subfigure[\datasetc]{
\begin{minipage}[b]{0.28\textwidth}
\centering
\includegraphics[width=1\textwidth]{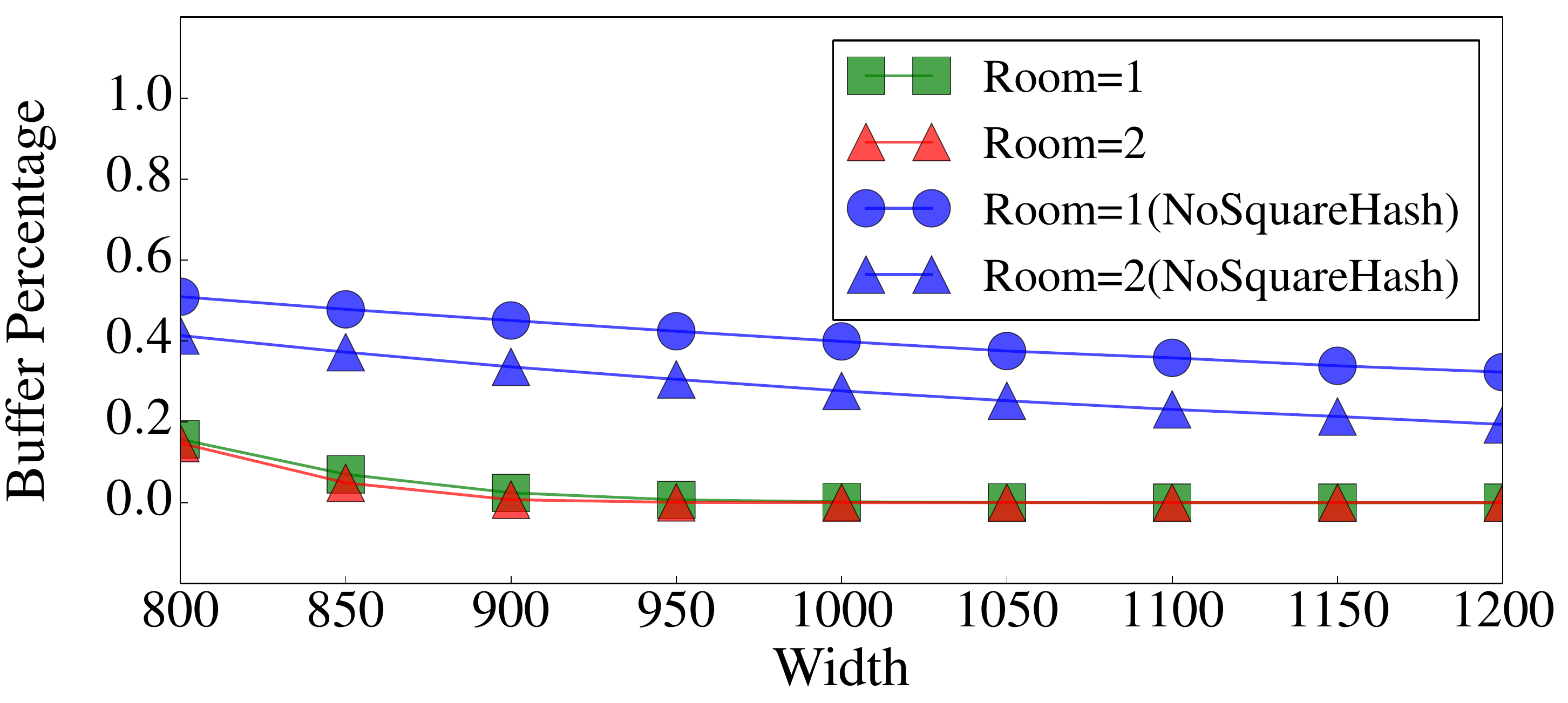}
\label{WNbuff}
\vspace{-0.4cm}
\end{minipage}
}
\subfigure[\datasetd]{
\begin{minipage}[b]{0.28\textwidth}
\centering
\includegraphics[width=1\textwidth]{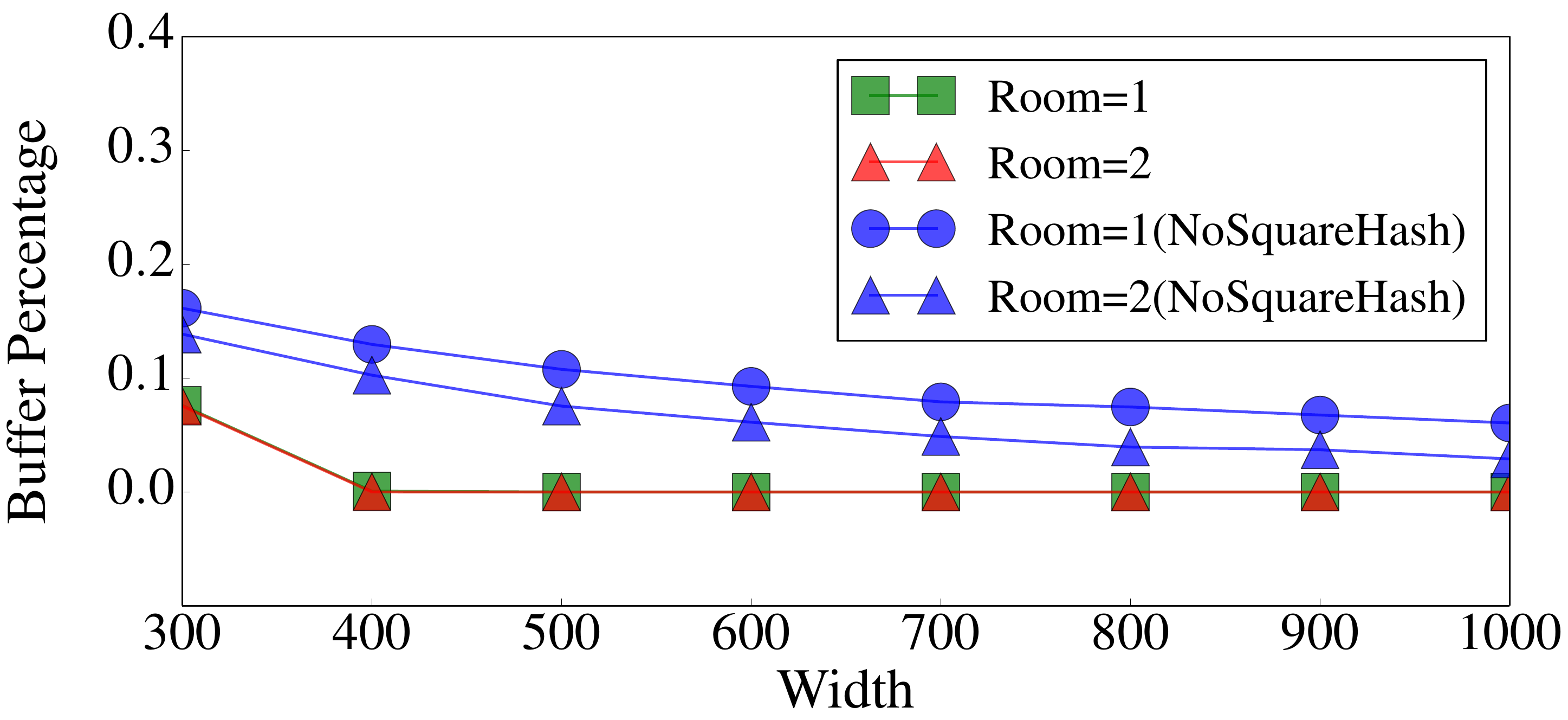}
\label{lkmlbuff}
\vspace{-0.4cm}
\end{minipage}
}
\subfigure[\datasete]{
\begin{minipage}[b]{0.28\textwidth}
\centering
\includegraphics[width=1\textwidth]{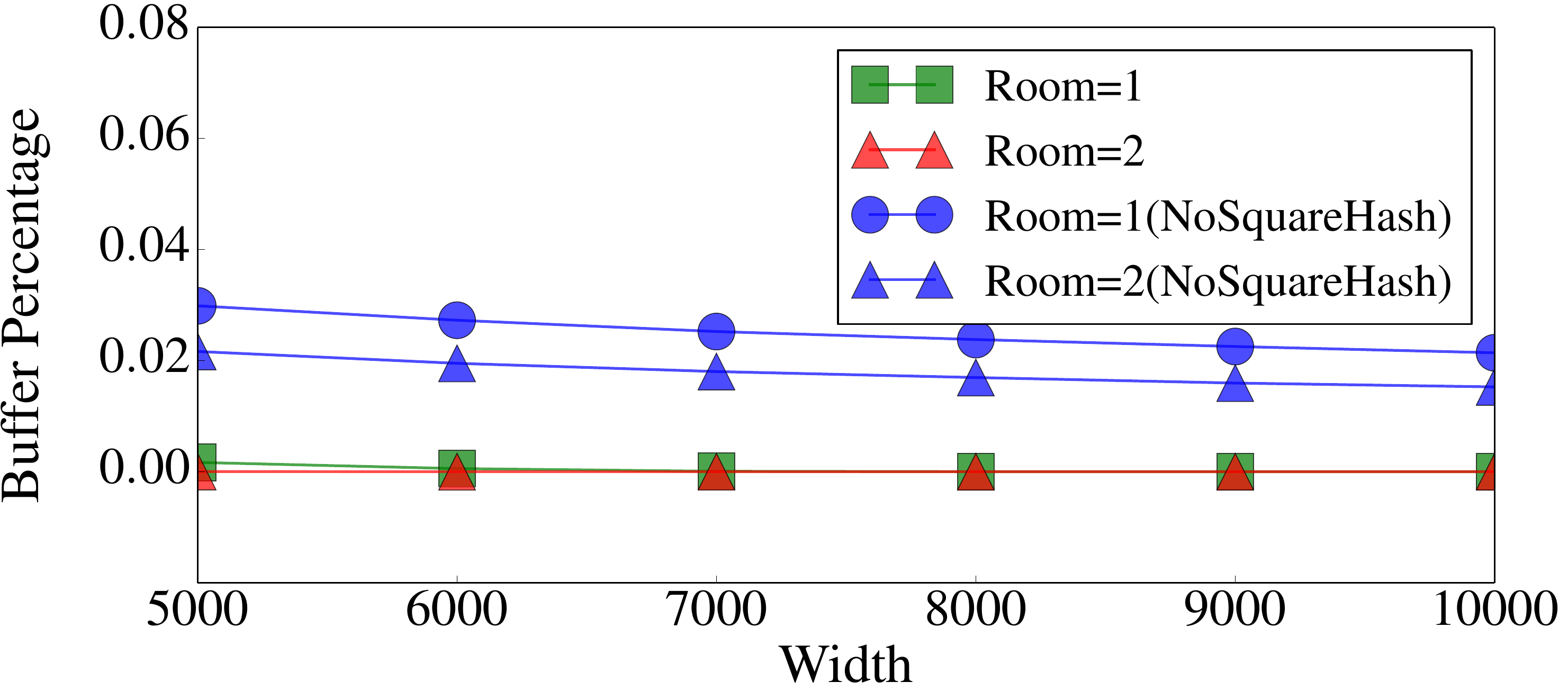}
\label{Caidabuff}
\vspace{-0.4cm}
\end{minipage}
}
\caption{Buffer Percentage}\label{caidabuff}
\end{figure*}
\presub
\subsection{Experiments on Buffer Size}\label{buffesize}
\postsub
Figure \ref{caidabuff} shows the buffer percentage for the three larger data sets \datasetc,\datasetd and \datasete. The four curves in the figure represent 1)GSS with 1 room in each bucket and no square hashing. 2) GSS with 2 rooms in each bucket and no square hashing. 3) GSS with 1 room in each bucket and square hashing. 4) GSS with 2 rooms in each bucket and square hashing. The x-label, $w$, is the side length of the matrices for the schemes with $2$ rooms in each bucket. When GSS has 1 room in each bucket, the width of the matrix is $2^{0.5}$ times larger to make the memory unchanged. The above results show that the decrement in buffer size brought by using square hashing and multiple rooms is significant, especially the square hashing. The results also show that the buffer percentage in the fully improved GSS (2 rooms each bucket, with square hashing) becomes $0$ in most experiments when the matrix size is close to $|E|$. In this case, the overhead brought by the insertion failure in the matrix is nearly 0.
\presub
\subsection{Experiment on update speed}
\label{EUS}
\vspace{-0.03in}
In this section we evaluate the update speed of \fname. We compare the update speed of \fname, TCM and adjacency lists, the result is shown in Table \ref{US}. The adjacency list is accelerated using a map that records the position of the list for each node. Because the update speed changes little with the matrix size, we only show the average speed here. The fingerprint size is 16-bit. TCM is still implemented with the same settings as above experiments. In each data set we insert all the edges into the data structure, repeat this procedure 100 times and calculate the average speed. The unit we use is Million Insertions per Second (Mips). From the figure we can see that the speed of \fname\ is similar to TCM, because though more memory accesses are needed, \fname\ computes less hash functions. Both of them are much higher than the adjacency list. We also show the speed of GSS without candidate bucket sampling. We can see that the speed without candidate sampling is lower than the full optimized one. The gap is not very large because most edges find empty bucket in few searches.
\vspace{-0.1in}
\begin{small}
\begin{center}
\begin{table}
\centering
\caption{Update Speed (Mips)}
\label{US}
\begin{tabular}{|p{2.1cm}|p{1.8cm}|p{1.8cm}|p{1.8cm}|}
\hline 
Data Structure & \dataseta & \datasetb & \datasetc \\
\hline
\fname & $2.2887$ & $2.61057$ &$2.40976$\\
\hline
\fname(no sampling) & $2.1245$ & $2.49191$ &$2.32015$\\
\hline
TCM & $2.10417$ & $2.5498$ &$2.07403$\\
\hline
Adjacency Lists & $0.578596$ & $0.3384$ &$0.52147$\\
\hline
\end{tabular}
\end{table}
\end{center}
\vspace{-0.1in}
\vspace{-0.2in}
\end{small}

\presub
\vspace{0.05in}
\subsection{Experiment on Other Compound Queries}
\label{ECQ}
\begin{figure}[ht]
\centering
\includegraphics[width=5.2cm,height=2.1cm]{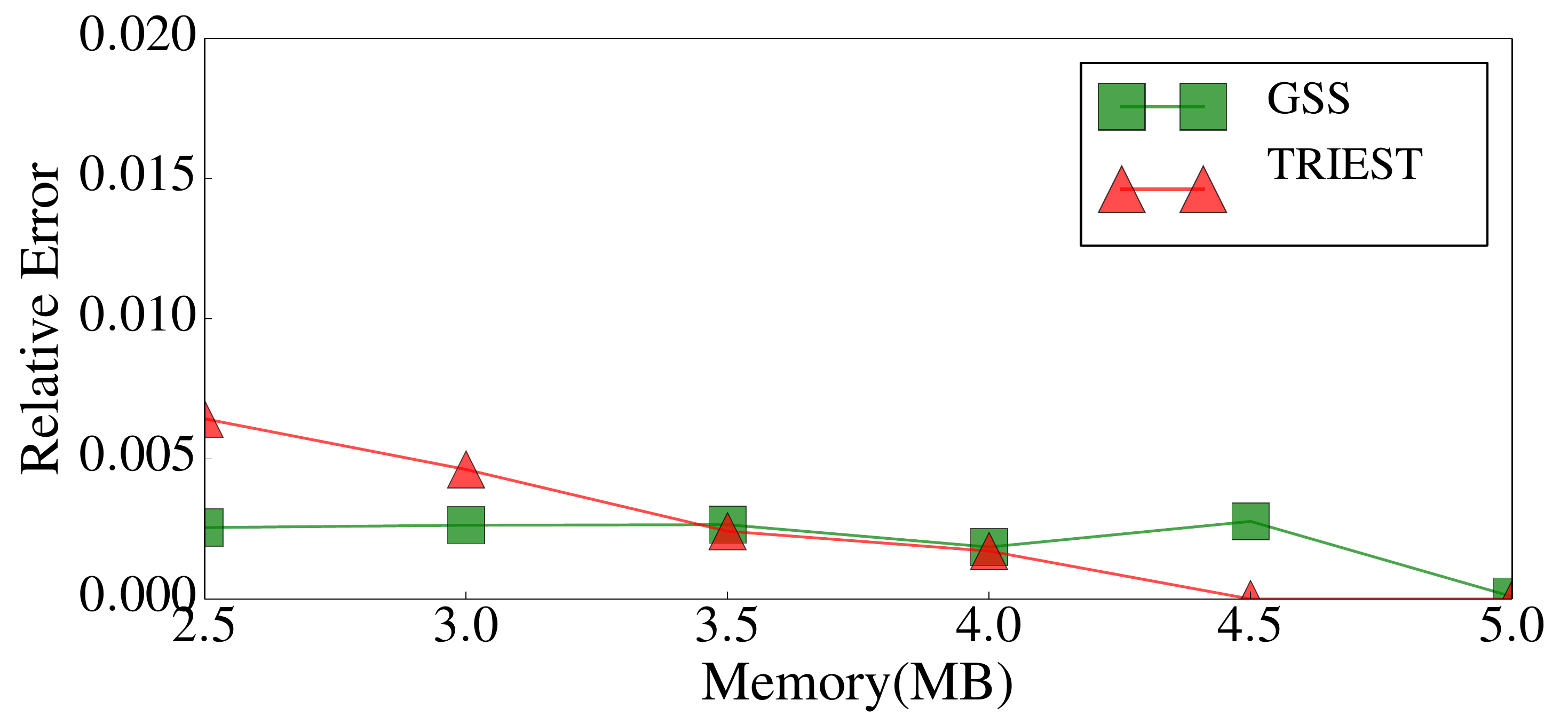}
\vspace{-0.1cm}
%
%
\caption{Triangle count in \datasetb}
\label{lkml-triangle}
\vspace{-0.4cm}
\end{figure}

\begin{figure}[ht]
\centering
\includegraphics[width=5.2cm,height=2.1cm]{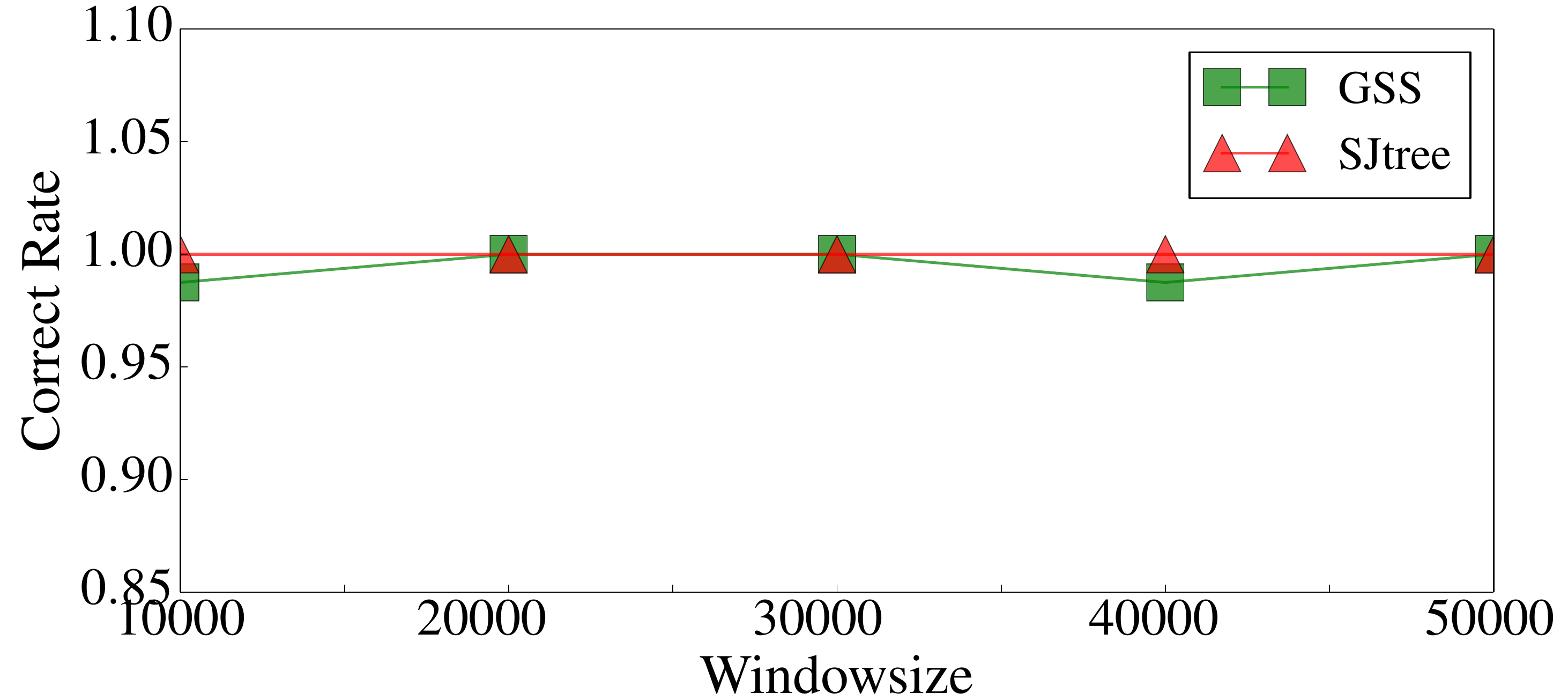}
\caption{Subgraph matching in \datasetc}
\label{gmatch}
\vspace{-0.1cm}
\end{figure}

\postsub
We compare \fname\ with state-of-the-art graph processing algorithms in triangle counting and subgraph matching in this Section. 
We compare \fname\ with TRIEST \cite{Stefani2016TRI} in triangle counting with the same memory. We use relative error between the reported results and the true value as evaluation metrics. TRIEST does not support multiple edges. Therefore we unique the edges in the dataset for it. The results are shown in Figure \ref{lkml-triangle}. The results show that they achieve similarly high accuracy with relative error less than $1\%$. 
\nop{The average throughput o1f TRIEST is $0.178$M edges per second, much lower than \fname\, which is shown in table \ref{US}}
We compare \fname\ with SJ-tree\cite{Choudhury2015A} in subgraph matching. As SJ-tree is an accurate algorithm, we set \fname\ to $\frac{1}{10}$ of its memory. We use VF2 algorithm when querying in GSS, other algorithms can also be used.  We use \datasetc\ and search for subgraphs in windows of the data stream. The edges in the graph are labeled by the ports and the protocol. We carry out experiment on $5$ window sizes, and for each window size, we randomly select $5$ windows in the stream. In each window, we generate $4$ kinds of subgraphs with $6$, $9$, $12$ and $15$ edges and $5$ instances in each kind by random walk. We use the correct rate as evaluation metrics, which means the percentage of correct matches in the $100$ matches for each window size. Experimental results are shown in Figure \ref{gmatch}. We can see that \fname\ has nearly $100\%$ correct rate. Both TRIEST an SJ-tree have throughput less than $2\times 10^5$ edges per second, much lower than the update speed of GSS, and high update speed is important in high speed streams.
\nop{The average throughput of SJ-tree is less than $0.187$M edges per second and falls greatly as window size enlarges. Our work outperforms it in throughput.}

\postfig
\nop{
\presub
\subsection{Summary of Experimental Studies}\label{summary}
\postsub
After extensive experiments of \fname\ and TCM on three data sets, we can make following conclusions:

1)\fname\ can perform  better than the state-of-the-art by orders of magnitudes in supporting basic query primitives and compound queries.

\nop{2)\fname\ is more resistent to the uneven distribution of node degrees by introducing fingerprint and mapping each edge into more than one buckets. So \fname\  can be a better solution to support most kinds of queries on very large graph streams.}

2)The buffer size of \fname\ is very small. The memory and time cost of \fname are barely influenced by it.

3) \fname has high update speed which is beyond $2$ million insertions per second. }

\section {Conclusion}

\label{sec:con}
Graph stream summarization is a problem rising in many fields. However, as far as we know, there are no prior work that can support all kinds of queries with high accuracy.
In this paper, we propose graph stream summarization data structure Graph Stream Sketch (GSS).
It has $O(|E|)$ memory usage where $|E|$ is the number of edges in the graph stream, and $O(1)$ update speed. 
It supports almost queries based on graphs and has accuracy which is higher than state-of-the-art by magnitudes.
Both mathematical analysis and experiment results confirm the superiority of our work.


\balance

\bibliographystyle{ieeetr}
\bibliography{reference.bib}  


\end{document}